\documentclass[aps,10pt,twocolumn,superscriptaddress,longbibliography]{revtex4-2}

\usepackage{amsmath,amsfonts,amssymb,amsthm}
\usepackage{mathtools}
\usepackage{physics}
\usepackage{braket}
\usepackage{mathrsfs}
\usepackage{graphicx}
\usepackage{booktabs}
\usepackage{xcolor}
\usepackage{enumitem}
\usepackage{fancyhdr}

\usepackage{algorithm}
\usepackage{algpseudocode}

\usepackage{tikz}
\usetikzlibrary{arrows.meta}
\usepackage{quantikz}
\makeatletter
\renewcommand{\ALG@name}{ALGORITHM}
\makeatother

\usepackage[colorlinks]{hyperref}
\usepackage[capitalize]{cleveref}

\newtheorem{theorem}{Theorem}
\newtheorem{lemma}{Lemma}
\newtheorem{corollary}{Corollary}
\newtheorem{definition}{Definition}
\newtheorem{proposition}{Proposition}
\newtheorem{problem}{Problem}

\begin{document}

\title{Efficient Quantum Algorithm for Mixed-State Quantum Geometry}
\author{Chenyang Wang}
\affiliation{Center on Frontiers of Computing Studies, Peking University, Beijing 100871, China}
\affiliation{School of Physics, Peking University, Beijing 100871, China}
\author{Qi Zhao}
\email{zhaoqi@cs.hku.hk}
\affiliation{QICI Quantum Information and Computation Initiative, School of Computing and Data Science,
The University of Hong Kong, Pokfulam Road, Hong Kong}
\author{Xiao Yuan}
\email{xiaoyuan@pku.edu.cn}
\affiliation{Center on Frontiers of Computing Studies, Peking University, Beijing 100871, China}
\affiliation{School of Computer Science, Peking University, Beijing 100871, China}
\author{Xiaoxia Cai}
\email{xxcai@ihep.ac.cn}
\affiliation{Institute of High Energy Physics, Chinese Academy of Sciences, Beijing 100049, China}

\date{\today}

\begin{abstract}

Mixed-state quantum geometry plays a fundamental role in quantum metrology, multiparameter estimation, and the characterization of quantum many-body systems. Yet, unlike its pure-state counterpart, accessing mixed-state geometry requires solving an ill-conditioned Lyapunov equation for the symmetric logarithmic derivative (SLD), posing a major challenge for resource-efficient quantum computation. Here we address the challenge and develop a novel quantum algorithmic framework for efficiently accessing the quantum Fisher information matrix (QFIM) and quantum geometric tensor (QGT) of mixed states.
Our key insight is to uncover a previously unexplored connection between the positive semidefinite structure of density matrices and exponentially convergent Gauss–Laguerre quadrature, enabling an efficient integral-based construction of the SLD.
A rigorous complexity analysis further establishes the resource efficiency of our algorithm, demonstrating logarithmic precision dependence and double-logarithmic ancillary-qubit overhead at fixed conditioning.
Building on this construction, we achieve near-Heisenberg precision scaling for diagonal QFI estimation without quantum phase estimation and construct unified block encodings of the full QFIM and QGT under a bounded spectral-gap assumption. A complementary regularized algorithm extends efficient diagonal estimation to exponentially small spectral gaps under a bounded-response condition.
Our results establish a resource-efficient route to coherent mixed-state quantum geometry, enabling its direct integration into quantum linear-algebra algorithms without classical reconstruction and opening new possibilities for quantum-enhanced multiparameter metrology.

\end{abstract}

\maketitle

\section{Introduction}\label{sec:introduction}

Computing the quantum Fisher information (QFI) or the quantum geometric tensor (QGT) of an unknown mixed quantum state is a task that sits at the intersection of quantum metrology, many-body physics, and quantum algorithm design~\cite{helstromQuantumDetectionEstimation1969,holevoProbabilisticStatisticalAspects2011,braunsteinStatisticalDistanceGeometry1994,liuQuantumFisherInformation2020,guFidelityApproachQuantum2010,houLocalGeometryQuantum2024,wangQuantumGeometricTensor2025}. Operationally, the QFI quantifies the sensitivity of a state as a probe for a given parameter and serves as a witness of multipartite entanglement~\cite{tothQuantumMetrologyQuantum2014,haukeMeasuringMultipartiteEntanglement2016}, while the full QFI matrix (QFIM) supplies the Cram\'er--Rao matrix bound. In multiparameter problems this bound need not be jointly attainable: compatibility conditions and tighter operational bounds, notably the Holevo bound, enter the ultimate precision limit~\cite{demkowiczBeyondQFI2020}. The QGT, whose real part is the Bures metric on the manifold of parameterized quantum states and whose imaginary part encodes the Uhlmann curvature up to convention-dependent normalization~\cite{wangQuantumGeometricTensor2025,wangxiaoguangMixedStateBerry2025}, captures this compatibility structure. For pure states, the QGT and QFIM admit closed-form expressions in terms of overlaps of the wavefunction and its derivatives. For mixed states, however, no comparable shortcut exists: these two quantities are defined through the symmetric logarithmic derivative (SLD), which is itself defined only implicitly by a Lyapunov equation, \(\frac{1}{2}(\rho \mathcal{L}_j + \mathcal{L}_j\rho) = \partial_j \rho\)~\cite{helstromQuantumDetectionEstimation1969,holevoProbabilisticStatisticalAspects2011}. Therefore, accessing mixed-state quantum geometry reduces to a support-sensitive matrix inversion problem.

This paper provides an efficient quantum algorithm for this task. The key insight is that the SLD equation is not an arbitrary linear system: it carries the specific structure of a density matrix and its derivative---the positive semidefiniteness of \(\rho\) and the absence of a kernel--kernel component in \(\partial_j\rho\) under local rank preservation. Crucially, we show how this structure converts the inversion problem into an integral representation naturally matched to exponentially convergent Gauss--Laguerre quadrature~\cite{davis1984methods,szego1939orthogonal}. We work within the block-encoding framework, assuming a block encoding of the probe state \(\rho\)---obtained, for instance, from a purification oracle \(O_\rho\)---and, for each parameter direction \(x_j\), a block encoding \(U_{\partial_j\rho}\) of the derivative \(\partial_j\rho\). We quantify the resource requirements through oracle query complexity and explicitly track the ancillary-qubit overhead. The complexity bounds are controlled by conditioning parameters that capture the spectral structure of \(\rho\) and the alignment of its derivative with poorly conditioned spectral directions.

Prior approaches to accessing quantum geometry use several distinct access models. Variational and overlap-based methods estimate QFI, QGT, or their approximations and bounds through repeated circuit evaluations~\cite{Stokes2020,Gacon2021,Beckey2022,tanPurityMinimization2021,Halla2025,GomezLurbe2025}.
Randomized-measurement methods instead extract QFI estimates or lower bounds from measurements on independent copies of the state and classical postprocessing~\cite{Rath2021,Vitale2024}.
Krylov shadow tomography provides a hierarchy of non-polynomial lower bounds whose highest member equals the QFI~\cite{ZhangTong2025}; subsequent work establishes exponential convergence with the Krylov order and low-order exactness for low-rank states~\cite{WangZhang2026Krylov}.
This exactness removes the gap between the bound and the QFI, while statistical uncertainty and the increasing measurement and classical postprocessing costs at higher orders remain relevant. These methods return classical QFI estimates from measurements on independent copies, complementing our framework, which also constructs block encodings of the SLD, QGT, and QFIM for subsequent quantum processing.
Ref.~\cite{wojtowiczLyapunovTensorNetwork2025} use the same Lyapunov-integral representation with classical tensor-network time evolution and numerical integration, whose computational cost depends on the required integration time and the bond dimensions needed to represent the state and evolving operators.

The closest coherent-access estimation algorithms are those of Ref.~\cite{ZhangYuan2025} and Ref.~\cite{Zhang2026thesis}. The former estimates fidelity susceptibility, equivalently one quarter of the QFI, for a non-degenerate pure ground-state family, using ground-state preparation and block encodings of the shifted Hamiltonian and its parameter derivative. The latter considers mixed-state single-parameter QFI using purified access and generator block encodings under unitary encoding. Table~\ref{tab:qfim_prior_work} separates these input assumptions from the associated query bounds. Our derivative-oracle interface covers general smooth, locally rank-preserving parameterizations; a common-input specialization to unitary encoding is discussed below the table.

The present work extends purified-access QFI algorithms from a single generator-induced parameter direction to general smooth, locally rank-preserving parameterizations supplied through derivative block encodings. When the smallest nonzero eigenvalue of the state is inverse-polynomially bounded, the algorithm retains near-Heisenberg precision scaling for diagonal QFI estimation. A complementary regularized construction replaces this spectral assumption with a bound on the parameter response, allowing polynomial-resource estimation for structured families whose smallest nonzero eigenvalue may be exponentially small. Most importantly, it provides, to our knowledge, the first explicit unified block encoding of the full mixed-state QGT and QFIM.

The key step in this work is the Gauss--Laguerre quadrature reduction~\cite{davis1984methods}. On the support of the density matrix, the Lyapunov equation admits a Laplace-transform solution expressing the canonical SLD as an exponentially filtered derivative~\cite{parisQUANTUMESTIMATIONQUANTUM2009}. We discretize this integral representation with Gauss--Laguerre quadrature, obtaining an exponentially convergent finite-sum approximation whose convergence rate is controlled by the spectral gap (Theorem~\ref{theorem:quadrature-main}). Crucially, the compensation factors required for coherent LCU assembly do not overwhelm this convergence: their total coefficient norm grows only linearly with the quadrature order, as summarized in the same theorem. From this approximation we construct block encodings of the approximate SLDs with logarithmic dependence on the inverse target precision, while the number of additional ancilla qubits required scales only double-logarithmically with the inverse precision at fixed conditioning and input-oracle resources. Quantum algorithms for more general Lyapunov and Sylvester equations provide alternative matrix-solution block encodings~\cite{claytonDifferentiableQuantumComputing2024,sommaQuantumAlgorithmLinear2025}. Their applicability and resource requirements depend on the matrix assumptions and input oracles. Table~\ref{tab:sld_solver_comparison} compares the construction stage separately from QFI estimation, including output normalization and LCU index-register requirements.

Based on the SLD construction, we develop two complementary algorithmic capabilities. The first estimates diagonal QFI entries via amplitude estimation that deliberately avoids quantum phase estimation, preserving the ancilla efficiency of the SLD construction. The second, conceptually the most significant, yields a unified block encoding of the full quantum geometric tensor and QFIM. Once block-encoded, the entire matrix becomes a substrate for downstream quantum linear algebra: coherent access to the inverse QFIM entering the SLD Cram\'er--Rao matrix bound, eigenvalue processing for identifying candidate high-information parameter directions, and threshold-based detection of multiparameter incompatibility from the antisymmetric imaginary part of the QGT, without first reconstructing its individual matrix elements classically.

These algorithms operate under two alternative promises on the probe state, reflecting different physical regimes. The first, which we call the bounded-gap promise, assumes the minimal nonzero eigenvalue of the density matrix is lower-bounded by a parameter, which is at least inverse polynomial in the number of qubits---a natural condition for well-conditioned probe states. Under this promise, the query complexity to $O_\rho$ for estimating QFI admits near-Heisenberg scaling with respect to the prescribed error tolerance $\epsilon$, and the number of ancilla qubits scales only double-logarithmically with $1/\epsilon$. The second, which we call the bounded-response promise, relaxes the gap condition and instead bounds the Frobenius norm of the product of the pseudoinverse of $\rho$ and its derivative, i.e. $\|\rho^{+} \partial_j \rho\|_F^2 \le M$ where $M$ is polynomial in the number of qubits. It applies, for example, to truncated commuting Gibbs ensembles with polynomial-size support and polynomially bounded inverse temperature, even when their smallest nonzero eigenvalue is exponentially small (Appendix~\ref{app:reg:physical}). Under this promise, we achieve polynomial query complexity in $M$ and $\epsilon^{-1}$, trading sharper $\epsilon$-scaling for applicability to a wider range of structured parameterized families.

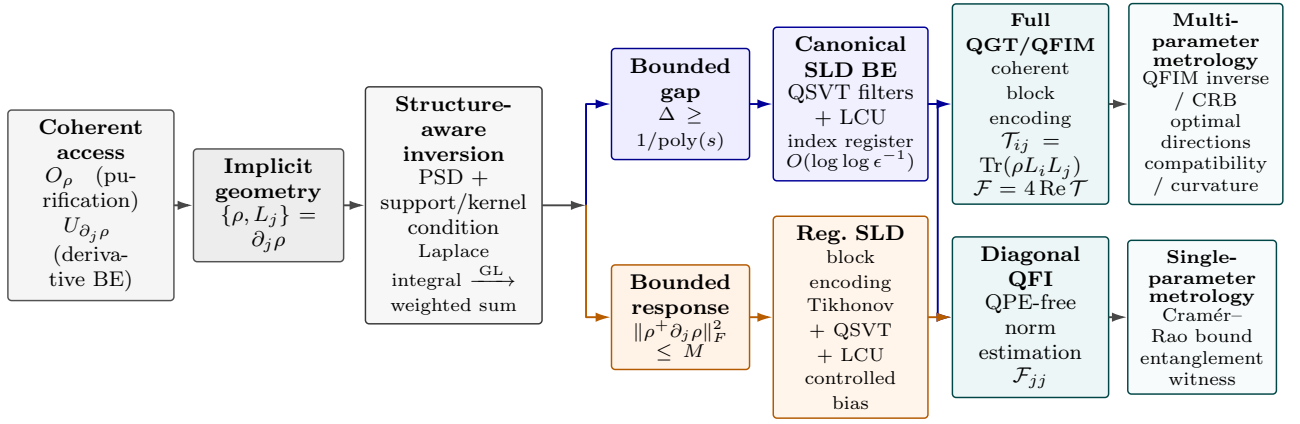
\begin{figure*}[t]
\centering

\begin{tikzpicture}[
  font=\footnotesize,
  >=Latex,
  line cap=round,
  line join=round,
  every node/.style={align=center},
  module/.style={
    draw=black!68, rounded corners=1.4pt, line width=.55pt,
    fill=white, minimum height=1.08cm, inner xsep=4.5pt, inner ysep=3.5pt
  },
  input/.style={module, fill=black!4, minimum width=2.05cm, text width=1.88cm},
  problem/.style={module, fill=black!7, minimum width=1.85cm, text width=1.67cm},
  pivot/.style={module, draw=black!78, fill=black!3, minimum width=2.20cm, text width=2.02cm,
                minimum height=1.38cm},
  gap/.style={module, draw=blue!55!black, fill=blue!7, minimum width=1.68cm, text width=1.51cm},
  response/.style={module, draw=orange!70!black, fill=orange!10, minimum width=1.68cm, text width=1.51cm},
  canonical/.style={module, draw=blue!60!black, fill=blue!6, minimum width=1.90cm, text width=1.73cm,
                    minimum height=1.28cm},
  regularized/.style={module, draw=orange!70!black, fill=orange!8, minimum width=1.90cm, text width=1.73cm,
                      minimum height=1.28cm},
  output/.style={module, draw=teal!55!black, fill=teal!7, minimum width=1.93cm, text width=1.75cm},
  application/.style={module, draw=teal!55!black, fill=teal!3, minimum width=1.82cm, text width=1.65cm,
                      minimum height=1.28cm, font=\scriptsize},
  matrixapplication/.style={application, minimum width=1.90cm, text width=1.72cm},
  flow/.style={-{Latex[length=2.0mm,width=1.25mm]}, line width=.72pt, draw=black!72},
  gapflow/.style={flow, draw=blue!60!black},
  responseflow/.style={flow, draw=orange!72!black},
]
  \node[input] (access) at (0.75,0.70) {\textbf{Coherent access}\\[-1pt]
    $O_\rho$ \; (purification)\\
    $U_{\partial_j\rho}$ \; (derivative BE)};
  \node[problem] (sld) at (3.10,0.70) {\textbf{Implicit geometry}\\[-1pt]
    $\{\rho,L_j\}=\partial_j\rho$};
  \node[pivot] (pivot) at (5.55,0.70) {\textbf{Structure-aware inversion}\\[-1pt]
    PSD $+$ support/kernel condition\\[-1pt]
    {\scriptsize Laplace integral $\xrightarrow{\,\mathrm{GL}\,}$ weighted sum}};

  \node[gap] (gap) at (8.55,2.04) {\textbf{Bounded gap}\\[-1pt]
    {\scriptsize $\Delta\geq1/\mathrm{poly}(s)$}};
  \node[response] (response) at (8.55,-0.78) {\textbf{Bounded response}\\[-1pt]
    {\scriptsize $\|\rho^+\partial_j\rho\|_F^2$}\\[-2pt]
    {\scriptsize $\leq M$}};
  \node[canonical] (canonical) at (10.80,2.04) {\textbf{Canonical SLD BE}\\[-1pt]
    QSVT filters $+$ LCU\\[-1pt]
    {\scriptsize index register}\\[-2pt]
    {\scriptsize $O(\log\log\epsilon^{-1})$}};
  \node[regularized] (regularized) at (10.80,-0.78) {\textbf{Reg.\ SLD}\\[-1pt]
    {\scriptsize block encoding}\\[-1pt]
    {\scriptsize Tikhonov $+$ QSVT $+$ LCU}\\[-1pt]
    {\scriptsize controlled bias}};

  \node[output] (qgt) at (13.18,2.04) {{\scriptsize\textbf{Full QGT/QFIM}}\\[-1pt]
    {\scriptsize coherent block encoding}\\[-1pt]
    $\mathcal T_{ij}=\operatorname{Tr}(\rho L_iL_j)$\\[-1pt]
    $\mathcal F=4\operatorname{Re}\mathcal T$};
  \node[output] (qfi) at (13.18,-0.78) {\textbf{Diagonal QFI}\\[-1pt]
    QPE-free norm estimation\\[-1pt]
    $\mathcal F_{jj}$};
  \node[matrixapplication] (apps) at (15.50,2.04) {\textbf{Multi-parameter}\\[-1pt]
    \textbf{metrology}\\[-1pt]
    QFIM inverse / CRB\\
    optimal directions\\
    compatibility / curvature};
  \node[application] (scalarapps) at (15.45,-0.78) {\textbf{Single-parameter}\\[-1pt]
    \textbf{metrology}\\[-1pt]
    Cram\'er--Rao bound\\
    entanglement witness};

  \draw[flow] (access) -- (sld);
  \draw[flow] (sld) -- (pivot);
  \coordinate (split) at (7.30,0.70);
  \draw[flow] (pivot.east) -- (split);
  \draw[gapflow] (split) |- (gap.west);
  \draw[responseflow] (split) |- (response.west);
  \draw[gapflow] (gap) -- (canonical);
  \draw[responseflow] (response) -- (regularized);

  \draw[gapflow] (canonical.east) -- (qgt.west);
  \draw[gapflow] (canonical.east) -- ++(.12,0) |- (qfi.west);
  \draw[responseflow] (regularized.east) -- (qfi.west);
  \draw[flow] (qgt) -- (apps);
  \draw[flow] (qfi) -- (scalarapps);
\end{tikzpicture}
\caption{Overview of the algorithmic pipeline. Starting from purified access to the probe state $\rho$ and block-encoding access to its parameter derivatives $\partial_j\rho$ (Sec.~\ref{sec:access-model}), the Lyapunov equation defining the SLD is converted through its Laplace integral representation into a Gauss--Laguerre quadrature sum---an exponentially convergent finite combination of filtered derivatives (Sec.~\ref{sec:gapped-quadrature-outline}). Block-encoding the quadrature terms yields the SLD as an algorithmic primitive with a single ancilla register whose width grows only double-logarithmically with the target precision (Sec.~\ref{algorithm-design-for-solving-the-lyapunov-equation}). Two capabilities are built on this primitive: near-Heisenberg estimation of diagonal QFI entries via QPE-free amplitude estimation, and a single block encoding of the full QGT and QFIM (Sec.~\ref{sec:accessing-qfim}), which becomes a substrate for quantum linear algebra---coherent access to the inverse QFIM entering the SLD Cram\'er--Rao matrix bound, eigenvalue processing for identifying candidate high-information parameter directions, and threshold-based detection of multiparameter incompatibility from the antisymmetric imaginary part of the QGT. Two alternative conditioning promises, the bounded-gap promise ($\Delta\ge 1/\operatorname{poly}(s)$) and the bounded-response promise ($M=\mathcal{O}(\operatorname{poly}(s))$), extend the framework across physical regimes.}
\label{fig:pipeline-overview}
\end{figure*}

The rest of this paper is organized as follows. Section~\ref{sec:preliminaries} formulates the problem and access model. Section~\ref{sec:main-results} introduces the algorithmic toolkit and states the main results. Section~\ref{applying-the-idea-of-gaussian-laguerre-quadrature-to-quantum-algorithm-design} develops the Gauss--Laguerre quadrature reduction and its block-encoding realization under both promise regimes. Section~\ref{sec:accessing-qfim} presents the algorithms for QFI estimation and full QGT/QFIM block encoding, and Section~\ref{sec:discussion} discusses the scope and limitations. The overall pipeline is summarized in Fig.~\ref{fig:pipeline-overview}.

\section{Problem Formulation and Access Model}\label{sec:preliminaries}

\subsection{Symmetric Logarithmic Derivative, QFIM, and QGT}
\label{sec:preliminaries-sld-qfim-qgt}
Let \(\rho(\boldsymbol{x})\) be a family of density operators parameterized by \(\boldsymbol{x} \in \mathcal{X} \subset \mathbb{R}^m\) (\(m \ge 1\)), twice continuously differentiable in \(\boldsymbol{x}\). Let the spectral decomposition of \(\rho\) be
\begin{equation}
  \rho(\boldsymbol{x}) = \sum_{n=1}^N p_n(\boldsymbol{x}) \ketbra{\phi_n(\boldsymbol{x})}{\phi_n(\boldsymbol{x})},
  \label{eq:spectral-decomposition-of-rho}
\end{equation}
with \(p_n(\boldsymbol{x}) > 0\) for \(n \in \{1,\dots,r\}\) and \(p_n(\boldsymbol{x}) = 0\) for \(n \in \{r+1,\dots,N\}\), where \(r = \mathrm{rank}(\rho)\) and \(N = \dim\mathcal{H}\) (taken as \(N=2^s\) where \(s\) is a positive integer, by embedding into a larger Hilbert space if required). The support and the kernel of \(\rho\) are denoted by \(\mathcal{S} = \mathrm{span}\{\ket{\phi_n}: p_n>0\}\) and \(\mathcal{K} = \mathrm{span}\{\ket{\phi_n}: p_n=0\}\). Denote the support and kernel projectors by \(P_{\mathcal{S}}\) and \(P_{\mathcal{K}}\) respectively.

The symmetric logarithmic derivative (SLD) associated with the parameter \(x_j\) is defined implicitly by the Lyapunov equation
\begin{equation}
  \{ \rho(\boldsymbol{x}), L_j \} = \partial_{j}\rho(\boldsymbol{x}),
  \label{eq:definition-of-sld}
\end{equation}
where \(\partial_j \equiv \partial/\partial x_j\). Eq.~\eqref{eq:definition-of-sld} absorbs the conventional factor \(1/2\) into \(L_j\) (i.e., \(L_j = \mathcal{L}_j/2\) in the standard normalization~\cite{helstromQuantumDetectionEstimation1969,holevoProbabilisticStatisticalAspects2011}), which simplifies the quantum algorithmic treatment.
The SLD organizes the two central geometric objects of this paper. The quantum Fisher information matrix (QFIM) is the \(m\times m\) real symmetric positive-semidefinite matrix
\begin{equation}
  \mathcal{F}_{jk}(\boldsymbol{x}) = 2\,\mathrm{Tr}\!\left[\rho(\boldsymbol{x})\{L_j, L_k\}\right] = 4\,\Re\!\left(\mathrm{Tr}\!\left[\rho(\boldsymbol{x}) L_j L_k\right]\right),
  \label{eq:definition-of-qfim}
\end{equation}
whose diagonal entry \(\mathcal{F}_{jj}\) is the quantum Fisher information for the parameter \(x_j\), governing the optimal sensing sensitivity via the quantum Cram\'er--Rao bound \(\mathrm{Var}_{\rho}[\hat{x}_j] \ge 1/(\nu\mathcal{F}_{jj})\), where \(\hat{x}_j\) is any unbiased estimator and \(\nu\) the number of independent measurements~\cite{tothQuantumMetrologyQuantum2014}.

In the SLD-product convention, the quantum geometric tensor (QGT) for mixed states is~\cite{wangxiaoguangMixedStateBerry2025}
\begin{equation}
  \mathcal{T}_{jk} = \mathrm{Tr}\!\left[\rho L_j L_k\right].
  \label{eq:qgt-definition}
\end{equation}
With our normalization \(L_j=\mathcal{L}_j/2\), this is \(\mathcal{T}_{jk}=\tfrac14\mathrm{Tr}[\rho\mathcal{L}_j\mathcal{L}_k]\) in terms of the conventional SLD. Its real part is the symmetric Bures metric \(g_{jk}=\Re(\mathcal{T}_{jk})=\mathcal{F}_{jk}/4\), while its imaginary part gives the mean Uhlmann curvature \(\Omega_{jk}:=-2\,\Im(\mathcal{T}_{jk})\) in the convention adopted here~\cite{wangxiaoguangMixedStateBerry2025}. When \(\Im(\mathcal{T}_{jk}) \neq 0\), the parameters \(x_j\) and \(x_k\) cannot be simultaneously estimated with optimal precision. For pure states, the QGT reduces to the Provost--Vallee form~\cite{provostRiemannianStructureManifolds1980} \(\mathcal{T}_{jk} = \langle\partial_j\psi|\partial_k\psi\rangle - \langle\partial_j\psi|\psi\rangle\langle\psi|\partial_k\psi\rangle\) and the SLD takes the closed form \(L_j = \ket{\partial_j\psi}\bra{\psi} + \ket{\psi}\bra{\partial_j\psi}\). For mixed states, no comparably direct formula exists---this is the fundamental difficulty that the present algorithms address.

Note that the solution to Eq.~\eqref{eq:definition-of-sld} is well-defined only if the kernel condition
\begin{equation}
  P_{\mathcal{K}}\, \partial_j \rho(\boldsymbol{x})\, P_{\mathcal{K}} = 0
  \label{eq:condition-for-sld-to-be-well-defined}
\end{equation}
is satisfied. Under this condition, the solution is unique on the support of \(\rho\); the kernel-kernel block is gauge freedom and is set to zero throughout this work. We assume that \(\rho(\boldsymbol{x})\) has constant rank in a neighborhood of the evaluation point \(\boldsymbol{x}_0\) (which we refer to as local rank preservation). This condition implies Eq.~\eqref{eq:condition-for-sld-to-be-well-defined}---in particular, \(\partial_j p_n(\boldsymbol{x}_0)=0\) for all kernel eigenvalues (\(p_n(\boldsymbol{x}_0)=0\))---and guarantees that the SLD is well defined on the entire neighborhood. The derivation is given in Appendix~\ref{appendix:promises-imply-local-rank-preservation}. Under the same condition the kernel Hessian contribution to the Bures metric vanishes~\cite{safranekDiscontinuitiesQuantumFisher2017}, so that \(g_{jk} = \mathcal{F}_{jk}/4\) holds exactly. The algorithm therefore targets the canonical SLD-based QFIM in the locally rank-preserving regime; rank-changing points, where the Bures metric and the SLD-based QFIM diverge, lie outside the scope of the present work, and details are discussed in Appendix~\ref{appendix:bures-discontinuities}.

For algorithmic construction we require the spectral forms of the SLD and QFIM. Under the canonical gauge choice, the solution to Eq.~\eqref{eq:definition-of-sld} is given by
\begin{equation}
  L_j(\boldsymbol{x}) = \sum_{\substack{m,n=1 \\ p_m + p_n > 0}}^{N} \frac{\langle \phi_m | \partial_j \rho(\boldsymbol{x}) | \phi_n \rangle}{p_m(\boldsymbol{x}) + p_n(\boldsymbol{x})} \ketbra{\phi_m(\boldsymbol{x})}{\phi_n(\boldsymbol{x})}.
  \label{eq:sld-spectral-form}
\end{equation}
Substituting into Eq.~\eqref{eq:definition-of-qfim} and Eq.~\eqref{eq:qgt-definition} yields the QFIM and QGT spectral formula
\begin{equation}
  \mathcal{F}_{jk}(\boldsymbol{x}) = 2 \sum_{\substack{m,n=1 \\ p_m + p_n > 0}}^{N} \frac{\Re\bigl( \langle \phi_m | \partial_j \rho(\boldsymbol{x}) | \phi_n \rangle\! \langle \phi_n | \partial_k \rho(\boldsymbol{x}) | \phi_m \rangle \bigr)}{p_m(\boldsymbol{x}) + p_n(\boldsymbol{x})}.
  \label{eq:qfim-spectral-formula}
\end{equation}

\begin{equation}
  \mathcal{T}_{jk}(\boldsymbol{x}) = \sum_{\substack{m,n=1 \\ p_m + p_n > 0}}^{N} \frac{p_m(\boldsymbol{x})\,\langle \phi_m | \partial_j \rho(\boldsymbol{x}) | \phi_n \rangle\! \langle \phi_n | \partial_k \rho(\boldsymbol{x}) | \phi_m \rangle}{\bigl(p_m(\boldsymbol{x}) + p_n(\boldsymbol{x})\bigr)^2},
  \label{eq:qgt-spectral-formula}
\end{equation}

The analytical solution of the Lyapunov equation can also be expressed as the Lyapunov integral~\cite{parisQUANTUMESTIMATIONQUANTUM2009,liuQuantumFisherInformation2016}
\begin{equation}
  L_j = \int_0^\infty e^{-t\rho}(\partial_j\rho)\,e^{-t\rho}\,dt,
  \label{eq:integral-expression-for-sld}
\end{equation}
obtained by vectorizing the equation by row stacking to \(\mathcal{L}_\rho\ket{L_j} = \ket{\partial_j\rho}\) with \(\mathcal{L}_\rho = \rho\otimes I + I\otimes\rho^T\) and applying the Laplace representation of the pseudoinverse.  The integral converges under the local rank-preservation condition, which ensures that \(\ket{\partial_j\rho}\) has no component in the kernel of~\(\mathcal{L}_\rho\).  This representation is the analytical starting point for the Gauss--Laguerre quadrature construction and its block-encoding realization developed in Section~\ref{applying-the-idea-of-gaussian-laguerre-quadrature-to-quantum-algorithm-design}.

The spectral expressions also reveal the intrinsic difficulty of the SLD inversion and QFI estimation, which can be ill-posed when the minimum nonzero eigenvalue of \(\rho\) (denoted as \(p^+_{\min}\)) is exponentially small.  Such regimes are not pathological: for a finite-temperature \(s\)-qubit Gibbs state \(\rho_\beta = e^{-\beta H}/Z\) with a local Hamiltonian \(H\), the spectral width is \(\mathcal{O}(s)\) and the smallest Boltzmann weight scales as \(e^{-\beta \mathcal{O}(s)}/Z\), giving \(p^+_{\min} = e^{-\Theta(s)}\).  A standard remedy for ill-posed inverse problems is Tikhonov regularization~\cite{tikhonovSolutionsIllPosedProblems1977} (ridge regression): add a constant shift to the singular operator.  Introducing a parameter \(\theta>0\), we solve the regularized Lyapunov equation
\begin{equation}
  (\rho(\boldsymbol{x}) + \tfrac{\theta}{2} I) L_j^{(\theta)}
  + L_j^{(\theta)} (\rho(\boldsymbol{x}) + \tfrac{\theta}{2} I)
  = \partial_j \rho(\boldsymbol{x}),
  \label{eq:reg:sld_def}
\end{equation}
or equivalently \((\mathcal{L}_\rho + \theta I)\ket{L_j^{(\theta)}} = \ket{\partial_j\rho}\).  The spectrum of \(\mathcal{L}_\rho + \theta I\) is bounded below by \(\theta\), making the operator invertible.  The spectral solution is
\begin{equation}
  L_j^{(\theta)} = \sum_{m,n}
  \frac{\langle \phi_m | \partial_j \rho | \phi_n \rangle}
       {p_m + p_n + \theta} \,
  \ket{\phi_m}\bra{\phi_n},
  \label{eq:reg:L_theta}
\end{equation}
with every denominator protected by at least~\(\theta\).  Equivalently, vectorizing and applying the Laplace transform gives the regularized integral representation
\begin{equation}
\begin{aligned}
  L_j^{(\theta)} &= \int_0^\infty e^{-t(\rho + \frac{\theta}{2}I)} (\partial_j\rho)\, e^{-t(\rho + \frac{\theta}{2}I)}\, dt \\
  &= \frac{1}{\theta} \int_0^\infty e^{-x}\, e^{-\frac{x}{\theta}\rho} (\partial_j\rho)\, e^{-\frac{x}{\theta}\rho}\, dx.
\end{aligned}
  \label{eq:reg:integral}
\end{equation}
Unlike the canonical integral~\eqref{eq:integral-expression-for-sld}, the regularized version converges unconditionally---the factor \(e^{-t\theta}\) provides exponential damping on \emph{all} spectral sectors, including the kernel---and requires no rank-preservation condition.

Substituting into the QFIM trace formula gives the regularized QFIM entries
\begin{equation}
  \mathcal{F}_{jk}^{(\theta)}
  = 2\sum_{m,n}
    \frac{\Re\bigl(\langle \phi_m | \partial_j \rho | \phi_n \rangle
                  \langle \phi_n | \partial_k \rho | \phi_m \rangle\bigr)
          (p_m + p_n)}
         {(p_m + p_n + \theta)^2}.
  \label{eq:reg:F_jk}
\end{equation}

Or specifically, the regularized QFI for parameter \(x_j\) is
\begin{equation}
  \mathcal{F}_{jj}^{(\theta)}
  = 2\sum_{m,n}
    \frac{|\langle \phi_m | \partial_j \rho | \phi_n \rangle|^2 (p_m + p_n)}
         {(p_m + p_n + \theta)^2}.
  \label{eq:reg:F_theta}
\end{equation}

We now bound the bias introduced by the regularization.
Expanding \((p_m+p_n)/(p_m+p_n+\theta)^2 = 1/(p_m+p_n) - 2\theta/(p_m+p_n)^2 + \mathcal{O}(\theta^2/(p_m+p_n)^3)\) and substituting into the spectral expressions for \(\mathcal{F}_{jj}^{(\theta)}\) and \(\mathcal{F}_{jj}\) yields the first-order relation \(\mathcal{F}_{jj}^{(\theta)} - \mathcal{F}_{jj} = -4\theta\,\Xi_{jj} + \mathcal{O}(\theta^2)\), where
\begin{equation}
  \Xi_{jj} := \sum_{\substack{m,n \\ p_m+p_n>0}}
  \frac{|\langle\phi_m|\partial_j\rho|\phi_n\rangle|^2}
       {(p_m + p_n)^2}.
  \label{eq:reg:Xi_def}
\end{equation}
To bound \(\Xi_{jj}\), we use the elementary inequality
\(1/(p_m+p_n)^2 \le \mathbf{1}_{p_m>0}/p_m^2 + \mathbf{1}_{p_n>0}/p_n^2\) for \(p_m,p_n\ge 0\) with \(p_m+p_n>0\).  Summing over \(m,n\) and using the
Hermiticity of \(\partial_j\rho\) (which gives
\(|\langle\phi_m|\partial_j\rho|\phi_n\rangle|
 = |\langle\phi_n|\partial_j\rho|\phi_m\rangle|\)) to symmetrize the
second term,
\begin{equation}
  \Xi_{jj} \le 2\sum_{\substack{m,n \\ p_m>0}}
  \frac{|\langle\phi_m|\partial_j\rho|\phi_n\rangle|^2}{p_m^2}
  = 2\,\|\rho^+ \partial_j \rho\|_F^2,
  \label{eq:reg:Xi_bound}
\end{equation}
Where \(\|A\|_F^2 = \mathrm{Tr}[A^\dagger A]\) denotes the square of the Frobenius norm.
Defining the bounded-response parameter
\begin{equation}
  M := \max_{j\in[m]} \|\rho^{+} \partial_j \rho\|_F^2,
  \label{eq:reg:M_def}
\end{equation}
we obtain the core inequality for the regularization bias as
\begin{equation}
  |\mathcal{F}_{jj}^{(\theta)} - \mathcal{F}_{jj}| \le 8 M \theta.
  \label{eq:reg:bias_bound}
\end{equation}

Here \(\rho^{+}\) denotes the Moore--Penrose pseudoinverse of \(\rho\). The quantity \(M\) serves as a metric of the alignment between the derivative and the support of the density matrix, which emerges naturally from this analysis as the single parameter controlling the regularization error.  It penalizes overlap between \(\partial_j\rho\) and the soft spectral tail of~\(\rho\): matrix elements along small eigenvalue directions are weighted by \(p_m^{-2}\).  Bounded \(M\) indicates that the derivative is well-aligned with the support of \(\rho\) and avoids ill-conditioned directions, even when the spectral gap is exponentially small.  Since \(\partial_j\rho\) is the linear response of the state to a variation of~\(x_j\), we refer to the condition \(M = \mathcal{O}(\operatorname{poly}(s))\) as the \textbf{bounded-response promise}, formalized in Section~\ref{sec:access-model}.

The regularized framework is not restricted to the regime of an exponentially small spectral gap.  When the bounded-gap promise holds (\(\Delta \ge 1/\operatorname{poly}(s)\)), the trivial bound
\(\|\rho^+\partial_j\rho\|_F \le \|\rho^+\|\,\|\partial_j\rho\|_F = \|\partial_j\rho\|_F/\Delta\) implies \(M = \mathcal{O}(\|\partial_j\rho\|_F^2/\Delta^2) = \mathcal{O}(\operatorname{poly}(s))\); that is, the bounded-gap promise is a strictly stronger condition that implies the bounded-response promise.  The regularized algorithm serves as a universal fallback: it applies whenever the bounded-response promise holds---including the gapped regime.  The selection criterion is discussed in Section~\ref{sec:discussion}.


\subsection{Algorithmic Access Model and Task}\label{sec:access-model}\label{sec:bounded-response}

We now state the quantum input model. The algorithm is formulated at a fixed parameter point \(\boldsymbol{x}_0\). Let the system register \(S\) contain \(s\) qubits and let \(A\) denote an ancilla register of \(a \ge \log_2 r\) qubits. We assume purified access to \(\rho(\boldsymbol{x}_0)\), namely an oracle \(O_{\rho(\boldsymbol{x}_0)}\) (abbreviated \(O_\rho\) when the parameter point is clear from context) such that
\begin{equation}
  O_{\rho(\boldsymbol{x}_0)} \ket{0}_S \ket{0}_A = \sum_{n=1}^{r} \sqrt{p_n} \ket{\phi_n}_S \ket{\pi_n}_A =: \ket{\rho}_{SA},
  \label{eq:purification-oracle-intro}
\end{equation}
where \(\{\ket{\pi_n}\}\) is an orthonormal basis in the ancilla space. For each parameter direction \(x_j\) we further assume an \((\alpha_j,m_j,0)\)-block-encoding \(U_{\partial_j \rho}\) of \(\frac{\partial}{\partial x_j} \rho(\boldsymbol{x})|_{\boldsymbol{x}=\boldsymbol{x}_0}\) (or abbreviated as \(\partial_j \rho\) in the rest of this paper)
\begin{equation}
    \partial_j \rho = \alpha_j (\langle 0|^{\otimes m_j} \otimes I_S) U_{\partial_j \rho} (|0\rangle^{\otimes m_j} \otimes I_S),
    \label{eq:derivative-block-encoding-intro}
\end{equation}
with \(\alpha_j \ge \|\partial_j \rho\|\).
We note that the purified-access assumption on the state itself provides a distinct alternative access model: the standard density-matrix-exponentiation pipeline~\cite{kimmelHamiltonianSimulationOptimal2017,gilyenQuantumSingularValue2019,gilyenImprovedQuantumAlgorithms2022} can in principle convert copies of \(\rho\) into an approximate block encoding at a constant normalization overhead, so that sampling access suffices for the \(\rho\)-portion of the construction; the derivative block encoding \(U_{\partial_j \rho}\), by contrast, does not follow from samples and remains an independent ingredient of the access model.
The derivative-access assumption is not intended to imply that \(U_{\partial_j\rho}\) follows automatically from the purification oracle.
Rather, it covers parameterized families for which \(\partial_j\rho\) has a known analytic representation in terms of elementary operators admitting efficient quantum access, with controlled block-encoding normalization and composition cost.
The cost of instantiating this oracle is model dependent and is separate from the oracle-query complexities reported below.
For instance, when the parameter enters through a known Hamiltonian generator ($\partial_j\rho = i[\rho, H_j]$), the block encoding of \(\partial_j\rho\) can be constructed from the block encoding of \(H_j\) and the purification oracle of \(\rho\).

Moreover, the complexity bounds are expressed in terms of the standard Frobenius norm $\|\partial_j\rho\|_F$, which quantifies the overall magnitude of the parameter response.
Unlike the subnormalization $\alpha_j$, which bounds the spectral norm and is determined by the block-encoding construction, the Frobenius norm is a property of the specific parameterized family and can be bounded analytically in the physical settings of interest.
In the unitary-parameter case, submultiplicativity of the Frobenius norm with respect to the spectral norm yields $\|\partial_j\rho\|_F \le 2\|H_j\|$.
More generally, $\|\partial_j\rho\|_F$ is treated as a known constant in the analysis.

For the conditioning promise of the input model, we consider two alternative assumptions. The first is a strong \textbf{bounded-gap} promise: there exists an open neighborhood \(\mathcal{X} \ni \boldsymbol{x}_0\) and a known parameter \(\Delta>0\) such that
\begin{equation}
    \inf_{\boldsymbol{x} \in \mathcal{X}} \min_{n:\, p_n(\boldsymbol{x})>0} p_n(\boldsymbol{x}) \ge \Delta,
    \qquad
    \Delta^{-1} = \mathcal{O}(\operatorname{poly}(s)).
    \label{eq:vartheta-intro}
\end{equation}
This promise controls the worst-case amplification of the inverse map underlying the SLD equation. Under this promise, a numerical lower bound on the smallest positive eigenvalue of \(\rho(\boldsymbol{x}_0)\) can be certified from the purification oracle \(O_\rho\). The purification state \(\ket{\rho}_{SA}\) has unit overlap with the support subspace \(\mathcal{S}\) of \(\rho\) (its Schmidt coefficients are precisely \(\sqrt{p_n}\)), so detecting the smallest positive eigenvalue reduces to probing the entire support subspace rather than a single eigenstate. Concretely, eigenstate filtering~\cite{linNearoptimalGroundState2020} combined with the thresholding primitive of Lemma~\ref{lemma:thresholding} yields a lower bound within a constant factor of \(\min_{p_n>0}p_n(\boldsymbol{x}_0)\). The construction and its preprocessing cost are detailed in Appendix~\ref{appendix:self-certifying-delta}. For simplicity of presentation, we henceforth treat \(\Delta\) as known.

The second promise is more instance-dependent and captures the effective conditioning of the physically generated \(\partial_j\rho\), called the \textbf{bounded-response} promise. We assume that there exists an open neighborhood \(\mathcal{X} \ni \boldsymbol{x}_0\) and a known parameter \(M\) such that
\begin{equation}
  \max_{j \in [m]} \sup_{\boldsymbol{x} \in \mathcal{X}}\left\|\rho(\boldsymbol{x})^{+} \, \partial_j \rho(\boldsymbol{x})\right\|_F^2 \le M,
  \qquad
  M = \mathcal{O}(\operatorname{poly}(s)).
  \label{eq:derivative-bound-intro}
\end{equation}
This bound measures how strongly the parameter derivative aligns with poorly conditioned spectral directions, and therefore serves as an effective conditioning parameter for the SLD inversion problem. As with an \textit{a priori} upper bound on the condition number in HHL-type quantum linear-system algorithms~\cite{harrowQuantumAlgorithmLinear2009}, \(M\) enters the access model as an input parameter rather than a quantity certified by the algorithm: unlike the numerical spectral-gap lower bound, which can be certified from \(O_\rho\) under the bounded-gap promise, no procedure is known to certify \(M\) from the oracles of the access model. For the structured families of Appendix~\ref{app:reg:physical} the bound \(M = \mathcal{O}(\operatorname{poly}(s))\) can be derived analytically from the model.
Both conditioning promises imply that the support structure is locally stable near the target point. More precisely, under either Eq.~\eqref{eq:vartheta-intro} or Eq.~\eqref{eq:derivative-bound-intro}, the rank of \(\rho(\boldsymbol{x})\) is constant on \(\mathcal{X}\). In particular, for every parameter direction \(j\), the kernel-kernel component of the derivative vanishes throughout \(\mathcal{X}\):
\begin{equation}
  P_{\mathcal K(\boldsymbol{x})}\, \partial_j \rho(\boldsymbol{x})\, P_{\mathcal K(\boldsymbol{x})} = 0,
  \qquad \forall \boldsymbol{x}\in \mathcal{X}, \forall j \in [m].
  \label{eq:kernel-condition-from-promises}
\end{equation}
Hence the SLD is well defined on \(\mathcal{X}\), and the algorithm never encounters the rank-changing regime in which the standard SLD-based QFIM is not the appropriate local geometric quantity. We defer the proof of this implication to Appendix~\ref{appendix:promises-imply-local-rank-preservation}.


Under this model the problem addressed by this paper is the following.

\begin{problem}[Mixed-state QFIM and QGT estimation]\label{prob:qfim-estimation}
Given:
\begin{enumerate}
  \item a parameter point \(\boldsymbol{x}_0 \in \mathcal{X}\);
  \item a unitary \(O_\rho\) preparing a purification \(\ket{\rho}_{SA}\) of \(\rho(\boldsymbol{x}_0)\);
  \item for each parameter direction \(j \in [m]\), an \((\alpha_j, m_j, 0)\)-block-encoding \(U_{\partial_j}\) of \(\partial_j\rho(\boldsymbol{x}_0)\);
  \item a lower bound \(\Delta\) on the minimum nonzero eigenvalue of \(\rho(\boldsymbol{x}_0)\) (or, alternatively, an effective conditioning parameter \(M\) as in Eq.~\eqref{eq:derivative-bound-intro});
  \item a target precision \(\epsilon \in (0,1)\) and a confidence parameter \(\vartheta \in (0,1)\);
\end{enumerate}
The target is to provide algorithmic access to the QFIM and QGT at \(\boldsymbol{x}_0\) through the following complementary capabilities:
\begin{enumerate}
  \item[(i)] additive estimates \(\hat{\mathcal{F}}_{jj}\) of diagonal QFIM entries satisfying \(|\hat{\mathcal{F}}_{jj} - \mathcal{F}_{jj}(\boldsymbol{x}_0)| \le \epsilon\) under both promises, with success probability at least \(1-\vartheta\);
  \item[(ii)]  a unitary block encoding of the full \(m \times m\) QGT \(\mathcal{T} = (\Tr[\rho L_j L_k])_{j,k=1}^m\) and of the QFIM \(\mathcal{F} = 4\,\Re(\mathcal{T})\) under the bounded-gap promise, whose subnormalization factor and ancilla count are controlled by the input parameters.
\end{enumerate}
The cost of each capability is measured by the number of queries to \(O_\rho\), \(O_\rho^\dagger\), \(U_{\partial_j}\), and \(U_{\partial_j}^\dagger\), as well as the number of ancilla qubits required, and is expressed in terms of parameters mentioned above.
\end{problem}

\section{Main Results}\label{sec:main-results}

The key resource guarantee comes from the Gauss--Laguerre approximation of the canonical SLD: exponential convergence in the quadrature order is accompanied by only linear growth of the LCU coefficient norm. QSVT implements the exponential filters, and block-encoding products and LCU combine them with the state derivatives~\cite{gilyenQuantumSingularValue2019}. Section~\ref{applying-the-idea-of-gaussian-laguerre-quadrature-to-quantum-algorithm-design} develops this construction; Appendix~\ref{app:implementation-tools} collects the required quantum primitives.

The following theorem summarizes diagonal QFI estimation under the bounded-gap and bounded-response promises, together with full QGT and QFIM block encodings under the bounded-gap promise.

\begin{theorem}[Main results, informal]\label{thm:informal-main}
Under the access model of Section~\ref{sec:access-model}, assume that the local rank-preservation condition in Eq.~\eqref{eq:kernel-condition-from-promises} holds for the relevant parameter directions. The following guarantees hold, with polylogarithmic factors suppressed in the query complexities.
\begin{enumerate}
    \item Under the bounded-gap promise, a diagonal entry \(\mathcal{F}_{jj}\) can be estimated to additive error \(\epsilon\), with success probability at least \(1-\vartheta\), using
    \[
        Q_\rho^{jj}=\widetilde{\mathcal{O}}\!\left(\frac{\alpha_j \|\partial_j \rho\|_F}{\epsilon \Delta^{7/2}}\right)
    \]
    purification-oracle queries and
    \[
        Q_{\partial_j}^{jj}=\widetilde{\mathcal{O}}\!\left(\frac{\alpha_j \|\partial_j \rho\|_F}{\epsilon \Delta^{5/2}}\right)
    \]
    derivative-oracle queries.

    \item Under the bounded-response promise, the same diagonal entry can be estimated to additive error \(\epsilon\), with success probability at least \(1-\vartheta\), using
    \[
        Q_\rho^{\mathrm{reg}}=\widetilde{\mathcal{O}}\!\left(\frac{\alpha_j M^{7/2}}{\epsilon^4}\right),\qquad
        Q_{\partial_j}^{\mathrm{reg}}=\widetilde{\mathcal{O}}\!\left(\frac{\alpha_j M^{3/2}}{\epsilon^2}\right)
    \]
    purification- and derivative-oracle queries, respectively.

    \item Under the bounded-gap construction, the full \(m\times m\) QGT admits an \((m\alpha^2,a_{\mathrm{tot}},\epsilon_{\mathrm{tot}})\)-block encoding, and the QFIM admits a \((4m\alpha^2,a_{\mathrm{tot}}+1,4\epsilon_{\mathrm{tot}})\)-block encoding, where
    \[
    a_{\mathrm{tot}} = \lceil\log_2 m\rceil + 2a_L + a + s,
    \]
    \[
    \epsilon_{\mathrm{tot}} = \Bigl(\sum_{i,j=1}^{m} \bigl[\alpha(\epsilon_i + \epsilon_j) + \epsilon_i\epsilon_j\bigr]^2\Bigr)^{1/2},
    \]
    and each \(\epsilon_j\) is the per-parameter SLD block-encoding error. In the uniform-error case \(\epsilon_j=\epsilon\), this reduces to \(\epsilon_{\mathrm{tot}}=m(2\alpha\epsilon+\mathcal{O}(\epsilon^2))\).
\end{enumerate}
Here \(\widetilde{\mathcal{O}}(\cdot)\) suppresses polylogarithmic factors in the parameters appearing in the corresponding bound. On nontrivial branches, resource bounds are understood to include the cost of at least one invocation of each required primitive; internal approximation tolerances may always be decreased to fixed admissible constants. The quantities \(a_L\) and \(\epsilon_j\) are, respectively, the largest ancilla count and the approximation error among the parameter-specific SLD block encodings.
\end{theorem}

These three guarantees separate the end-to-end capabilities from the mechanism used to realize them. Sections~\ref{sec:gapped-quadrature-outline} and~\ref{sec:regularized-quadrature-outline} derive finite filtered-derivative expansions of the canonical and regularized SLDs, and Section~\ref{algorithm-design-for-solving-the-lyapunov-equation} outlines their block-encoding realizations. Section~\ref{sec:accessing-qfim} then reduces diagonal QFI estimation to a norm-estimation problem under the two respective promises and, under the bounded-gap promise, constructs the full QGT/QFIM block encoding. Detailed gate counts, intermediate errors, and error-budget allocations are retained in the appendices.

Table~\ref{tab:qfim_prior_work} summarizes coherent-access algorithms for scalar QFI or fidelity-susceptibility estimation. Each row retains its own input oracles and conditioning assumption; a diagonal QFIM entry is itself a single-parameter QFI.

\begin{table*}[t]
\centering
\caption{Coherent-access algorithms for QFI and fidelity-susceptibility estimation to additive error $\epsilon$ at fixed success probability. Each query count refers to the indicated oracle, including its inverse and the controlled access required by the algorithm; $\widetilde{\mathcal{O}}$ suppresses polylogarithmic factors. The Hamiltonian gap $\Delta_H$ and the density-matrix gap $\Delta_\rho$ are different parameters. Native access models are retained, so the rows do not constitute a common-input ranking.}
\label{tab:qfim_prior_work}
\small
\setlength{\tabcolsep}{4pt}
\begin{tabular}{@{}p{2.25cm} p{4.65cm} p{5.45cm} p{3.15cm}@{}}
\toprule
\raggedright Work & \raggedright Task and access assumptions & \raggedright Oracle queries & \raggedright Main ingredients \tabularnewline
\midrule
\raggedright Zhang--Yuan~\cite{ZhangYuan2025}
& \raggedright Pure ground-state $\chi_F=\mathcal F/4$; state-preparation oracle $U_0$, shifted-Hamiltonian BE $U_H$, perturbation BE $U_I$; gap $\Delta_H$
& \raggedright $Q_H=\widetilde{\mathcal{O}}(\alpha_H\alpha_I^2/(\epsilon\Delta_H^3))$;\newline
$Q_I,Q_0=\widetilde{\mathcal{O}}(\alpha_I^2/(\epsilon\Delta_H^2))$
& \raggedright Resolvent; amplitude estimation \tabularnewline
\addlinespace
\raggedright Zhang~\cite{Zhang2026thesis}
& \raggedright Mixed-state QFI; purified access $O_\rho$, generator BE $U_A$; unitary encoding; gap $\Delta_\rho$
& \raggedright $Q_\rho=\widetilde{\mathcal{O}}(\alpha_A^2/(\epsilon\Delta_\rho^6))$;\newline
$Q_A=\widetilde{\mathcal{O}}(\alpha_A^2/(\epsilon\Delta_\rho^3))$
& \raggedright Rational kernel; amplitude estimation \tabularnewline
\addlinespace
\raggedright This work (bounded gap)
& \raggedright Mixed-state QFI; purified access $O_\rho$, derivative BE $U_{\partial_j\rho}$; gap $\Delta_\rho$
& \raggedright $Q_\rho=\widetilde{\mathcal{O}}(\alpha_j\|\partial_j\rho\|_F/(\epsilon\Delta_\rho^{7/2}))$;\newline
$Q_{\partial_j}=\widetilde{\mathcal{O}}(\alpha_j\|\partial_j\rho\|_F/(\epsilon\Delta_\rho^{5/2}))$
& \raggedright Gauss--Laguerre quadrature; QPE-free amplitude estimation \tabularnewline
\addlinespace
\raggedright This work (bounded response)
& \raggedright Mixed-state QFI; same oracles; $\|\rho^+\partial_j\rho\|_F^2\le M$; no gap lower bound
& \raggedright $Q_\rho=\widetilde{\mathcal{O}}(\alpha_j M^{7/2}/\epsilon^4)$;\newline
$Q_{\partial_j}=\widetilde{\mathcal{O}}(\alpha_j M^{3/2}/\epsilon^2)$
& \raggedright Tikhonov regularization; Gauss--Laguerre quadrature; QPE-free amplitude estimation \tabularnewline
\bottomrule
\end{tabular}
\par\smallskip\raggedright\footnotesize
\textit{Note.} BE denotes block encoding, and $\Delta_\rho$ is denoted by $\Delta$ in our theorems. The factors $\alpha_H$, $\alpha_I$, $\alpha_A$, and $\alpha_j$ are the normalizations of the supplied block encodings. Query counts exclude the cost of constructing these input oracles.
\end{table*}

Under unitary encoding, our algorithm can be instantiated with the same purification and generator oracles used in Ref.~\cite{Zhang2026thesis}. Let $\rho(x)=e^{-ixA}\rho_0e^{ixA}$, and suppose that $O_\rho$ and an $\alpha_A$-normalized block encoding $U_A$ are supplied. A two-term commutator construction implements $\partial_x\rho=-i[A,\rho]$ with normalization $2\alpha_A$, using a constant number of queries to each input oracle. Under the bounded-gap promise, applying Theorem~\ref{thm:diagonal-qfi-estimation} with $\|\partial_x\rho\|_F\le2\|A\|\le2\alpha_A$ gives
\begin{equation}
\begin{aligned}
Q_\rho^{\rm unitary}&=\widetilde{\mathcal{O}}\!\left(\frac{\alpha_A^2}{\epsilon\Delta_\rho^{7/2}}\right),\\
Q_A^{\rm unitary}&=\widetilde{\mathcal{O}}\!\left(\frac{\alpha_A^2}{\epsilon\Delta_\rho^{5/2}}\right).
\end{aligned}
\label{eq:unitary-common-input-queries}
\end{equation}
These counts include the purification queries needed to construct the derivative block encoding. With the same input normalization, the resulting bounds have smaller inverse-gap powers than those of Ref.~\cite{Zhang2026thesis} listed in Table~\ref{tab:qfim_prior_work}, while retaining near-Heisenberg precision scaling.

This specialization also illustrates the role of the Frobenius response norm in our general complexity bounds. Converting an operator-norm bound through $\|\partial_x\rho\|_F\le\sqrt{N}\|\partial_x\rho\|$ would introduce a dimension factor. The commutator structure instead gives $\|\partial_x\rho\|_F\le2\|A\|\|\rho\|_F\le2\|A\|$, using the normalization of the density matrix. Retaining the response norm until a probe family is specified therefore allows structure-specific analytical bounds to enter the resource estimate; tighter available bounds can further sharpen the specialization above.

\section{Quadrature Construction and Block-Encoding Realization of the Symmetric Logarithmic Derivative}\label{applying-the-idea-of-gaussian-laguerre-quadrature-to-quantum-algorithm-design}

The analytical reduction exploits the spectrum of the Lyapunov map instead of treating the SLD equation as an unstructured linear system. Both conditioning promises lead to a finite linear combination of two-sided exponential filters acting on \(\partial_j\rho\), but they place the conditioning cost in different parameters. We first derive the canonical construction under a support gap and then show how Tikhonov regularization produces a parallel expansion when that gap is too small. Appendix~\ref{appendix:laguerre-gl-background} collects the Laguerre quadrature background and the asymptotic analysis of its LCU coefficient norm.

\subsection{Bounded-gap quadrature}\label{sec:gapped-quadrature-outline}

Vectorizing the Lyapunov equation (by row stacking) gives
\begin{equation}
  \begin{aligned}
    \mathcal{L}_\rho&=\rho\otimes I+I\otimes\rho^T,\\
    \ket{L_j}&=\mathcal{L}_\rho^{+}\ket{\partial_j\rho}\\
    &=\int_0^\infty e^{-t\mathcal{L}_\rho}
      \ket{\partial_j\rho}\,dt.
  \end{aligned}
  \label{eq:main-vectorized-sld-integral}
\end{equation}
The kernel condition in Eq.~\eqref{eq:condition-for-sld-to-be-well-defined} implies \(\ket{\partial_j\rho}\in\operatorname{Ran}(\mathcal L_\rho)\), so only the support-restricted inverse is relevant. Under the bounded-gap promise, the nonzero spectrum of \(\mathcal L_\rho\) lies in \([\Delta,2]\). After the change of variables \(x=\Delta \cdot t\), define
\begin{equation}
  \mathcal M=\frac{\mathcal L_\rho-\Delta I}{\Delta},
  \qquad
  \operatorname{spec}\!\left(\mathcal M\big|_{\operatorname{Ran}(\mathcal L_\rho)}\right)
  \subseteq\left[0,\frac{2-\Delta}{\Delta}\right].
  \label{eq:main-rescaled-lyapunov-map}
\end{equation}
Equation~\eqref{eq:main-vectorized-sld-integral} then becomes
\begin{equation}
  \ket{L_j}=\frac{1}{\Delta}\int_0^\infty e^{-x}e^{-x\mathcal M}\ket{\partial_j\rho}\,dx.
  \label{eq:main-rescaled-sld-integral}
\end{equation}

Let \(\{\ket{\varphi_\ell}\}\) be an orthonormal eigenbasis of \(\mathcal M\) on \(\operatorname{Ran}(\mathcal L_\rho)\), with \(\mathcal M\ket{\varphi_\ell}=\mu_\ell\ket{\varphi_\ell}\). Expanding the derivative in this basis makes the scalar functions to be integrated explicit:
\begin{equation}
  \ket{L_j}
  =\frac{1}{\Delta}\sum_\ell\int_0^\infty
  e^{-x}e^{-\mu_\ell x}
  \braket{\varphi_\ell|\partial_j\rho}\,dx\,
  \ket{\varphi_\ell}.
  \label{eq:expression-for-sld-in-eigenbasis-of-M}
\end{equation}

This is precisely the form addressed by Gauss--Laguerre quadrature~\cite{davis1984methods}. Let \(L_n(x)\) be the standard Laguerre polynomial normalized by \(L_n(0)=1\). The \(K\)-point rule uses the positive roots \(x_{k,K}\) of \(L_K\) and positive weights \(w_{k,K}\) to write
\begin{equation}
  \int_0^\infty e^{-x}f(x)\,dx
  =\sum_{k=1}^K w_{k,K}f(x_{k,K})+E_K[f].
  \label{eq:gauss-laguerre-rule-basic}
\end{equation}
Its defining property is exactness for every polynomial of degree at most \(2K-1\):
\begin{equation}
  \int_0^\infty e^{-x}p(x)\,dx
  =\sum_{k=1}^K w_{k,K}p(x_{k,K}),
  \qquad \deg p\le 2K-1.
  \label{eq:gauss-laguerre-exactness}
\end{equation}
The relevant exponential has the Laguerre expansion
\begin{equation}
\begin{aligned}
  e^{-\mu x}&=\sum_{n=0}^\infty a_n(\mu)L_n(x),\\
  a_n(\mu)
  &=\int_0^\infty e^{-(1+\mu)x}L_n(x)\,dx
  =\frac{\mu^n}{(1+\mu)^{n+1}}.
\end{aligned}
  \label{eq:laguerre-expansion-exponential-ingredients}
\end{equation}
The last identity follows directly by inserting the finite power-series representation of \(L_n\); the standard Laguerre facts and the full coefficient calculation are collected in Appendix~\ref{appendix:laguerre-expansion-exponential}. Because Eq.~\eqref{eq:gauss-laguerre-exactness} integrates every term with \(n<2K\) exactly, the scalar remainder is entirely the Laguerre tail
\begin{equation}
  E_K[e^{-\mu x}]
  =-\sum_{n=2K}^\infty
  \frac{\mu^n}{(1+\mu)^{n+1}}
  \sum_{k=1}^K w_{k,K}L_n(x_{k,K}).
  \label{eq:error-term-of-gauss-laguerre-quadrature}
\end{equation}

Applying the rule mode by mode in Eq.~\eqref{eq:expression-for-sld-in-eigenbasis-of-M} and recombining the spectral projectors gives
\begin{align}
  \ket{\widehat L_j}
  &=\frac{1}{\Delta}\sum_{k=1}^K w_{k,K}
    e^{-x_{k,K}\mathcal M}\ket{\partial_j\rho}\nonumber\\
  &=\frac{1}{\Delta}\sum_{k=1}^K w_{k,K}e^{x_{k,K}}
    \exp\!\left(-\frac{x_{k,K}}{\Delta}\mathcal L_\rho\right)
    \ket{\partial_j\rho}.
  \label{eq:expression-for-sld-after-gauss-laguerre-quadrature}
\end{align}
Unvectorizing the Lyapunov exponential produces the central finite-sum representation
\begin{equation}
  \widehat L_j
  =\frac{1}{\Delta}\sum_{k=1}^K w_{k,K}e^{x_{k,K}}
  e^{-x_{k,K}\rho/\Delta}(\partial_j\rho)e^{-x_{k,K}\rho/\Delta}.
  \label{eq:approximate-solution-to-sld-in-matrix-form}
\end{equation}

We next quantify how rapidly this expression converges. For every relevant spectral mode, the coefficients in Eq.~\eqref{eq:laguerre-expansion-exponential-ingredients} have geometric ratio
\begin{equation}
  q(\mu_\ell)=\frac{\mu_\ell}{1+\mu_\ell}
  \le 1-\frac{\Delta}{2}.
  \label{eq:main-gapped-geometric-ratio}
\end{equation}
Combining this decay with the uniform bound on the discrete Laguerre moments gives
\begin{equation}
  \|L_j-\widehat L_j\|
  \le \frac{2}{\Delta}\left(1-\frac{\Delta}{2}\right)^{2K}\|\partial_j\rho\|_F.
  \label{eq:main-gapped-quadrature-remainder}
\end{equation}
If \(\partial_j\rho=0\), both the canonical and regularized SLDs vanish, and no quadrature is needed. Otherwise, throughout the quadrature-order and resource formulas below, logarithmic factors are bounded below by one (including nested logarithms); in particular, the chosen quadrature orders are positive integers. This convention leaves the high-precision scalings unchanged. Consequently, a truncation error at most \(\varepsilon\) is obtained by taking \(K\ge K_\varepsilon\), where
\begin{equation}
  K_\varepsilon
  :=\left\lceil\frac{1}{\Delta}
  \log\!\left(\frac{2\|\partial_j\rho\|_F}{\varepsilon\Delta}\right)\right\rceil
  =\Theta\!\left(\frac{1}{\Delta}
  \log\!\left(\frac{\|\partial_j\rho\|_F}{\varepsilon\Delta}\right)\right).
  \label{eq:quadrature-main-scaling}
\end{equation}
At fixed conditioning, the quadrature order therefore grows only logarithmically with the inverse target error. The complete tail expansion, the uniform discrete-moment bound, and an explicit sufficient integer choice of \(K\) are given in Appendix~\ref{truncation-error-analysis}.

\begin{figure}[t]
\centering
\includegraphics[width=\columnwidth]{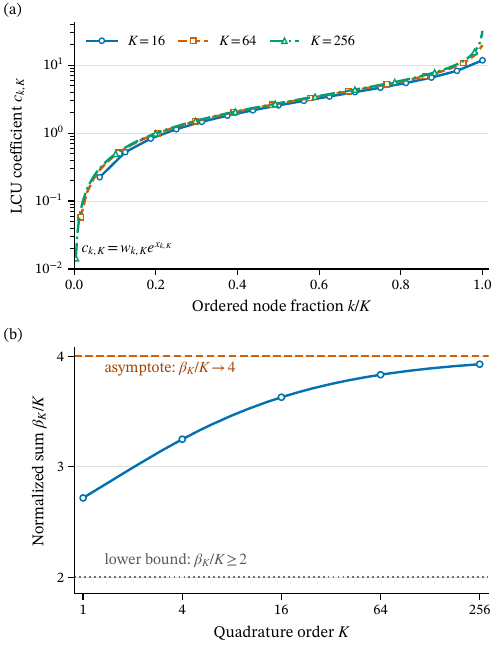}
\caption{Numerical structure of the compensated Gauss--Laguerre coefficients. (a) The individual LCU coefficients \(c_{k,K}=w_{k,K}e^{x_{k,K}}\), ordered by their nodes and plotted against \(k/K\). The bulk profiles remain of comparable scale as \(K\) increases, while the enhancement near the largest nodes is confined to the soft edge. (b) The normalized sum \(\beta_K/K\) for \(1\le K\le256\). The data remain above the rigorous lower bound \(2\) and approach the asymptotic value \(4\), illustrating the third statement of Theorem~\ref{theorem:quadrature-main}. The coefficients were evaluated through the exponentially scaled Laguerre recurrence, avoiding the underflow incurred by directly forming \(w_{k,K}e^{x_{k,K}}\) at large \(K\).}
\label{fig:lcu-coefficient-scaling}
\end{figure}

Exponential convergence is necessary but not by itself sufficient for a resource-efficient coherent implementation. Each term in Eq.~\eqref{eq:approximate-solution-to-sld-in-matrix-form} contains the compensation factor \(e^{x_{k,K}}\), which is exponentially large in the node value. The LCU normalization, however, depends on the aggregate coefficient sum
\begin{equation}
  \beta_K:=\sum_{k=1}^K |w_{k,K}e^{x_{k,K}}| = \sum_{k=1}^K w_{k,K}e^{x_{k,K}}
  \label{eq:main-beta-definition}
\end{equation}
rather than on its largest summand. Remarkably, the decay of the Gauss--Laguerre weights compensates the exponential factors strongly enough that
\begin{equation}
  \beta_K\ge 2K,
  \qquad
  \lim_{K\to\infty}\frac{\beta_K}{4K}=1,
  \label{eq:gauss-laguerre-coefficient-norm-theta-K}
\end{equation}
and hence \(\beta_K=\Theta(K)\). Thus the LCU coefficient normalization is only \(\beta_K/\Delta=\Theta(K/\Delta)\): the exponential quadrature convergence is not erased by coherent assembly. At fixed conditioning, both \(K\) and this normalization are logarithmic in \(1/\varepsilon\), while the quadrature-index register has width \(\lceil\log_2K\rceil=\mathcal{O}(\log\log(1/\varepsilon))\). Figure~\ref{fig:lcu-coefficient-scaling} illustrates both the distribution of the individual compensated coefficients and the linear growth of their sum. The asymptotic analysis establishing Eq.~\eqref{eq:gauss-laguerre-coefficient-norm-theta-K} is given in Appendix~\ref{appendix:lcu-coefficient-norm}.

\begin{theorem}[Gauss--Laguerre approximation of the canonical SLD]
\label{theorem:quadrature-main}
Assume the local rank-preservation condition in Eq.~\eqref{eq:kernel-condition-from-promises}, and let \(\Delta\) lower-bound the minimum nonzero eigenvalue of \(\rho\). For the \(K\)-point Gauss--Laguerre nodes and weights \(\{x_{k,K},w_{k,K}\}_{k=1}^K\), the following statements hold.
\begin{enumerate}
  \item The finite filtered-derivative approximation \(\widehat L_j\) is given by
  \[
  \widehat L_j
  =\frac{1}{\Delta}\sum_{k=1}^K w_{k,K}e^{x_{k,K}}
  e^{-x_{k,K}\rho/\Delta}(\partial_j\rho)e^{-x_{k,K}\rho/\Delta}.
  \]
  \item Its truncation error obeys  \(\|L_j-\widehat L_j\|
  \le \frac{2}{\Delta}\left(1-\frac{\Delta}{2}\right)^{2K}\|\partial_j\rho\|_F\); in particular, \(\|L_j-\widehat L_j\|\le\varepsilon\) for any \(K\ge K_\varepsilon\), with \(K_\varepsilon=\left\lceil\frac{1}{\Delta}
  \log\!\left(\frac{2\|\partial_j\rho\|_F}{\varepsilon\Delta}\right)\right\rceil
  =\Theta\!\left(\frac{1}{\Delta}
  \log\!\left(\frac{\|\partial_j\rho\|_F}{\varepsilon\Delta}\right)\right)\).
  \item The compensated coefficient sum \(\beta_K = \sum_{k=1}^K w_{k,K}e^{x_{k,K}}\) satisfies
  \[\beta_K\ge 2K,\qquad
  \lim_{K\to\infty}\frac{\beta_K}{4K}=1.\]Consequently, the LCU coefficient normalization is \(\Theta(K/\Delta)\).
\end{enumerate}
\end{theorem}

\subsection{Regularized quadrature}\label{sec:regularized-quadrature-outline}

When the support gap is too small, we first shift the Lyapunov map and then discretize its inverse. The regularized equation in Eq.~\eqref{eq:reg:sld_def} gives
\begin{equation}
  \ket{L_j^{(\theta)}}=(\mathcal L_\rho+\theta I)^{-1}\ket{\partial_j\rho}
  =\frac{1}{\theta}\int_0^\infty e^{-x}e^{-x\mathcal L_\rho/\theta}\ket{\partial_j\rho}\,dx.
  \label{eq:main-regularized-vectorized-integral}
\end{equation}
Here the shift supplies the Gauss--Laguerre weight directly, including on the kernel sector. Applying the same quadrature rule and unvectorizing yields
\begin{equation}
  \widehat L_j^{(\theta)}=\frac{1}{\theta}\sum_{k=1}^K
  w_{k,K}e^{-x_{k,K}\rho/\theta}(\partial_j\rho)e^{-x_{k,K}\rho/\theta}.
  \label{eq:main-regularized-quadrature-sum}
\end{equation}
The corresponding truncation bound is
\begin{equation}
  \|L_j^{(\theta)}-\widehat L_j^{(\theta)}\|
  \le \frac{2}{\theta}\left(1+\frac{\theta}{2}\right)^{-2K}\|\partial_j\rho\|_F,
  \label{eq:main-regularized-quadrature-remainder}
\end{equation}
so a truncation error at most \(\varepsilon\) is obtained with
\begin{equation}
  K\ge K_{\varepsilon} = \left\lceil\frac{1 + \theta/2}{\theta}
  \log\!\frac{2\|\partial_j\rho\|_F}{\varepsilon\theta}\right\rceil
  =\Theta\!\left(\frac{1}{\theta}
  \log\!\frac{\|\partial_j\rho\|_F}{\varepsilon\theta}\right).
  \label{eq:main-regularized-quadrature-order}
\end{equation}

Regularization removes the spectral-gap dependence from the construction but introduces a bias in the target quantity. Under the bounded-response promise, Eq.~\eqref{eq:reg:bias_bound} gives \(|\mathcal F_{jj}^{(\theta)}-\mathcal F_{jj}|\le8M\theta\); hence \(\theta\) simultaneously sets the numerical conditioning and the admissible statistical bias. Unlike the bounded-gap sum, Eq.~\eqref{eq:main-regularized-quadrature-sum} needs no \(e^{x_{k,K}}\) compensation. Since \(\sum_k w_{k,K}=1\), its LCU coefficient norm is \(1/\theta\), rather than \(\beta_K/\Delta\). Appendix~\ref{sec:reg:quadrature} contains the complete regularized truncation analysis. The two routes therefore share the same filtered-derivative architecture while paying for conditioning in different parameters.

\begin{figure*}[t]
\centering
\begin{quantikz}[row sep=0.45cm, column sep=0.85cm]
\lstick{$\ket{0^b}_{\mathrm{idx}}$} & \qw \qwbundle{b} & \gate{\mathcal{P}} & \ctrl{2} & \qw & \ctrl{3} & \gate{\mathcal{P}^\dagger} & \qw & \qw \\
\lstick{$\ket{0^{m_j}}_{A_\partial}$} & \qw \qwbundle{m_j} & \qw & \qw & \gate[3]{U_{\partial_j\rho}^{(A_\partial S)}} & \qw & \qw & \qw & \qw \\
\lstick{$\ket{0^{a_{\exp}}}_{A_1}$} & \qw \qwbundle{a_{\exp}} & \qw & \gate[2]{U_k} & & \qw & \qw & \qw & \qw \\
\lstick{$\ket{0^{s}}_{S}$} & \qw & \qw & & & \gate[2]{U_k} & \qw & \qw & \qw \\
\lstick{$\ket{0^{a_{\exp}}}_{A_2}$} & \qw \qwbundle{a_{\exp}} & \qw & \qw & \qw & & \qw & \qw & \qw
\end{quantikz}
\caption{Block encoding of the approximate SLD. The index register loads the compensated quadrature coefficients, the two controlled QSVT blocks implement \(e^{-x_{k,K}\rho/\Delta}\) on either side of \(U_{\partial_j\rho}\), and unpreparing the index register assembles the Gauss--Laguerre sum. Postselection on the zero-ancilla subspace gives normalization \(\alpha_j\beta_K/\Delta\) for \(\widehat L_j\).}
\label{fig:sld-block-encoding}
\end{figure*}
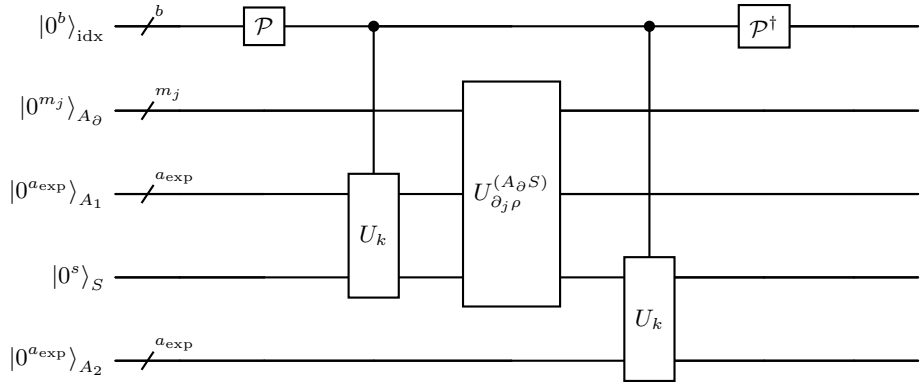

\subsection{Block-encoding realization}\label{algorithm-design-for-solving-the-lyapunov-equation}

The two finite expansions derived above have the same modular form: a weighted sum of two-sided exponential filters acting on \(\partial_j\rho\). We first describe the bounded-gap construction explicitly. Purified access supplies a \((1,a+s,0)\)-block encoding of \(\rho\) through the standard SWAP construction~\cite{gilyenQuantumSingularValue2019},
\begin{equation}
  U_\rho=(O_\rho^\dagger\otimes I_S)\operatorname{SWAP}_{S,S'}(O_\rho\otimes I_S),
  \label{eq:block-encoding-of-rho-using-purification}
\end{equation}
where \(S'\) is an additional \(s\)-qubit system register. For the \(k\)-th quadrature node, set \(t_{k,K}=x_{k,K}/\Delta\). Lemma~\ref{lem:gapped_exponential_anchor_tt} provides an even polynomial \(P_t\), bounded by one on \([-1,1]\), that approximates \(e^{-t|x|}\) to error \(\epsilon_{\mathrm p}\) for \(|x|\in[\Delta,1]\) and approximates one to the same error at \(x=0\), with degree
\begin{equation}
  d=\mathcal O\!\left(\frac{1}{\Delta}\log\frac{1}{\epsilon_{\mathrm p}}\right),
  \label{eq:main-gapped-filter-degree}
\end{equation}
independently of \(t\). Since \(\operatorname{spec}(\rho)\subseteq\{0\}\cup[\Delta,1]\), applying QSVT with \(P_{t_{k,K}}\) to \(U_\rho\) produces a \((1,a_{\exp},\epsilon_{\mathrm p})\)-block encoding \(U_k\) of \(e^{-t_{k,K}\rho}\), where \(a_{\exp}:=a+s+1\). The approximation at \(x=0\) is needed because \(\partial_j\rho\) can connect the support to the kernel even though its kernel--kernel block vanishes.

Let \(A_1\) and \(A_2\) denote independent \(a_{\exp}\)-qubit ancilla registers for the two filters, and let \(A_\partial\) denote the \(m_j\)-qubit ancilla register of \(U_{\partial_j\rho}\). For each node, the unitary
\[
  V_{j,k}:=U_k^{(A_2S)}U_{\partial_j\rho}^{(A_\partial S)}U_k^{(A_1S)},
\]
where each factor acts as the identity on the registers not shown in its superscript, has a postselected block proportional to
\begin{equation}
  A_{j,k}:=e^{-t_{k,K}\rho}(\partial_j\rho)e^{-t_{k,K}\rho}.
  \label{eq:main-node-filtered-derivative}
\end{equation}
Two applications of the block-encoding product rule make \(V_{j,k}\) an \((\alpha_j,2a_{\exp}+m_j,2\alpha_j\epsilon_{\mathrm p})\)-block encoding of \(A_{j,k}\). To assemble the nodes coherently, define
\begin{equation}
  c_{k,K}:=w_{k,K}e^{x_{k,K}},\qquad
  \beta_K=\sum_{k=1}^K c_{k,K},\qquad
  b=\lceil\log_2K\rceil,
  \label{eq:main-lcu-coefficients}
\end{equation}
where \(\ket{k}\) denotes a fixed \(b\)-bit encoding of node \(k\). Let the ideal coefficient-preparation unitary satisfy
\begin{equation}
  \mathcal P\ket{0^b}
  =\sum_{k=1}^K\sqrt{\frac{c_{k,K}}{\beta_K}}\ket{k}.
  \label{eq:main-lcu-prepare}
\end{equation}
The remaining \(2^b-K\) index states have zero amplitude, so no power-of-two assumption on \(K\) is required. Define their projector by \(\Pi_\perp:=I_{\mathrm{idx}}-\sum_{k=1}^K\ket{k}\!\bra{k}\), let \(I_{\mathrm{work}}\) denote the identity on all non-index registers, and extend the two controlled filter operations by the identity on the unused index subspace:
\begin{equation}
\begin{aligned}
  \operatorname{SELECT}_R
  &:=\sum_{k=1}^K\ket{k}\!\bra{k}\otimes U_k^{(A_1S)}
    +\Pi_\perp\otimes I_{\mathrm{work}},\\
  \operatorname{SELECT}_L
  &:=\sum_{k=1}^K\ket{k}\!\bra{k}\otimes U_k^{(A_2S)}
    +\Pi_\perp\otimes I_{\mathrm{work}}.
\end{aligned}
  \label{eq:main-lcu-select}
\end{equation}
The circuit shown in Fig.~\ref{fig:sld-block-encoding} is therefore
\begin{equation}
\begin{aligned}
  U_{L_j}:={}&(\mathcal P^\dagger\otimes I_{\mathrm{work}})
  \operatorname{SELECT}_L
  \bigl(I_{\mathrm{idx}}\otimes U_{\partial_j\rho}^{(A_\partial S)}\bigr)\\
  &\times\operatorname{SELECT}_R
  (\mathcal P\otimes I_{\mathrm{work}}).
\end{aligned}
  \label{eq:block-encoding-unitary-for-L}
\end{equation}
In the idealized limit of exact coefficient preparation and exact node filters, projecting the index and block-encoding ancillas onto zero gives
\[
\begin{aligned}
  &(\bra{0^{b+2a_{\exp}+m_j}}\otimes I_S)
  U_{L_j}
  (\ket{0^{b+2a_{\exp}+m_j}}\otimes I_S)\\
  &\qquad=\frac{1}{\alpha_j\beta_K}
  \sum_{k=1}^K c_{k,K}A_{j,k}
  =\frac{\Delta}{\alpha_j\beta_K}\widehat L_j.
\end{aligned}
\]
This identity gives the normalization \(\alpha_j\beta_K/\Delta\) directly. The different nodes share the same \(U_\rho\) query backbone and differ only through multiplexed QSVT phases. Consequently, the two SELECT operations together use \(\mathcal{O}(d)\), rather than \(\mathcal{O}(Kd)\), queries to \(U_\rho\), although their elementary-gate count retains the cost of multiplexing over the \(K\) phase sequences.

Approximating the exponential filters by QSVT polynomials and preparing the LCU coefficients with finite precision perturb this postselected block. Together with the error incurred by replacing the Laplace integral with the finite quadrature sum, these contributions determine the total operator-norm error, denoted by \(\epsilon_j\), of the resulting SLD block encoding. Appendix~\ref{app:sld-implementation} derives the complete error bound and gives an explicit allocation of the internal precisions.

\begin{proposition}[Block encoding of the canonical SLD]
\label{prop:canonical-sld-block-encoding}
Under the access model of Section~\ref{sec:access-model} and the bounded-gap promise, for any target accuracy \(\epsilon_j>0\), choose
\begin{equation}
  K=\left\lceil\frac{1}{\Delta}
  \log\!\left(\frac{6\|\partial_j\rho\|_F}{\epsilon_j\Delta}\right)\right\rceil.
  \label{eq:main-sld-block-encoding-order}
\end{equation}
The construction above gives an \((\alpha_{L_j},a_{L_j},\epsilon_j)\)-block encoding of \(L_j\), where
\begin{equation}
  \alpha_{L_j}=\frac{\alpha_j\beta_K}{\Delta},\qquad
  a_{L_j}=2a+2s+m_j+2+\lceil\log_2K\rceil.
  \label{eq:main-sld-block-encoding-parameters}
\end{equation}
One application uses
\begin{align}
  Q_\rho^{(j)}
  &=\mathcal O\!\left(\frac{1}{\Delta}\left[
  \log\!\left(\frac{\alpha_j}{\epsilon_j\Delta}\right)
  +\log\log\!\left(\frac{\|\partial_j\rho\|_F}{\epsilon_j\Delta}\right)
  \right]\right), \nonumber\\
  Q_{\partial_j}&=\mathcal{O}(1)
  \label{eq:main-sld-block-encoding-queries}
\end{align}
purification- and derivative-oracle queries, respectively.
\end{proposition}

Substituting the quadrature order in Eq.~\eqref{eq:main-sld-block-encoding-order} into Eq.~\eqref{eq:main-sld-block-encoding-parameters}, the total ancilla count scales as
\begin{equation}
  a_{L_j}
  =2a+2s+m_j+2
  +\mathcal O\!\left(
    \log\frac{1}{\Delta}
    +\log\log\frac{\|\partial_j\rho\|_F}{\epsilon_j\Delta}
  \right).
  \label{eq:main-sld-ancilla-scaling}
\end{equation}
Thus, at fixed conditioning and fixed oracle-register sizes, the auxiliary-qubit overhead grows double-logarithmically with the inverse target precision.

The regularized construction reuses the circuit in Eq.~\eqref{eq:block-encoding-unitary-for-L} with two modifications. The quadrature times become \(t_{k,K}=x_{k,K}/\theta\), while the LCU coefficients become \(w_{k,K}/\theta\). Since \(\sum_k w_{k,K}=1\), their sum is \(1/\theta\), with no \(e^{x_{k,K}}\) compensation or \(\beta_K\) factor. In addition, without a spectral gap the filters must approximate \(e^{-tx}\) across the full spectrum \([0,1]\). We therefore use the arbitrary-parity eigenvalue transformation of Lemma~\ref{lemma:qsvt-arbitrary-parity}. Under its normalization convention, each filter has normalization \(\mathcal{O}(1)\) and degree \(\mathcal{O}(\log(2/\epsilon_{\mathrm p}))\) for \(0\le t<1\), or \(\mathcal{O}(t\log(t/\epsilon_{\mathrm p}))\) for \(t\ge1\). The largest quadrature time satisfies \(t_{\max}=x_{K,K}/\theta\ge1\) and therefore falls in the second case, giving maximum degree \(\mathcal{O}((K/\theta)\log(K/(\theta\epsilon_{\mathrm p})))\), in contrast to the bounded-gap construction, whose filter degree is independent of the quadrature time. We summarize the resulting block-encoding parameters and query counts in the following proposition.

\begin{proposition}[Block encoding of the regularized SLD]
\label{prop:regularized-sld-block-encoding}
Under the access model of Section~\ref{sec:access-model}, let \(0<\theta\le1\) and let \(\epsilon_j>0\) be the target block-encoding error. Choose
\begin{equation}
  K=\left\lceil\frac{1+\theta/2}{\theta}
  \log\!\left(\frac{6\|\partial_j\rho\|_F}{\epsilon_j\theta}\right)\right\rceil.
  \label{eq:main-regularized-sld-block-encoding-order}
\end{equation}
The construction above gives an \((\alpha_{L_j^{(\theta)}},a_{L_j^{(\theta)}},\epsilon_j)\)-block encoding of the regularized SLD \(L_j^{(\theta)}\), where
\begin{equation}
  \alpha_{L_j^{(\theta)}}=\mathcal{O}\!\left(\frac{\alpha_j}{\theta}\right),\qquad
  a_{L_j^{(\theta)}}=2a+2s+m_j+4+\lceil\log_2K\rceil.
  \label{eq:main-regularized-sld-block-encoding-parameters}
\end{equation}
One application uses
\begin{equation}
  Q_\rho^{(j,\theta)}=\widetilde{\mathcal O}(\theta^{-2}),\qquad
  Q_{\partial_j}^{(j,\theta)}=\mathcal{O}(1)
  \label{eq:main-regularized-sld-block-encoding-queries}
\end{equation}
purification- and derivative-oracle queries, respectively, where the tilde suppresses polylogarithmic factors in \(\theta^{-1}\), \(\epsilon_j^{-1}\), \(\alpha_j\), and \(\|\partial_j\rho\|_F\).
\end{proposition}

The operator constructed in Proposition~\ref{prop:regularized-sld-block-encoding} is \(L_j^{(\theta)}\), rather than an operator-norm approximation to the canonical SLD \(L_j\). Its connection to the target diagonal QFI is instead supplied by the bounded-response promise at the scalar level: Eq.~\eqref{eq:reg:bias_bound} gives \(\lvert\mathcal F_{jj}^{(\theta)}-\mathcal F_{jj}\rvert\le8M\theta\). Appendix~\ref{sec:reg:be_construction} gives the polynomial construction, error allocation, and complete resource accounting.

\paragraph{Comparison with matrix-equation solvers.}
Table~\ref{tab:sld_solver_comparison} compares our construction with existing quantum algorithms for general Lyapunov and Sylvester equations~\cite{claytonDifferentiableQuantumComputing2024,sommaQuantumAlgorithmLinear2025} when applied to solving the SLD equation.
Clayton et al.~\cite{claytonDifferentiableQuantumComputing2024} approximate a Lyapunov integral using the trapezoidal rule. For the SLD equation, the coefficient matrix is $-\rho$ and the source matrix is $\partial_j\rho$, giving exactly the integral in Eq.~\eqref{eq:integral-expression-for-sld}. Under the bounded-gap promise and the kernel condition, its tail decays exponentially, so an integration interval of length $\widetilde{\mathcal{O}}(\Delta_\rho^{-1})$ suffices. Specializing their discretization and block-encoding assembly to this integral gives the Clayton row of Table~\ref{tab:sld_solver_comparison}. At fixed input normalization and gap, this construction uses $\widetilde{\mathcal{O}}(\epsilon_L^{-1/2})$ state-oracle queries and $\mathcal{O}(\log(1/\epsilon_L))$ LCU index qubits, compared with our logarithmic query dependence and double-logarithmic index width.

Somma et al.'s normal-matrix construction~\cite{sommaQuantumAlgorithmLinear2025} specializes to the SLD equation with an effective conditioning parameter bounded by $\mathcal{O}(\Delta_\rho^{-1})$. It also accommodates rank-deficient probes under the kernel condition of Eq.~\eqref{eq:condition-for-sld-to-be-well-defined}: the symmetric Fourier--LCU sum vanishes on the zero spectrum and approximates the inverse on the nonzero spectrum, yielding the canonical SLD without an additional support projection.

\begin{table*}[t]
\centering
\caption{Construction of an $(\alpha_L,a_L,\epsilon_L)$-block encoding of the canonical SLD under the bounded-gap promise and kernel condition: $\|L_j-\alpha_L\langle0^{a_L}|U|0^{a_L}\rangle\|\le\epsilon_L$. The Clayton and Somma rows are specializations derived here from their respective constructions.}
\label{tab:sld_solver_comparison}
\small
\setlength{\tabcolsep}{4pt}
\begin{tabular}{@{}p{2.15cm} p{4.05cm} p{3.7cm} p{2.9cm} p{2.7cm}@{}}
\toprule
\raggedright Work & \raggedright Construction Method & \raggedright Oracle queries & \raggedright $
\alpha_L$ & \raggedright LCU index qubits \tabularnewline
\midrule
\raggedright Clayton et al.~\cite{claytonDifferentiableQuantumComputing2024}
& \raggedright trapezoidal quadrature-based LCU
& \raggedright $\widetilde{\mathcal{O}}(\Delta_\rho^{-5/2}\sqrt{\alpha_j/\epsilon_L})$ to $O_\rho$; $\mathcal{O}(1)$ to $U_{\partial_j\rho}$
& \raggedright $\widetilde{\mathcal{O}}(\alpha_j\Delta_\rho^{-3})$
& \raggedright $\mathcal{O}(\log(1/\epsilon_L))$ \tabularnewline
\addlinespace
\raggedright Somma et al.~\cite{sommaQuantumAlgorithmLinear2025}
& \raggedright Fourier/Riemann quadrature-based LCU
& \raggedright $\widetilde{\mathcal{O}}(\Delta_\rho^{-1})$ to $O_\rho$; $\mathcal{O}(1)$ to $U_{\partial_j\rho}$
& \raggedright $\widetilde{\mathcal{O}}(\alpha_j\Delta_\rho^{-1})$
& \raggedright $\mathcal{O}(\log(1/\epsilon_L))$ \tabularnewline
\addlinespace
\raggedright This work (bounded gap)
& \raggedright Gauss--Laguerre quadrature-based LCU
& \raggedright $\widetilde{\mathcal{O}}(\Delta_\rho^{-1})$ to $O_\rho$; $\mathcal{O}(1)$ to $U_{\partial_j\rho}$
& \raggedright $\widetilde{\mathcal{O}}(\alpha_j\Delta_\rho^{-2})$
& \raggedright $\mathcal{O}(\log\log(1/\epsilon_L))$ \tabularnewline
\bottomrule
\end{tabular}
\par\smallskip\raggedright\footnotesize
\textit{Note.} $\Delta_\rho=\Delta$, and $\alpha_j$ is the derivative block-encoding normalization. Somma's construction uses block encodings of $\rho/2$ and $\partial_j\rho/2$. Our row substitutes the quadrature order with $\epsilon_j=\epsilon_L$ and uses $\beta_K=\Theta(K)$. The Clayton row retains their separate implementation of each controlled exponential and its time-dependent normalization. $\widetilde{\mathcal{O}}$ suppresses polylogarithmic factors; the index-qubit column displays precision dependence at fixed gap, dimension, and input normalizations, excluding oracle workspace.
\end{table*}

\section{QFIM and QGT Estimation}\label{sec:accessing-qfim}

The block-encoding of each SLD \(L_j\) established in Proposition~\ref{prop:canonical-sld-block-encoding} is the atomic primitive from which all subsequent algorithms are built. We now show how this single primitive unlocks two complementary modes of accessing mixed-state geometry, corresponding to the two tasks formulated in Problem~\ref{prob:qfim-estimation}. The key observation is that the SLD block encoding gives direct quantum access to a family of states whose inner products are the QGT.

Concretely, from the block encoding \(U_{L_j}\) we construct unnormalized states
\begin{equation}
  \ket{\psi_j} = (L_j \otimes I_A) \ket{\rho}_{SA},
  \label{eq:definition-of-psi-j}
\end{equation}
where \(\ket{\rho}_{SA}=O_\rho\ket{0}_S\ket{0}_A\) is a purification of \(\rho\). Their inner products directly encode the QGT:
\[
\langle\psi_j|\psi_k\rangle = \Tr(\rho L_j L_k) = \mathcal{T}_{jk},
\qquad
\mathcal{F}_{jk} = 4\,\Re(\mathcal{T}_{jk}).
\]
The diagonal QFIM entries are encoded in the norms \(\mathcal{N}_j \equiv \|\ket{\psi_j}\| = \sqrt{\mathcal{F}_{jj}}/2\). In this language, the two tasks of Problem~\ref{prob:qfim-estimation} reduce to estimating state norms (diagonal entries) and encoding the full inner-product matrix \(\mathcal{T}_{jk}\) in a single block-encoded unitary.

These two tasks define complementary algorithmic constructions with different hardware requirements. The norm-estimation route (Section~\ref{estimation-of-the-norms}) requires only the individual SLD block encoding and amplitude estimation---it is the lightest-weight approach, sufficient when only diagonal sensitivities are needed. The full-matrix block encoding (Sections~\ref{sec:block-encoding-qgt}--\ref{sec:block-encoding-qgt}) requires controlled multiplexed SLD operations over all \(m\) parameter directions, but provides the entire QGT and QFIM as a single block-encoded matrix, enabling downstream QSVT-based linear algebra: coherent access to the inverse QFIM entering the SLD Cram\'er--Rao matrix bound, eigenvalue processing for identifying candidate high-information parameter directions, and threshold-based detection of multiparameter incompatibility from the antisymmetric imaginary part of the QGT, without first reconstructing its individual matrix elements classically.

Section~\ref{sec:reg:qfi_estimation} extends the diagonal estimation to the bounded-response regime using the regularized SLD block encoding of Section~\ref{sec:reg:be_construction}, offering a more robust alternative to estimate QFI at the cost of a worsened query complexity scaling. The regularized construction is analyzed here only for diagonal QFI estimation; its extension to block encodings of the full QGT and QFIM is beyond the scope of this work.
\subsection{Estimation for Diagonal Entries}\label{estimation-of-the-norms}

For a fixed direction \(j\), the QFI is a norm: \(\mathcal F_{jj}=4\mathcal N_j^2\), where \(\mathcal N_j=\|(L_j\otimes I_A)|\rho\rangle\|\). Let \(\widetilde L_j\) denote the operator encoded by the approximate SLD block encoding of Proposition~\ref{prop:canonical-sld-block-encoding}, and define
\begin{equation}
  \mathcal W_j
  :=(U_{L_j}\otimes I_A)(I_{a_{L_j}}\otimes O_\rho).
  \label{eq:main-success-amplitude-unitary}
\end{equation}
Acting on the all-zero input, this unitary has the decomposition
\begin{equation}
\begin{aligned}
  \mathcal W_j\ket{0^{a_{L_j}}}\ket{0}_{SA}
  &=\gamma_j\ket{0^{a_{L_j}}}\ket{\widehat\psi_j}\\
  &\quad+\sqrt{1-\gamma_j^2}\ket{\perp_j},\\
  \gamma_j&=\frac{\widetilde{\mathcal N}_j}{\alpha_{L_j}},
\end{aligned}
  \label{eq:main-qfi-success-amplitude}
\end{equation}
where \(\widetilde{\mathcal N}_j:=\|(\widetilde L_j\otimes I_A)\ket{\rho}_{SA}\|\), \(\ket{\widehat\psi_j}\) is normalized, and \(\ket{\perp_j}\) is orthogonal to the zero state of the block-encoding ancillas. Thus diagonal QFI estimation reduces to estimating the success amplitude \(\gamma_j\).

A conventional QPE-based amplitude estimator~\cite{brassardQuantumAmplitudeAmplification2002} would add a coherent \(\mathcal{O}(\log(1/\epsilon_{\mathrm{amp}}))\)-qubit phase register, which would erase the double-logarithmic precision dependence established in Proposition~\ref{prop:canonical-sld-block-encoding}. We instead adapt the binary-decision strategy of Ref.~\cite{linNearoptimalGroundState2020} and its QPE-free polynomial implementation in Ref.~\cite{dongGroundStatePreparationEnergy2022} to obtain the following additive estimator.

\begin{lemma}[QPE-free additive amplitude estimation]
\label{lem:qpe-free-additive-amplitude-estimation}
Let \(U\) be a unitary satisfying
\begin{equation}
  U\ket{0^a}\ket{0}_S
  =\gamma\ket{0^a}\ket{\phi}
  +\sqrt{1-\gamma^2}\ket{\perp},
  \label{eq:main-generic-success-amplitude}
\end{equation}
where \(\gamma\in[0,1]\), \(\ket{\phi}\) is normalized, and \(\ket{\perp}\) is orthogonal to the zero state of the first register. For any \(\epsilon_{\mathrm{amp}}\in(0,1/2)\) and \(\vartheta_{\mathrm{amp}}\in(0,1)\), there is a QPE-free, classically controlled quantum algorithm that outputs \(\widehat\gamma\) satisfying
\begin{equation}
  |\widehat\gamma-\gamma|\le\epsilon_{\mathrm{amp}}
  \label{eq:main-additive-amplitude-guarantee}
\end{equation}
with probability at least \(1-\vartheta_{\mathrm{amp}}\). It uses
\begin{equation}
  Q_{\mathrm{amp}}
  =\mathcal O\!\left(
    \frac{1}{\epsilon_{\mathrm{amp}}}
    \left(
      \log\frac{1}{\vartheta_{\mathrm{amp}}}
      +\log\log\frac{1}{\epsilon_{\mathrm{amp}}}
    \right)
  \right)
  \label{eq:query-complexity-for-binary-amplitude-estimation}
\end{equation}
queries to \(U\) and \(U^\dagger\) and one additional ancilla qubit.
\end{lemma}

Applying Lemma~\ref{lem:qpe-free-additive-amplitude-estimation} with \(U=\mathcal W_j\) gives \(\widehat{\mathcal N}_j=\alpha_{L_j}\widehat\gamma_j\) and hence \(|\widehat{\mathcal N}_j-\widetilde{\mathcal N}_j|\le\alpha_{L_j}\epsilon_{\mathrm{amp}}\). The SLD block-encoding error controls \(|\widetilde{\mathcal N}_j-\mathcal N_j|\), while the bounded-gap norm estimate controls the propagation through \(\mathcal F_{jj}=4\mathcal N_j^2\). Optimizing these internal precisions under the resulting error constraint yields the following result.

\begin{theorem}[QFI estimation]
\label{thm:diagonal-qfi-estimation}
Under the access model of Section~\ref{sec:access-model} and the local rank-preservation condition, provided that the bounded-gap promise holds, there exists a quantum algorithm that, for any parameter index \(j\), estimates the QFI \(\mathcal{F}_{jj}\) to additive error \(\epsilon\) with success probability at least \(1-\vartheta\).  The query complexities to \(O_\rho\) and \(U_{\partial_j \rho}\), up to polylogarithmic factors in \(\epsilon^{-1}, \vartheta^{-1}, \Delta^{-1}, \alpha_j\) and the derivative norm, are
\[
Q_\rho^{jj} = \widetilde{\mathcal{O}}\!\left(\frac{\alpha_j\|\partial_j\rho\|_F}{\epsilon\,\Delta^{7/2}}\right),\qquad
Q_{\partial_j}^{jj} = \widetilde{\mathcal{O}}\!\left(\frac{\alpha_j\|\partial_j\rho\|_F}{\epsilon\,\Delta^{5/2}}\right),
\label{eq:Q-diag-oracles}
\]
respectively. The circuit uses \(3a + 3s + 3 + m_j + \mathcal{O}\!\left(\log\frac{1}{\Delta} + \log\log\frac{1}{\epsilon}\right)\) qubits in total and requires \(\widetilde{\mathcal{O}}(a+s+m_j+\Delta^{-1})\) additional elementary gates per input-oracle query, where \(a\) and \(s\) are the purification ancilla and system qubit counts, and \(m_j\) is the ancilla count of the derivative block encoding.
\end{theorem}

The small-response case can be handled directly: since \(\mathcal F_{jj}\le 2\|\partial_j\rho\|_F^2/\Delta\), outputting zero suffices whenever this upper bound is at most \(\epsilon\). This test may use an analytically available upper bound on the derivative norm; it does not require estimating that norm from its block encoding. The nontrivial branch is analyzed in Appendix~\ref{app:diagonal-analysis}.

Appendix~\ref{app:qpe-free-amplitude-estimation} proves Lemma~\ref{lem:qpe-free-additive-amplitude-estimation}, while Appendix~\ref{app:diagonal-analysis} gives the detailed proof of Theorem~\ref{thm:diagonal-qfi-estimation}, including the full derivation of the query complexity and complete error allocation.
The Frobenius norm in the complexity is a property of the parameterized family, and retaining it preserves sensitivity to the effective rank of the parameter response rather than replacing it by a worst-case factor. Indeed, for any operator \(X\), \(\|X\|_F \le \|X\|\sqrt{\operatorname{rank}(X)}\). Under the kernel condition, \(\operatorname{rank}(\partial_j\rho)\le\min\{2r,N\}\), so \(\|\partial_j\rho\|_F\) can be bounded by \(\sqrt{\min\{2r,N\}}\alpha_j\) when only the derivative normalization and rank are available (Lemma~\ref{lemma:rank-bound-derivative}). Consequently, whenever the probe rank or response rank is polynomially bounded---as in low-rank probe families and explicitly truncated thermal or low-energy subspaces---the Frobenius-norm dependence can be far smaller than a Hilbert-space-dimension worst-case bound.

\subsection{Block Encoding of the Full Quantum Geometric Tensor and QFIM}\label{sec:block-encoding-qgt}

The norm-estimation approach above processes each diagonal QFIM entry independently, requiring separate amplitude-estimation rounds for each parameter direction. The full multiparameter structure, however, is encoded in the off-diagonal entries, and processing all \(\mathcal{O}(m^2)\) entries individually would be wasteful when the entire QFIM is needed. We now present a complementary approach: a single unitary circuit that encodes the \emph{entire} \(m \times m\) quantum geometric tensor as a block-encoded matrix, from which the QFIM follows by linear combination.

\begin{theorem}[Block encoding of the quantum geometric tensor]\label{thm:qgt-block-encoding}
Let \(\rho\) be an \(s\)-qubit density operator with purification unitary \(O_\rho\) as in Eq.~\eqref{eq:purification-oracle-intro}. For each parameter direction \(j = 1,\dots,m\) (let \(w = \lceil \log_2 m\rceil\)), assume we are given an \((\alpha, a_{L_j}, \epsilon_j)\)-block-encoding \(U_{L_j}\) of the SLD operator \(L_j\), i.e.,
\[
\|L_j - \alpha (\langle 0^{a_{L_j}}| \otimes I_s) U_{L_j} (|0^{a_{L_j}}\rangle \otimes I_s)\| \le \epsilon_j,
\]
with \(\|L_j\| \le \alpha\) for all \(j\). We embed \(\mathcal T\) in the first \(m\) basis states of the \(w\)-qubit matrix register, with zeros on the unused subspace. Let \(a_L = \max_j a_{L_j}\). We use independent ancilla registers for the two SLD block encodings. Then there exists a unitary \(U_{\mathcal{T}}\) that is an \((m\alpha^2, a_{\mathrm{tot}}, \epsilon_{\mathrm{tot}})\)-block-encoding of the quantum geometric tensor
\[
\mathcal{T}_{ij} = \Tr(\rho L_i L_j),
\]
where
\[
a_{\mathrm{tot}} = w + 2a_L + a + s, \qquad
\epsilon_{\mathrm{tot}} = \Bigl(\sum_{i,j=1}^{m} \delta_{ij}^2\Bigr)^{1/2},
\]
and \(\delta_{ij} = \alpha(\epsilon_i + \epsilon_j) + \epsilon_i\epsilon_j\).  If all block encodings share a common error \(\epsilon_j = \epsilon\) (or one conservatively bounds each \(\epsilon_j \le \epsilon\)), this reduces to \(\epsilon_{\mathrm{tot}} = m(2\alpha\epsilon + \epsilon^2)\).
\end{theorem}

\begin{proof}
Denote \(\tilde L_j = \alpha (\langle 0^{a_L}| \otimes I_s) U_{L_j} (|0^{a_L}\rangle \otimes I_s)\) (with ancilla padding where needed), so \(\|\tilde L_j - L_j\| \le \epsilon_j\).
The construction uses the registers depicted in Fig.~\ref{fig:qgt-circuit}: an auxiliary row-index register \(X\) of \(w\) qubits, a matrix register \(R\) of \(w\) qubits, two independent SLD block-encoding ancillas \(A_{L,R}\) and \(A_{L,L}\) of \(a_L\) qubits each, and the purification registers \(S\) and \(A\) of \(s\) and \(a\) qubits, respectively.

We first define the uniform superposition gate \(\mathcal{D}\) on register \(X\) as
\[
\mathcal{D}|0\rangle_X = \frac{1}{\sqrt{m}}\sum_{k=1}^m |k\rangle_X .
\]

Set \(U_{L_k}=I\) for \(m<k\le 2^w\), so that the controlled operations act as the identity on unused index states. Denote the two controlled multiplexed block encodings
\[
\begin{aligned}
\mathcal{U}_R &= \sum_{k=1}^{2^w} I_X \otimes |k\rangle\langle k|_R \otimes U_{L_k}^{(A_{L,R}S)} \otimes I_{A_{L,L}A}, \\
\mathcal{U}_L &= \sum_{k=1}^{2^w} |k\rangle\langle k|_X \otimes I_R\otimes U_{L_k}^{(A_{L,L}S)} \otimes I_{A_{L,R}A} .
\end{aligned}
\]

The resulting circuit, shown in Fig.~\ref{fig:qgt-circuit}, is
\begin{equation}
\begin{aligned}
U_{\mathcal{T}} &= (O_\rho^\dagger \otimes I)
\cdot (\mathcal{D}^\dagger \otimes I)
\cdot \mathrm{SWAP}_{X,R} \\
&\quad \cdot \mathcal{U}_L \cdot \mathcal{U}_R
\cdot (\mathcal{D} \otimes I) \\
&\quad \cdot (O_\rho \otimes I),
\end{aligned}
\label{eq:qgt-circuit}
\end{equation}

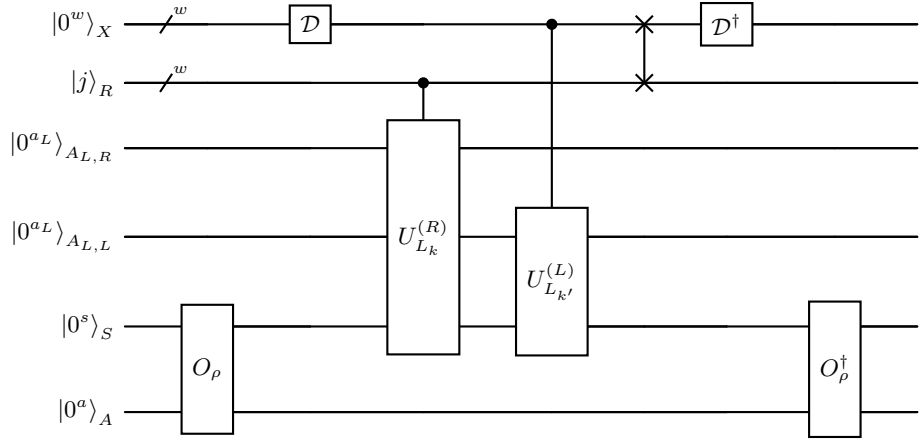
\begin{figure*}[htbp]
\centering
\begin{quantikz}[row sep=0.42cm, column sep=0.75cm]
\lstick{$\ket{0^w}_X$} & \qw \qwbundle{w} & \gate{\mathcal{D}} & \qw & \ctrl{3} & \swap{1} & \gate{\mathcal{D}^\dagger} & \qw & \qw \\
\lstick{$\ket{j}_R$} & \qw \qwbundle{w} & \qw & \ctrl{1} & \qw & \targX{} & \qw & \qw & \qw \\
\lstick{$\ket{0^{a_L}}_{A_{L,R}}$} & \qw & \qw & \gate[3]{U_{L_k}^{(R)}} & \qw & \qw & \qw & \qw & \qw \\
\lstick{$\ket{0^{a_L}}_{A_{L,L}}$} & \qw & \qw & \qw & \gate[2]{U_{L_{k'}}^{(L)}} & \qw & \qw & \qw & \qw \\
\lstick{$\ket{0^{s}}_S$} & \gate[2]{O_\rho} & \qw & & & \qw & \qw & \gate[2]{O_\rho^\dagger} & \qw \\
\lstick{$\ket{0^{a}}_A$} & & \qw & \qw & \qw & \qw & \qw & & \qw
\end{quantikz}
\caption{Quantum circuit for the block encoding of the quantum geometric tensor $\mathcal{T}_{ij}=\Tr(\rho L_i L_j)$. The two controlled-multiplexed SLD operations use independent ancilla registers $A_{L,R}$ and $A_{L,L}$, so that nonzero ancilla components produced by the first block encoding cannot be erased by the second. Projecting $X$, $A_{L,R}$, and $A_{L,L}$ onto their zero states and the purification registers onto $\bra{\rho}_{SA}$ extracts the desired block. The register roles are: $X$ and $R$ ($w$ qubits each), $A_{L,R}$ and $A_{L,L}$ ($a_L$ qubits each), $S$ ($s$ qubits), and $A$ ($a$ qubits).}
\label{fig:qgt-circuit}
\end{figure*}
Projecting every ancilla register onto zero gives
\begin{equation}
\begin{aligned}
 &(\bra{0^{a_{\mathrm{tot}}}}\otimes I_R)
 U_{\mathcal T}
 (\ket{0^{a_{\mathrm{tot}}}}\otimes I_R) \\
 &\qquad=
 \frac{1}{m\alpha^2}
 \sum_{i,j=1}^m
 \Tr(\rho\,\widetilde L_i\widetilde L_j)
 \ket{i}\!\bra{j}_R
 =\frac{\widetilde{\mathcal T}}{m\alpha^2}.
\end{aligned}
\label{eq:qgt-projected-block-main}
\end{equation}
Appendix~\ref{app:qgt-block-encoding-details} gives the detailed derivation of Eq.~\eqref{eq:qgt-projected-block-main}.

Using the identity \(\tilde{L}_i \tilde{L}_j - L_i L_j = \tilde{L}_i(\tilde{L}_j - L_j) + (\tilde{L}_i - L_i)L_j\) and the fact that \(\Tr\rho = 1\) with \(\rho \ge 0\), we obtain the elementwise bound
\[
\begin{aligned}
|\widetilde{\mathcal{T}}_{ij} - \mathcal{T}_{ij}|
&\le \|\tilde{L}_i\| \cdot \|\tilde{L}_j - L_j\| + \|\tilde{L}_i - L_i\| \cdot \|L_j\| \\
&\le (\alpha + \epsilon_i)\,\epsilon_j + \epsilon_i\,\alpha = \alpha(\epsilon_i + \epsilon_j) + \epsilon_i\epsilon_j .
\end{aligned}
\]
Converting to the spectral norm via the Frobenius norm,
\[
\|\mathcal{T} - \tilde{\mathcal{T}}\| \le \|\mathcal{T} - \tilde{\mathcal{T}}\|_F
\le \Bigl(\sum_{i,j=1}^{m} \bigl[\alpha(\epsilon_i + \epsilon_j) + \epsilon_i\epsilon_j\bigr]^2\Bigr)^{1/2}.
\]
If one conservatively bounds each \(\epsilon_j \le \epsilon\) , this simplifies to the uniform bound \(\|\mathcal{T} - \tilde{\mathcal{T}}\| \le m(2\alpha\epsilon + \epsilon^2)\).
\end{proof}

A useful property of the construction is that swapping the order of the two controlled block encodings in Eq.~\eqref{eq:qgt-circuit}---replacing \(\mathcal{U}_L \cdot \mathcal{U}_R\) with \(\mathcal{U}_R \cdot \mathcal{U}_L\)---yields a block encoding of the transposed QGT \(\mathcal{T}^\top_{ij} = \Tr(\rho L_j L_i)\).
Once block encodings of both \(\mathcal{T}\) and \(\mathcal{T}^\top\) are available, a block encoding of the QFIM follows directly from the symmetry relation \(\mathcal{F} = 4\Re(\mathcal{T}) = 2(\mathcal{T} + \mathcal{T}^\top)\), in the SLD convention of Eq.~\eqref{eq:definition-of-sld}. Each \(L_i\) is Hermitian and consequently \(\mathcal{T}\) is a Hermitian matrix (\(\mathcal{T}_{ji} = \mathcal{T}_{ij}^*\)). Formally, this is expressed in the following corollary.

\begin{corollary}[Block encoding of the QFIM]\label{thm:qfim-block-encoding}
Under the same assumptions as Theorem~\ref{thm:qgt-block-encoding}, there exists a unitary \(U_{\mathcal{F}}\) that is a \((4m\alpha^2, a_{\mathrm{tot}}+1, 4\epsilon_{\mathrm{tot}})\)-block-encoding of the quantum Fisher information matrix \(\mathcal{F}_{ij} = 4\Re \Tr(\rho L_i L_j)\). The circuit uses one additional control ancilla qubit, one controlled call to each of \(U_{\mathcal{T}}\) and \(U_{\mathcal{T}^\top}\), and two single-qubit Hadamard gates.
\end{corollary}

\begin{proof}
Let \(U_{\mathcal{T}}\) denote the block encoding of \(\mathcal{T}\) from Theorem~\ref{thm:qgt-block-encoding}, satisfying
\[
(\langle 0|^{a_{\mathrm{tot}}} \otimes I_R) \, U_{\mathcal{T}} \, (|0\rangle^{a_{\mathrm{tot}}} \otimes I_R) = \frac{\mathcal{T}}{m\alpha^2} + E_{\mathcal{T}},
\]
with \(\|E_{\mathcal{T}}\| \le \epsilon_{\mathrm{tot}}/(m\alpha^2)\). Let \(U_{\mathcal{T}^\top}\) be the block encoding of \(\mathcal{T}^\top\) with the same parameters. Following the standard LCU construction as described in Lemma~\ref{lem:lcu-symmetric}, the following quantum circuit

\begin{equation}
  U_{\mathcal{F}} = (H \otimes I) \cdot (|0\rangle\langle 0|_c \otimes U_{\mathcal{T}} + |1\rangle\langle 1|_c \otimes U_{\mathcal{T}^\top}) \cdot (H \otimes I),
\end{equation}
has projected block \((\mathcal{T}+\mathcal{T}^\top)/(2m\alpha^2)\) up to error at most \(\epsilon_{\mathrm{tot}}/(m\alpha^2)\). Since \(\mathcal{F}=2(\mathcal{T}+\mathcal{T}^\top)\), this is a \((4m\alpha^2, a_{\mathrm{tot}}+1, 4\epsilon_{\mathrm{tot}})\)-block-encoding of the QFIM. The factor of four changes the normalization and the absolute encoded-operator error, not the circuit or its query count.
\end{proof}

The block encodings of Theorem~\ref{thm:qgt-block-encoding} and Corollary~\ref{thm:qfim-block-encoding} elevate the geometric structure of a parameterized state family from a collection of individually estimated matrix elements into a single quantum-accessible matrix.  They thereby provide coherent matrix inputs for downstream QSVT-based processing~\cite{gilyenQuantumSingularValue2019}, without first reconstructing individual entries classically.  Subject to the spectral promises required by the corresponding linear-algebra routine, matrix inversion can provide coherent access to an inverse or regularized inverse of the QFIM associated with the multiparameter SLD Cram\'er--Rao matrix bound~\cite{helstromQuantumDetectionEstimation1969,holevoProbabilisticStatisticalAspects2011,parisQUANTUMESTIMATIONQUANTUM2009,liuQuantumFisherInformation2020,szczykulskaMultiParameterQuantumMetrology2016,demkowiczBeyondQFI2020}, while eigenvalue processing can identify candidate high-information parameter directions~\cite{liuQuantumFisherInformation2020,szczykulskaMultiParameterQuantumMetrology2016}.  The SLD matrix bound need not be jointly attainable for incompatible parameters, and the ultimate operational precision can instead require the Holevo bound or related formulations~\cite{demkowiczBeyondQFI2020}.  Likewise, because the imaginary part of the QGT encodes the Uhlmann curvature up to the normalization adopted here, singular-value thresholding of its antisymmetric form can certify multiparameter incompatibility when the relevant singular value is separated from zero by a promised finite gap~\cite{houLocalGeometryQuantum2024,wangxiaoguangMixedStateBerry2025,szczykulskaMultiParameterQuantumMetrology2016}.

\subsection{QFI Estimation under the Bounded-Response Promise}\label{sec:reg:qfi_estimation}
Under the bounded-response promise of Section~\ref{sec:bounded-response}, the regularized SLD block encoding of Proposition~\ref{prop:regularized-sld-block-encoding} can be used to estimate the diagonal QFIM entries with a similar routine as in~\ref{estimation-of-the-norms}.  The regularization parameter \(\theta\) controls the bias in the estimate, while the SLD block-encoding error \(\epsilon_j\) and amplitude-estimation error \(\epsilon_{\mathrm{amp}}\) control the other two contributions. For a prescribed additive error \(\epsilon\), we allocate the error budget among the three contributions. For the bias bound \(8M\theta\), one sets \(\theta=\Theta(\epsilon/M)\); controlling the other two contributions gives \(\epsilon_j=\Theta(\epsilon/\sqrt M)\) and \(\epsilon_{\mathrm{amp}}=\Theta(\epsilon^2/(\alpha_j M^{3/2}))\). Combining these allocations with the per-call filter cost yields the following result.

\begin{theorem}[QFI estimation under the bounded-response promise]
  \label{thm:reg:diagonal_qfi}
  Under the access model of Section~\ref{sec:access-model} and the bounded-response promise of Section~\ref{sec:bounded-response}, there exists a quantum algorithm that estimates \(\mathcal{F}_{jj}\) to additive error \(\epsilon\) with success probability at least \(1-\vartheta\).  With the parameter allocation stated above, the query complexities to the purification oracle \(O_\rho\) and the derivative block-encoding \(U_{\partial_j\rho}\), up to polylogarithmic factors in \(\epsilon^{-1}, \vartheta^{-1}, M, \alpha_j\) and the derivative norm, are
  \begin{equation}
    Q_\rho^{\mathrm{total}}
    = \widetilde{\mathcal{O}}\!\left(\frac{\alpha_j M^{7/2}}{\epsilon^4}\right),\qquad
    Q_{\partial_j}^{\mathrm{total}}
    = \widetilde{\mathcal{O}}\!\left(\frac{\alpha_j M^{3/2}}{\epsilon^2}\right).
    \label{eq:reg:final_complexity_qfi}
  \end{equation}
  The circuit uses \(3a + 3s + 5 + m_j + \mathcal{O}\!\left(\log\frac{M}{\epsilon}\right)\) qubits in total and requires \(\widetilde{\mathcal{O}}(a+s+m_j+M/\epsilon)\) additional elementary gates per input-oracle query, where \(a\) and \(s\) are the purification ancilla and system qubit counts, and \(m_j\) is the ancilla count of the derivative block encoding.
\end{theorem}

The regularized route replaces the bounded-gap assumption with a bound on the parameter response, at the cost of a weaker precision scaling and a \(\log(M/\epsilon)\) precision-dependent register width. Appendix~\ref{app:regularized-estimation} gives the intermediate norm bounds and complete error accounting; physical families with exponentially small gaps satisfying the promise are discussed in Appendix~\ref{app:reg:physical}.

Table~\ref{tab:reg:comparison} summarizes the two algorithmic routes for QFI estimation.

\begin{table}[htbp]
  \centering
  \caption{Systematic comparison of the two algorithmic routes for QFI estimation.}
  \label{tab:reg:comparison}
  {\small
  \setlength{\tabcolsep}{3pt}
  \begin{tabular}{@{}p{2.2cm} p{2.95cm} p{2.95cm}@{}}
    \toprule
    Aspect & bounded-gap promise (gapped) & bounded-response promise (regularized) \\
    \midrule
    Precondition &
    \(\Delta \ge 1/\operatorname{poly}(s)\) &
    \(M = \mathcal{O}(\operatorname{poly}(s))\) \\
    && (\(\Delta\) may be exponentially small) \\
    \(e^{x_k}\) compensation & Required & Not required \\
    QSVT degree &
    \(\mathcal{O}(\frac{1}{\Delta}\log\frac{1}{\epsilon_p})\), independent of \(t_k\) &
    \(\mathcal{O}(\frac{K}{\theta}\log\frac{K}{\theta\epsilon_p})\), \(\propto t_k\) \\
    Nodes \(K\) & \(\widetilde{\Theta}(1/\Delta)\) & \(\widetilde{\Theta}(M/\epsilon)\) \\
    \(Q_\rho^{\mathrm{total}}\) &
    \(\widetilde{\mathcal{O}}\!\left(\dfrac{\alpha_j\|\partial_j\rho\|_F}{\epsilon\Delta^{7/2}}\right)\) &
    \(\widetilde{\mathcal{O}}\!\left(\dfrac{\alpha_j M^{7/2}}{\epsilon^4}\right)\) \\
    \(\epsilon\)-scaling & Near-Heisenberg (\(\widetilde{\mathcal{O}}(1/\epsilon)\)) & Polynomial (\(\widetilde{\mathcal{O}}(1/\epsilon^4)\)) \\
    \bottomrule
  \end{tabular}}
\end{table}

\section{Discussion and Conclusion}\label{sec:discussion}

We have presented efficient quantum algorithms for accessing the geometry of mixed states through a coherent realization of the symmetric logarithmic derivative. The central construction uses Gauss--Laguerre quadrature to convert the support-sensitive Lyapunov inversion defining the SLD into a finite linear combination of exponentially filtered density-matrix derivatives. The quadrature converges exponentially with its order, while the normalization required for coherent LCU assembly grows only linearly, allowing this convergence to translate into a genuine quantum resource advantage. Under the bounded-gap promise, the resulting SLD primitive enables near-Heisenberg diagonal QFI estimation and unified block encodings of the full QGT and QFIM; a complementary regularized construction retains polynomial-resource diagonal estimation under the bounded-response promise even when the smallest nonzero eigenvalue is exponentially small. In this way, mixed-state quantum geometry becomes available as a coherent input to subsequent quantum processing rather than as a collection of matrix elements reconstructed classically.

The construction combines favorable precision dependence in oracle query complexity with a small precision-dependent ancillary-qubit overhead under the bounded-gap promise. At fixed conditioning, the exponential convergence of the quadrature makes the required number of nodes logarithmic in the inverse target precision, so the quadrature-index register uses only \(\mathcal{O}(\log\log(1/\epsilon))\) qubits. The QPE-free diagonal estimator preserves this advantage by avoiding the \(\mathcal{O}(\log(1/\epsilon_{\mathrm{amp}}))\)-qubit phase register of conventional QPE-based amplitude estimation and introducing only a constant number of additional ancillas. The complexity bounds also retain the response norm \(|\partial_j\rho|_F\), rather than replacing it by an operator-norm bound with a worst-case rank factor, so favorable low-rank or structured response remains visible. The finite quadrature constructions are also stable under imperfect state preparation: under either conditioning promise, the induced error is polynomially controlled by the preparation error (Appendix~\ref{appendix:robustness}). Table~\ref{tab:sld_solver_comparison} compares the constructions on the same SLD task. At fixed gap and input normalization, the trapezoidal construction of Ref.~\cite{claytonDifferentiableQuantumComputing2024} has an inverse-square-root precision dependence in its query bound, whereas the Somma and Gauss--Laguerre constructions have logarithmic dependence. The Somma and Gauss--Laguerre LCU index widths scale logarithmically and double-logarithmically, respectively, with inverse precision, with the output normalizations listed separately in the table. Whether these query and ancillary-space requirements can be improved remains open.

The two promises of the access model capture complementary notions of conditioning. The bounded-gap promise controls the worst-case spectral inversion and supports the near-Heisenberg diagonal estimator, whereas the bounded-response promise controls the alignment of the parameter derivative with poorly conditioned spectral directions and permits exponentially small spectral gaps at the price of a weaker polynomial dependence on the inverse target precision. The latter condition is restrictive: Appendix~\ref{app:reg:physical} gives sufficient conditions for low-rank unitary encodings and truncated commuting Gibbs families, which also extend to product encodings with a fixed pure environment. An explicit full commuting Gibbs example shows that the promise can fail when the relative response accumulates over exponentially many configurations. The bounded-response promise therefore captures response structure that is invisible to a purely spectral condition, but it does not provide a universal algorithm for small-gap states.

Several limitations of the present work motivate corresponding open questions. First, the framework operates in the locally rank-preserving regime, where the canonical SLD is well defined and the SLD-based QFIM coincides with the Bures metric; at rank-changing points the QFIM may be discontinuous, the regularity assumptions behind the standard Cram\'er--Rao theorem can fail, and the QFIM and Bures metric need not coincide~\cite{safranekDiscontinuitiesQuantumFisher2017,sevesoDiscontinuityQuantumFisher2020}. Whether the framework can be extended to such points through the Bures regularization that restores a well-defined metric there remains open~\cite{safranekDiscontinuitiesQuantumFisher2017}. In the regularized route the situation is subtler: the Tikhonov-regularized SLD remains mathematically well defined at rank-changing points, but the bounded-response promise itself fails there, and with it the complexity guarantees.

More generally, the bounded-response promise enters the analysis as an \textit{a~priori} assumption: we are not aware of a procedure that certifies the bound \(\|\rho^+\partial_j\rho\|_F^2 \le M\), and whether polynomially many queries to the purification and derivative oracles can do so is an open question of oracle separation. The polynomial scaling \(\widetilde{\mathcal{O}}(1/\epsilon^4)\) of the regularized route further raises the question whether it is optimal under the bounded-response promise alone, and whether additional tail-decay conditions---for instance a bound on \(\|(\rho^+)^2\partial_j\rho\|_F\)---improve it toward \(\widetilde{\mathcal{O}}(1/\epsilon^2)\) or even the near-Heisenberg scaling of the gapped route. A related question is whether a two-parameter promise---controlling the diagonal and off-diagonal derivative contributions independently---admits a natural algorithm whose complexity is not significantly worse than the single-parameter \(M\) version. Extending the full-matrix construction to the bounded-response regime is left for future work.

Second, the access model assumes a block encoding of each parameter derivative as a primitive. This oracle can be constructed efficiently for many standard parameterizations when the derivative has a known analytic decomposition into operators admitting efficient quantum access, but such a construction is not automatic. For instance, under a unitary encoding \(\rho(\boldsymbol{x}) = e^{-i\sum_k x_k H_k}\rho_0 e^{i\sum_k x_k H_k}\) with commutative generators, the derivative is the commutator \(\partial_j\rho = -i[H_j,\rho]\), implemented by the LCU construction whenever the generators admit efficient block encodings; for a thermal probe whose parameter of interest is the temperature itself, the derivative \(\partial_\beta\rho = -\rho\bigl(H - \langle H\rangle_\beta\bigr)\) of \(\rho(\beta) = e^{-\beta H}/Z(\beta)\) follows from the product rule and is again built from the state and the Hamiltonian alone. By contrast, when the parameter enters nonlinearly inside a matrix exponential, the derivative generally acquires a Duhamel integral representation~\cite{wilcoxExponentialOperatorsParameter1967}; realizing its block encoding then requires an additional approximation layer and may dominate the end-to-end cost.

Third, the full QFIM block encoding provides a coherent input interface for matrix inversion in quantum natural-gradient optimization, where iterative updates involve an inverse, pseudoinverse, or regularized inverse of the QFIM~\cite{Stokes2020,Gacon2021}; it does not by itself constitute an end-to-end optimization algorithm. The present QGT and QFIM constructions carry subnormalization factors \(m\alpha^2\) and \(4m\alpha^2\), respectively. When \(\alpha\) is independent of \(m\), both therefore scale linearly with the parameter dimension and enter the cost of subsequent QSVT or quantum-linear-system processing. Whether this linear dependence can be improved in a generic access model remains unclear. Under additional promises such as efficiently accessible sparsity, locality, low rank, or structured parameter couplings, constructing QGT/QFIM block encodings with smaller normalization---or another form of coherent quantum access suited to quantum natural-gradient processing---is an open problem.

Fourth, the Gauss--Laguerre weights were handled by a generic amplitude-loading circuit; whether their Christoffel--Darboux structure~\cite{szego1939orthogonal} enables a specialized preparation subroutine is another open question. The complexity bounds are also expressed in terms of generic parameters---\(\Delta\), \(M\), \(\|\partial_j\rho\|_F\). Extending the parameter estimates of Appendix~\ref{app:reg:physical} to broader classes of thermal states and unitary encodings is left for future work.

We also note a physical reading of the quadrature construction. Each kernel \(e^{-t\rho}\) in the SLD integral is, up to normalization, the Gibbs state of the effective Hamiltonian \(\rho\) at inverse temperature \(t\), so that in both the gapped and the regularized discretizations the quadrature nodes define a discrete family of effective temperatures---\(x_{k,K}/\Delta\) and \(x_{k,K}/\theta\), respectively---the two routes differing only in the temperature scale and in the weights multiplying the kernels. This reading suggests a possible complementary route to quantum Fisher information estimation under sampling access, where only independent copies of the probe state are available rather than purification oracles. Implementing this route would first require converting copies of \(\rho\), for example through density-matrix exponentiation, into sufficiently accurate effective-Hamiltonian access to \(\rho\), and then using this access to prepare the constituent Gibbs states with a quantum Gibbs sampler~\cite{poulinSamplingThermalQuantum2009,temmeQuantumMetropolisSampling2011,chowdhuryQuantumAlgorithmsGibbs2017,mottaDeterminingEigenstatesThermal2020,chenEfficientExactNoncommutative2023}. Both steps introduce additional precision-dependent costs, since the cited Gibbs samplers do not operate directly on copies of an unknown \(\rho\). If these costs can be controlled, one could attempt to assemble the quantum Fisher information from correlation functions of the filtered derivative operators, exploiting the same exponential convergence that underlies the block-encoding construction. This sampling route remains, at present, a conceptual direction rather than a complete algorithm, and its end-to-end sample complexity has not been established.

In summary, we have reduced the estimation of mixed-state quantum geometry to a quadrature problem. The implicit Lyapunov equation that defines the SLD---a support-sensitive matrix inversion---becomes, through the Gauss--Laguerre discretization, a finite linear combination of filtered derivative operators whose implementation reduces to standard block-encoding primitives with independently trackable costs, under either conditioning promise of the access model. The resulting end-to-end algorithms keep the mixed-state geometry visible from the integral representation through the block-encoding construction and into the final complexity bounds, expressed in terms of the parameters that govern the metrological value of the probe state: the derivative norms, the conditioning of the state, and the target precision. More broadly, the paper identifies the symmetric logarithmic derivative itself as the natural algorithmic interface for mixed-state geometry: once that interface is made explicit, norm estimation, state preparation, and the assembly of the full quantum geometric tensor become standard quantum-algorithmic subroutines, providing a rigorous baseline for quantum metrology protocols that interrogate the information content of an unknown mixed state without resorting to full tomography. By realizing this interface through an exponentially convergent quadrature whose coherent normalization grows only linearly with its order, the construction further identifies such quadrature as a resource-efficient design principle for structured quantum-information problems with analogous positive-semidefinite integral representations.

\begin{acknowledgments}
\noindent The authors thank Yukun Zhang for insightful discussions.
This work is supported by
Quantum Science and Technology-National Science and Technology Major Project (2023ZD0300200),
the National Natural Science Foundation of China Grant (No.~12361161602, 22303005),
NSAF (Grant No.~U2330201),
Beijing Natural Science Foundation Z250004,
Beijing Science and Technology Planning Project (Grant No.~Z25110100810000), the
Startup Fund of the Institute of High Energy Physics
and the High-performance Computing Platform of Peking University.

The authors used AI tools from OpenAI and DeepSeek to assist with developing and checking mathematical arguments, identifying relevant literature, revising and typesetting the manuscript. The tools were directed through author-specified questions and assumptions. All AI-assisted content retained in the manuscript was reviewed and verified by the authors. The authors made all final scientific decisions and take full responsibility for the accuracy, originality, and content of this work.
\end{acknowledgments}

\section*{Data Availability}
The numerical data and scripts used to generate the figures, together with the Lean~4 code for the formal verification of the asymptotic relation \(\beta_K/(4K)\to 1\) in Theorem~\ref{theorem:quadrature-main}, are available from the authors upon reasonable request.

\appendix

\section{Conditioning Promises Imply Local Rank Preservation}
\label{appendix:promises-imply-local-rank-preservation}

In this appendix we justify the claim made in Section~\ref{sec:access-model} that each of the two conditioning promises implies local rank preservation on the neighborhood \(\mathcal{X}\).

\begin{proposition}
\label{prop:promises-imply-local-rank-preservation}
Let \(\rho(\boldsymbol{x})\) be a \(C^1\) family of density operators on a finite-dimensional Hilbert space, parameterized on an open convex neighborhood \(\mathcal{X}\ni \boldsymbol{x}_0\).
\begin{enumerate}
    \item If the bounded-gap promise in Eq.~\eqref{eq:vartheta-intro} holds, then \(\rho(\boldsymbol{x})\) has constant rank on \(\mathcal{X}\).
    \item If the bounded-response promise in Eq.~\eqref{eq:derivative-bound-intro} holds, then \(\rho(\boldsymbol{x})\) has constant rank on \(\mathcal{X}\).
\end{enumerate}
Consequently, under either promise one has
\[
\begin{aligned}
P_{\mathcal K(\boldsymbol{x})}\, \partial_j\rho(\boldsymbol{x})\,
P_{\mathcal K(\boldsymbol{x})}&=0\qquad\forall \boldsymbol{x}\in \mathcal{X}, j\in[m].
\end{aligned}
\]
\end{proposition}

\begin{proof}
We prove the two parts separately.

\textit{Part 1: the bounded-gap promise.}
Suppose Eq.~\eqref{eq:vartheta-intro} holds. Then every nonzero eigenvalue of \(\rho(\boldsymbol{x})\) is bounded below by \(\Delta\) uniformly on \(\mathcal{X}\). Hence no positive eigenvalue can approach zero anywhere in \(\mathcal{X}\). Since the eigenvalues depend continuously on \(\boldsymbol{x}\) and \(\mathcal{X}\) is connected (being convex), the number of strictly positive eigenvalues is constant on \(\mathcal{X}\). Therefore the rank of \(\rho(\boldsymbol{x})\) is constant on \(\mathcal{X}\).

\textit{Part 2: the bounded-response promise.}
We work with the eigenvalues \(\lambda_1(\boldsymbol{x})\ge\cdots\ge\lambda_N(\boldsymbol{x})\) of \(\rho(\boldsymbol{x})\), ordered nonincreasingly so that they are continuous in \(\boldsymbol{x}\) (Weyl's inequality) and Lipschitz, hence absolutely continuous and differentiable almost everywhere, along any segment. Fix a unit direction \(\boldsymbol{v}=(v_1,\ldots,v_m)\) and write \(\partial_{\boldsymbol{v}}=\sum_j v_j\partial_j\). The bounded-response promise and the Cauchy--Schwarz inequality give the directional bound
\[
\begin{aligned}
\|\rho^{+}\partial_{\boldsymbol{v}}\rho\|_F
&\le \sum_{j=1}^{m}|v_j|\,\|\rho^{+}\partial_j\rho\|_F\\
&\le \left(\sum_{j=1}^{m}|v_j|^2\right)^{1/2}
\left(\sum_{j=1}^{m}\|\rho^{+}\partial_j\rho\|_F^2\right)^{1/2}\\
&\le \sqrt{mM}.
\end{aligned}
\]
Now fix \(\boldsymbol{x}^{\ast}\in\mathcal{X}\), let \(\ell=|\boldsymbol{x}^{\ast}-\boldsymbol{x}_0|\), and, when \(\ell>0\), set \(\boldsymbol{v}=(\boldsymbol{x}^{\ast}-\boldsymbol{x}_0)/\ell\) and \(\gamma(u)=\boldsymbol{x}_0+u\boldsymbol{v}\) for \(u\in[0,\ell]\). Convexity ensures that \(\gamma([0,\ell])\subset\mathcal{X}\). At a point where an ordered eigenvalue \(\lambda_k(\gamma(u))=\lambda>0\) is differentiable, the standard first-order perturbation formula for a possibly degenerate Hermitian eigenvalue states that \(\partial_u\lambda_k\) is an eigenvalue of the compressed derivative \(P_\lambda(\partial_{\boldsymbol{v}}\rho)P_\lambda\). Therefore
\[
\begin{aligned}
|\partial_u\lambda_k|
&\le\|P_\lambda\,\partial_{\boldsymbol{v}}\rho\,P_\lambda\|\\
&=\lambda\,\|P_\lambda\,\rho^{+}\partial_{\boldsymbol{v}}\rho\,P_\lambda\|\\
&\le \lambda\,\|\rho^{+}\partial_{\boldsymbol{v}}\rho\|\\
&\le \lambda\sqrt{mM},
\end{aligned}
\]
using \(P_\lambda\rho^{+}P_\lambda=\lambda^{-1}P_\lambda\) on the eigenspace, the contraction property of the operator norm under projections, and \(\|X\|\le\|X\|_F\). Hence
\[
\begin{aligned}
|\partial_u\log\lambda_k(\gamma(u))|
&\le\sqrt{mM}\\
&\qquad\text{almost everywhere on}\\
&\qquad\{u:\lambda_k(\gamma(u))>0\},
\end{aligned}
\]
so that \(\log\lambda_k\) is Lipschitz with constant \(\sqrt{mM}\) on every interval where \(\lambda_k>0\).

We now show that no ordered eigenvalue can pass between the positive spectrum and zero anywhere in \(\mathcal{X}\). Suppose \(\lambda_k(\boldsymbol{x}^\ast)=0\) while \(\lambda_k(\boldsymbol{x}_0)>0\), and let \(u_0=\min\{u\in[0,\ell]:\lambda_k(\gamma(u))=0\}\). For \(u<u_0\), \(\lambda_k(\gamma(u))>0\) and the Lipschitz bound integrates to
\[
\begin{aligned}
\log\lambda_k(\gamma(u))
&\ge\log\lambda_k(\gamma(0))
-\sqrt{mM}\,u,\\
&\qquad u<u_0,
\end{aligned}
\]
whose right-hand side stays finite as \(u\to u_0^-\), contradicting \(\lambda_k(\gamma(u_0))=0\). Thus no positive eigenvalue vanishes in \(\mathcal{X}\). Applying the same argument to the reversed segment shows that a zero eigenvalue cannot become positive either. Each ordered eigenvalue is therefore either positive throughout \(\mathcal{X}\) or zero throughout \(\mathcal{X}\), and the rank is constant on \(\mathcal{X}\).

Finally, the kernel condition follows directly from positivity and differentiability. Fix \(\boldsymbol{x}\in\mathcal{X}\), a parameter direction \(j\), and \(\ket{\psi}\in\mathcal{K}(\boldsymbol{x})\). Since \(\mathcal{X}\) is open, the function
\[
f_\psi(t)=\bra{\psi}\rho(\boldsymbol{x}+t\boldsymbol{e}_j)\ket{\psi}
\]
is defined for all sufficiently small positive and negative \(t\). It is nonnegative and satisfies \(f_\psi(0)=0\), so differentiability gives
\[
\bra{\psi}\partial_j\rho(\boldsymbol{x})\ket{\psi}=f_\psi'(0)=0.
\]
The Hermitian operator \(P_{\mathcal K}\partial_j\rho P_{\mathcal K}\) therefore has vanishing quadratic form on every vector in \(\mathcal{K}(\boldsymbol{x})\), and hence vanishes identically. This is exactly the kernel condition in Eq.~\eqref{eq:condition-for-sld-to-be-well-defined}.
This proves the claim.
\end{proof}

\section{Discontinuities of the SLD-based QFIM and the Bures Metric}
\label{appendix:bures-discontinuities}

The preceding proposition establishes that both conditioning promises guarantee local rank preservation. We now explain why this condition is necessary to identify the SLD-based QFIM with the Bures metric, and what happens outside the rank-preserving regime.

Under the twice-differentiability assumptions required for the eigenvalue Hessians below, \v{S}afr\'anek showed~\cite{safranekDiscontinuitiesQuantumFisher2017} that the Bures metric \(g_{jk}\) and the SLD-based QFIM \(\mathcal{F}_{jk}\) are related by
\begin{equation}
  g_{jk} = \frac{1}{4}\,\mathcal{F}_{jk} + \frac{1}{2}\sum_{n=r+1}^{N} \partial_{jk} p_n(\boldsymbol{x}),
  \label{eq:bures-qfim-relation-appendix}
\end{equation}
where \(\partial_{jk}p_n\) denotes the second derivative of the \(n\)th eigenvalue. The sum runs over the kernel eigenvalues---those that vanish at \(\boldsymbol{x}\). Writing \(\mathscr{H}_{jk}^{(n)} = \partial_{jk}p_n(\boldsymbol{x})\), each \(\mathscr{H}^{(n)}\) is positive semidefinite when \(p_n=0\)~\cite{safranekDiscontinuitiesQuantumFisher2017}, so the kernel contribution can only increase the metric relative to the QFIM term. One may also express the metric as

\begin{equation}
  g = \frac{1}{4} \mathcal{F} + \frac{1}{2} \sum_{n=r+1}^{N} \mathscr{H}^{(n)} .
  \label{eq:connection-between-bures-metric-and-qfim-with-hessian}
\end{equation}

The equality between Bures metric and SLD-based QFIM holds exactly when the kernel contribution vanishes. Concretely,
\begin{equation}
\begin{aligned}
g = \frac{1}{4} \mathcal{F}
\;&\Longleftrightarrow\;
\partial_{jj} p_n = 0 \\
&\text{for all} \quad j \in [m], \; n \in \{r+1, r+2, \cdots, N\}.
\end{aligned}
  \label{eq:condition-for-bures-metric-to-be-equal-to-qfim}
\end{equation}

Under local rank preservation, every eigenvalue that vanishes at \(\boldsymbol{x}\) remains identically zero in a neighborhood of \(\boldsymbol{x}\). Its first and second derivatives therefore vanish, and Eq.~\eqref{eq:bures-qfim-relation-appendix} gives \(g=\mathcal{F}/4\).

In this regime the SLD-based QFIM and the Bures metric contain the same information up to a constant factor, and the canonical SLD is the appropriate local geometric object.

At a rank-changing point, by contrast, the kernel Hessian terms in Eq.~\eqref{eq:bures-qfim-relation-appendix} need not vanish, and the QFIM may exhibit discontinuities while the Bures metric remains continuous~\cite{safranekDiscontinuitiesQuantumFisher2017,sevesoDiscontinuityQuantumFisher2020}. The usual regularity hypotheses behind the Cram\'er--Rao theorem can then fail, so the discontinuous QFI expression cannot be interpreted naively as an attainable precision bound~\cite{sevesoDiscontinuityQuantumFisher2020}. In such a regime the canonical SLD-based QFIM alone is no longer sufficient to capture the local geometry of the state manifold; a proper treatment requires the Bures-metric regularization framework developed in Ref.~\cite{safranekDiscontinuitiesQuantumFisher2017}. The present work is therefore confined to the locally rank-preserving setting, where the canonical SLD is well defined and the two geometric objects coincide.

\section{Block-Encoding Tools and Conventions}\label{app:implementation-tools}

We use the standard block-encoding framework for quantum algorithms~\cite{gilyenQuantumSingularValue2019}, which provides a convenient way to represent and manipulate operators in a quantum circuit. We briefly review the definition of block-encodings and the relevant primitives for our construction. For a comprehensive introduction to block-encodings, we refer the reader to~\cite{gilyenQuantumSingularValue2019}.

The block encoding is defined as follows:
\begin{definition}[Block Encoding]
\label{def:block-encoding}
Let \(A\) be an \(s\)-qubit operator. An \((\alpha, a, \epsilon)\)-block-encoding of \(A\) is a unitary \(U\) acting on (\(s+a\)) qubits, such that
\begin{equation}
\| A - \alpha (\langle 0|^{\otimes a} \otimes I) U (|0\rangle^{\otimes a} \otimes I) \| \le \epsilon,
\label{eq:block-encoding-definition}
\end{equation}
where \(\alpha \ge \|A\|\) is the normalization factor, \(a\) is the number of ancilla qubits, and \(\epsilon\) is the block encoding error.
\end{definition}
With block encodings of operators at hand, we can perform various arithmetic operations on them, such as addition and multiplication using standard techniques. For instance, given block encodings of two operators \(A\) and \(B\), we can construct a block encoding of their product \(AB\) using the following lemma:

\begin{lemma}[Product of two block-encoded matrices~{\cite[Lemma 53]{gilyenQuantumSingularValue2019}}]
  \label{lemma:block-encoding-product}Given \(U_A\)(an
\((\alpha_A, a, \epsilon_A)\)-block encoding of \(A\)) and \(U_B\)(a \((\alpha_B, b, \epsilon_B)\)-block encoding of \(B\)), \((I_b \otimes U_A)(U_b \otimes I_a)\) is an \((\alpha_A\alpha_B, a+b, \alpha_A\epsilon_B + \alpha_B\epsilon_A)\)-block encoding of \(AB\).
\end{lemma}

Besides the product, we will also use the linear combination of unitaries (LCU) technique to combine multiple block-encoded operators with classical coefficients~\cite{berrySimulatingHamiltonianDynamics2015,childsQuantumAlgorithmSystems2017,gilyenQuantumSingularValue2019}. The LCU construction allows us to create a block encoding of a linear combination of operators given block encodings of the individual operators and a state preparation oracle for the coefficients. In our setting the coefficients are non-negative and we only need the case where both preparation oracles are identical. We therefore state a streamlined version of the standard LCU lemma, adapted to a single symmetric state preparation unitary.

\begin{definition}[State preparation unitary]
\label{def:state-preparation-unitary}
Let \(y\in \mathbb{C}^m\) and \(\|y\|_1\le \beta\). A unitary \(P\) acting on \(b\) qubits is called a \((\beta, b, \varepsilon)\)-state-preparation unitary for \(y\) if \(P\ket{0}^{\otimes b}= \sum_{j=0}^{2^b-1} c_j \ket{j}\) such that \(\sum_{j=0}^{m-1} |\beta |c_j|^2 - y_j| \le \varepsilon\) and \(c_j = 0\) for all \(j\ge m\).
\end{definition}

\begin{lemma}[LCU with symmetric state preparation, adapted from~{\cite[Lemma 52]{gilyenQuantumSingularValue2019}}]\label{lem:lcu-symmetric}
Let \(A = \sum_{j=0}^{m-1} y_j A_j\) be an \(s\)-qubit operator, and let \(\varepsilon_1,\varepsilon_2 > 0\). Suppose that \(P\) is a \((\beta, b, \varepsilon_1)\)-state-preparation unitary for \(y\), and \(W = \sum_{j=0}^{m-1} \ket{j}\bra{j}\otimes U_j + \bigl((I-\sum_{j=0}^{m-1}\ket{j}\bra{j})\otimes I_a\otimes I_s\bigr)\) is an \((s+a+b)\)-qubit unitary such that each \(U_j\) is an \((\alpha, a, \varepsilon_2)\)-block-encoding of \(A_j\). Then \((P^\dagger \otimes I_a\otimes I_s)\, W\, (P \otimes I_a\otimes I_s)\) is an \((\alpha\beta, a+b, \alpha\varepsilon_1 + \beta\varepsilon_2)\)-block encoding of \(A\).
\end{lemma}
\begin{proof}
Let \(P\ket{0}^{\otimes b} = \sum_j c_j\ket{j}\). Then
\begin{align*}
&\Bigl\|A - \alpha\beta\bigl(\bra{0}^{\otimes b}\otimes\bra{0}^{\otimes a}\otimes I\bigr) \\
&\qquad \times (P^\dagger\otimes I\otimes I)\,W \\
&\qquad \times (P\otimes I\otimes I)\bigl(\ket{0}^{\otimes b}\otimes\ket{0}^{\otimes a}\otimes I\bigr)\Bigr\| \\
&= \Bigl\|A - \alpha\sum_{j=0}^{m-1}\beta |c_j|^2
   \bigl(\bra{0}^{\otimes a}\otimes I\bigr)U_j\bigl(\ket{0}^{\otimes a}\otimes I\bigr)\Bigr\| \\
&\le \alpha\!\sum_{j=0}^{m-1}|\beta |c_j|^2 - y_j| \\
&\quad + \Bigl\|A - \alpha\!\sum_{j=0}^{m-1}y_j
   \bigl(\bra{0}^{\otimes a}\otimes I\bigr)U_j\bigl(\ket{0}^{\otimes a}\otimes I\bigr)\Bigr\| \\
&\le \alpha\varepsilon_1 + \sum_{j=0}^{m-1} |y_j| \,\varepsilon_2
   \le \alpha\varepsilon_1 + \beta\varepsilon_2 .
\end{align*}
The first equality uses \(c_j=0\) for \(j\ge m\) to restrict the sum. The first inequality separates the coefficient mismatch from the operator error by the triangle inequality. The second inequality follows from
\(\bigl\|A - \alpha\sum_j y_j(\bra{0}^{\otimes a}\otimes I)U_j(\ket{0}^{\otimes a}\otimes I)\bigr\|
= \bigl\|\sum_j y_j\bigl(A_j - \alpha(\bra{0}^{\otimes a}\otimes I)U_j(\ket{0}^{\otimes a}\otimes I)\bigr)\bigr\|
\le \sum_j |y_j|\,\varepsilon_2\), and the final step uses \(\sum_j |y_j| \le \beta\).
\end{proof}

The Quantum Singular Value Transformation (QSVT)~\cite{gilyenQuantumSingularValue2019} applies a polynomial to the singular values of an arbitrary block-encoded operator. We state the version for real polynomials, which suffices for all constructions in this work.

\begin{lemma}[QSVT with real polynomials, adapted from~{\cite[Corollary~5]{gilyenQuantumSingularValue2019}}]
\label{lemma:qsvt-general}
Let \(U\) be an \((\alpha, a, 0)\)-block-encoding of an operator \(A\) (not necessarily Hermitian), and let \(P\in \mathbb{R}[x]\) be a real polynomial of degree \(d\) such that \(|P(x)| \le 1\) for all \(x\in [-1,1]\) and \(P\) has parity \((d\bmod 2)\). Then there exists a unitary circuit \(U_P\) that
\begin{itemize}
    \item uses \(d\) queries to \(U\) and \(U^\dagger\), one additional ancilla qubit, and \(\mathcal{O}(d(a+1))\) elementary gates,
    \item and satisfies
    \[
    (\bra{0}^{\otimes a+1}\otimes I)\,U_P\,(\ket{0}^{\otimes a+1}\otimes I)
    =
    P^{(SV)}(A/\alpha),
    \]
    where \(P^{(SV)}\) denotes singular value transformation: if \(A = \sum_i \sigma_i \ket{w_i}\!\bra{v_i}\) is a singular value decomposition (with \(\sigma_i \ge 0\)), then
    \[
    P^{(SV)}(A/\alpha) =
    \begin{cases}
    \sum_i P(\sigma_i/\alpha)\,\ket{w_i}\!\bra{v_i}, & d\text{ odd},\\[4pt]
    \sum_i P(\sigma_i/\alpha)\,\ket{v_i}\!\bra{v_i}, & d\text{ even}.
    \end{cases}
    \]
\end{itemize}
In particular, \(U_P\) is a \((1,a+1,0)\)-block-encoding of \(P^{(SV)}(A/\alpha)\).
\end{lemma}

For a Hermitian operator \(A\), the eigenvalues and singular values coincide up to sign, and quantum eigenvalue transformation is recovered as a special case of QSVT~\cite{linNearoptimalGroundState2020}.

The parity constraint in Lemma~\ref{lemma:qsvt-general} can be lifted when the target matrix is Hermitian, at the expense of a constant subnormalization factor. The following lemma, adapted from Ref.~\cite{gilyenQuantumSingularValue2019}, is used in the regularized construction of Section~\ref{sec:reg:be_construction}.

\begin{lemma}[Polynomial eigenvalue transformation of arbitrary parity, adapted from~{\cite[Theorem~56]{gilyenQuantumSingularValue2019}}]
\label{lemma:qsvt-arbitrary-parity}
Let \(U\) be an \((\alpha, a, 0)\)-block-encoding of a Hermitian matrix \(A\), and let \(P\in \mathbb{R}[x]\) be a degree-\(d\) polynomial such that \(|P(x)| \le 1/2\) for all \(x\in [-1,1]\).
Then there exists a unitary circuit \(\tilde{U}\) that
\begin{itemize}
    \item uses \(d\) queries to \(U\) and \(U^\dagger\), one controlled-\(U\) query, two additional ancilla qubits, and \(\mathcal{O}(d(a+1))\) elementary gates,
    \item and is a \((1, a+2, 0)\)-block-encoding of \(P(A/\alpha)\).
\end{itemize}
\end{lemma}

\section{QPE-free Amplitude Estimation}\label{app:qpe-free-amplitude-estimation}

The estimator in Section~\ref{estimation-of-the-norms} first converts a comparison of the unknown amplitude with a
threshold into a nearly deterministic measurement. The construction below
adapts the QPE-free binary-amplitude primitive of
Ref.~\cite{dongGroundStatePreparationEnergy2022} to the present oracle
interface.

\begin{lemma}[QSVT amplitude thresholding]
\label{lemma:thresholding}
Let \(U\) satisfy
\begin{equation}
  U\ket{0^a}\ket{0}_S
  =\gamma\ket{0^a}\ket{\phi}
  +\sqrt{1-\gamma^2}\ket{\perp},
\label{eq:block-encoding-like-relation-for-thresholding}
\end{equation}
where \(\gamma\in[0,1]\) and \(\ket{\perp}\) is orthogonal to the zero state of the first register. Let \(\mu,h,\eta\in(0,1)\), with \(0<h<\min\{\mu,1-\mu\}\). Under the promise that either \(\gamma\le\mu-h\) or \(\gamma\ge\mu+h\), there is a quantum procedure that outputs \(b=1\) in the first case and \(b=0\) in the second with probability at least \(1-\eta\). It uses
\begin{equation}
  \mathcal O\!\left(\frac{1}{h}\log\frac{1}{\eta}\right)
  \label{eq:appendix-threshold-query-complexity}
\end{equation}
queries to \(U\) and \(U^\dagger\) and one additional ancilla qubit.
\end{lemma}

\begin{proof}
According to Ref.~\cite[Lemma~29]{gilyenQuantumSingularValue2019}, for any
\(\epsilon'\in(0,1/2)\), there exists an even polynomial
\(P^{\mathrm{Th}}_{\epsilon',h,\mu}\) of degree
\begin{equation}
  \mathcal O\!\left(\frac{1}{h}\log\frac{1}{\epsilon'}\right)
  \label{eq:appendix-threshold-polynomial-degree}
\end{equation}
that is bounded by one on \([-1,1]\) and satisfies
\begin{equation}
\begin{aligned}
  P^{\mathrm{Th}}_{\epsilon',h,\mu}(x)&\in[1-\epsilon',1]
  &&\text{for }|x|\le\mu-h,\\
  P^{\mathrm{Th}}_{\epsilon',h,\mu}(x)&\in[0,\epsilon']
  &&\text{for }\mu+h\le|x|\le1.
\end{aligned}
\label{eq:appendix-threshold-polynomial-properties}
\end{equation}
To expose \(\gamma\) as an eigenphase, define
\begin{equation}
\begin{aligned}
  \Pi=\ket{0^a}\!\bra{0^a}\otimes I_S,
  \qquad& R_{\mathrm{good}}=2\Pi-I,\\
  &R_{\mathrm{in}}=2\ket{0^a0_S}\!\bra{0^a0_S}-I,
\end{aligned}
  \label{eq:appendix-amplitude-reflections}
\end{equation}
and the amplitude-amplification walk
\begin{equation}
  G=R_{\mathrm{good}}UR_{\mathrm{in}}U^\dagger.
  \label{eq:appendix-amplitude-walk}
\end{equation}
On the two-dimensional invariant subspace containing
\(\ket{\psi}=U\ket{0^a0_S}\), the eigenvalues of \(G\) are
\(e^{\pm2i\arccos\gamma}\).  Writing \(G=e^{-iH_G}\)
on this subspace therefore gives
\(\cos(H_G/2)=\gamma I\).  The unitary eigenvalue-transformation
construction of Ref.~\cite{dongGroundStatePreparationEnergy2022} realizes the even polynomial with one signal ancilla. For even degree \(d\), let \(C_G=\ket{0}\!\bra{0}_q\otimes I+\ket{1}\!\bra{1}_q\otimes G\). There are symmetric phases \(\phi_\ell=\phi_{d-\ell}\) such that
\begin{equation}
\begin{aligned}
 U_{\mathrm{Th}}={}&e^{i\phi_d X_q}C_G^\dagger e^{i\phi_{d-1}X_q}C_G\cdots\\
 &\qquad\cdots C_G^\dagger e^{i\phi_1X_q}C_G e^{i\phi_0X_q}.
\end{aligned}
\label{eq:appendix-threshold-qetu-sequence}
\end{equation}
On each walk eigenstate with eigenphase \(\lambda\), the projected signal block is \(P^{\mathrm{Th}}_{\epsilon',h,\mu}(\cos(\lambda/2))\). Since \(\cos(\lambda/2)=\gamma\) on the invariant subspace, this gives
\begin{equation}
  (\bra{0}_q\otimes I_{AS})U_{\mathrm{Th}}
  (\ket{0}_q\otimes\ket{\psi})
  =P^{\mathrm{Th}}_{\epsilon',h,\mu}(\gamma)\ket{\psi}.
  \label{eq:appendix-threshold-qsvt-block}
\end{equation}
Consequently, measuring the signal ancilla after applying
\(U_{\mathrm{Th}}\) produces the zero outcome with probability
\begin{equation}
  p_{\mathrm{Th}}
  =\left|P^{\mathrm{Th}}_{\epsilon',h,\mu}(\gamma)\right|^2.
  \label{eq:appendix-threshold-success-probability}
\end{equation}
For \(\epsilon'=1/10\), this probability is at least \(0.81\) when \(\gamma\le\mu-h\) and at most \(0.01\) when \(\gamma\ge\mu+h\). Assigning the zero measurement outcome to \(b=1\) and the nonzero outcome to \(b=0\), then repeating the experiment \(\mathcal{O}(\log(1/\eta))\) times and taking a majority vote, distinguishes the two promised cases with failure probability at most \(\eta\). Multiplying the repetition count by the polynomial degree gives Eq.~\eqref{eq:appendix-threshold-query-complexity}, because each query to \(G\) uses one query each to \(U\) and \(U^\dagger\).
\end{proof}

\begin{figure*}[t]
\centering
\begin{quantikz}[row sep=0.32cm, column sep=0.24cm]
\lstick{$\ket{0}_{q}$}
  & \qw & \gate{R_x(-2\phi_0)} & \ctrl{1}
  & \gate{R_x(-2\phi_1)} & \ctrl{1}
  & \cdots & \gate{R_x(-2\phi_d)} & \meter{} \\
\lstick{$\ket{0^a}_{A}$}
  & \gate[2]{U} & \qw & \gate[2]{G}
  & \qw & \gate[2]{G^\dagger}
  & \cdots & \qw & \qw \\
\lstick{$\ket{0}_{S}$}
  & & \qw &
  & \qw &
  & \cdots & \qw & \qw
\end{quantikz}
\caption{One QPE-free amplitude-threshold experiment. The input unitary prepares \(\ket{\psi}=U\ket{0^a0_S}\), after which the signal ancilla \(q\) drives the symmetric QETU sequence in Eq.~\eqref{eq:appendix-threshold-qetu-sequence}. Here \(G=R_{\mathrm{good}}UR_{\mathrm{in}}U^\dagger\), so each walk query uses \(U\), \(U^\dagger\), and the two projector reflections. Measuring \(q=0\) has probability \(p_{\mathrm{Th}}=|P^{\mathrm{Th}}_{\epsilon',h,\mu}(\gamma)|^2\), which is at least \(0.81\) for \(\gamma\le\mu-h\) and at most \(0.01\) for \(\gamma\ge\mu+h\) when \(\epsilon'=1/10\). Repetition and majority voting produce the threshold bit of Lemma~\ref{lemma:thresholding}.}
\label{fig:qsvt-amplitude-thresholding}
\end{figure*}
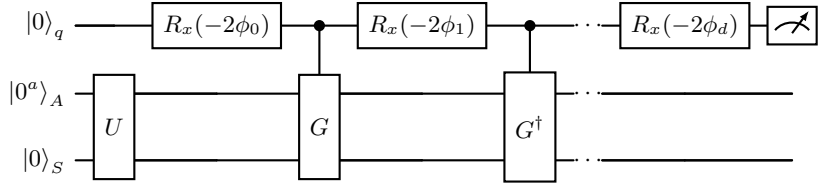

Figure~\ref{fig:qsvt-amplitude-thresholding} shows one threshold experiment. The input unitary first prepares \(\ket{\psi}\). Signal-qubit rotations \(R_x(-2\phi_\ell)=e^{i\phi_\ell X}\) with symmetric phases then alternate with controlled applications of \(G\) and \(G^\dagger\); after expanding Eq.~\eqref{eq:appendix-amplitude-walk}, these are alternating queries to \(U\) and \(U^\dagger\) separated by the two projector reflections. The phases encode \(P^{\mathrm{Th}}_{\epsilon',h,\mu}\), and the signal-qubit measurement samples the probability in Eq.~\eqref{eq:appendix-threshold-success-probability}.

The threshold has a fuzzy region \((\mu-h,\mu+h)\), within which either
output is admissible. Algorithm~\ref{alg:fuzzy-bisection} exploits this
freedom by retaining overlapping subintervals, following the fuzzy-bisection
strategy of Refs.~\cite{linNearoptimalGroundState2020,dongGroundStatePreparationEnergy2022}.

\begin{algorithm}[H]
  \caption{QPE-free amplitude estimation by fuzzy bisection}
  \label{alg:fuzzy-bisection}
  \begin{algorithmic}[1]
    \Require Unitary \(U\) satisfying
    Eq.~\eqref{eq:block-encoding-like-relation-for-thresholding}, target error
    \(\epsilon_{\mathrm{amp}}\), and failure probability
    \(\vartheta_{\mathrm{amp}}\).
    \Ensure Estimate \(\widehat\gamma\) with
    \(|\widehat\gamma-\gamma|\le\epsilon_{\mathrm{amp}}\).
    \State Set
    \(R_{\mathrm{amp}}\gets
    \lceil\log_{3/2}(1/(2\epsilon_{\mathrm{amp}}))\rceil\),
    \(L_0\gets0\), and \(R_0\gets1\).
    \For{\(r=0,\ldots,R_{\mathrm{amp}}-1\)}
      \State \(\mu_r\gets(L_r+R_r)/2\),
      \(h_r\gets(R_r-L_r)/6\), and
      \(\eta_r\gets\vartheta_{\mathrm{amp}}/R_{\mathrm{amp}}\).
      \State Apply Lemma~\ref{lemma:thresholding} with
      \((\mu,h,\eta)=(\mu_r,h_r,\eta_r)\) and obtain \(b_r\).
      \If{\(b_r=1\)}
        \State \([L_{r+1},R_{r+1}]\gets[L_r,\mu_r+h_r]\).
      \Else
        \State \([L_{r+1},R_{r+1}]\gets[\mu_r-h_r,R_r]\).
      \EndIf
    \EndFor
    \State \Return
    \(\widehat\gamma\gets(L_{R_{\mathrm{amp}}}+R_{R_{\mathrm{amp}}})/2\).
  \end{algorithmic}
\end{algorithm}

\begin{proof}[Proof of Lemma~\ref{lem:qpe-free-additive-amplitude-estimation}]
Consider the event that every threshold call obeys its certified implication: \(b_r=1\) implies \(\gamma<\mu_r+h_r\), while \(b_r=0\) implies \(\gamma>\mu_r-h_r\). Outside the fuzzy region these implications follow from Lemma~\ref{lemma:thresholding}; inside it they hold for either output. Since each call fails with probability at most \(\eta_r=\vartheta_{\mathrm{amp}}/R_{\mathrm{amp}}\), the union bound shows that all implications hold simultaneously with probability at least \(1-\vartheta_{\mathrm{amp}}\).

Condition on this event. We prove inductively that
\(\gamma\in[L_r,R_r]\). The claim is true for \([L_0,R_0]=[0,1]\). If
\(b_r=1\), then \(\gamma<\mu_r+h_r\), so the interval
\([L_r,\mu_r+h_r]\) retains \(\gamma\). If \(b_r=0\), then
\(\gamma>\mu_r-h_r\), so \([\mu_r-h_r,R_r]\) retains it. In either case,
\begin{equation}
  R_{r+1}-L_{r+1}
  =\frac{R_r-L_r}{2}+\frac{R_r-L_r}{6}
  =\frac{2}{3}(R_r-L_r).
  \label{eq:appendix-fuzzy-bisection-contraction}
\end{equation}
The final width is therefore \((2/3)^{R_{\mathrm{amp}}}\le
2\epsilon_{\mathrm{amp}}\), and its midpoint satisfies
\(|\widehat\gamma-\gamma|\le\epsilon_{\mathrm{amp}}\).

It remains to count queries. At round \(r\),
\begin{equation}
  h_r=\frac{1}{6}\left(\frac{2}{3}\right)^r,
  \label{eq:appendix-fuzzy-bisection-width}
\end{equation}
so Lemma~\ref{lemma:thresholding} uses
\begin{equation}
  \mathcal O\!\left(
    \left(\frac{3}{2}\right)^r
    \log\!\left(\frac{R_{\mathrm{amp}}}
    {\vartheta_{\mathrm{amp}}}\right)
  \right)
  \label{eq:appendix-fuzzy-bisection-round-cost}
\end{equation}
queries to \(U\) and \(U^\dagger\). Summing the geometric series gives
\begin{align}
  Q_{\mathrm{amp}}
  &=\mathcal O\!\left(
    \left(\frac{3}{2}\right)^{R_{\mathrm{amp}}}
    \left(
      \log\frac{1}{\vartheta_{\mathrm{amp}}}
      +\log R_{\mathrm{amp}}
    \right)
  \right) \nonumber\\
  &=\mathcal O\!\left(
    \frac{1}{\epsilon_{\mathrm{amp}}}
    \left(
      \log\frac{1}{\vartheta_{\mathrm{amp}}}
      +\log\log\frac{1}{\epsilon_{\mathrm{amp}}}
    \right)
  \right),
  \label{eq:appendix-fuzzy-bisection-total-cost}
\end{align}
which is Eq.~\eqref{eq:query-complexity-for-binary-amplitude-estimation}. Each threshold experiment uses the same single signal ancilla shown in Fig.~\ref{fig:qsvt-amplitude-thresholding}; the projector reflections can be compiled without another clean register, and the bisection intervals and majority votes are stored classically.
\end{proof}

\section{Self-Certifying the Spectral-Gap Parameter \texorpdfstring{\(\Delta\)}{Delta}}
\label{appendix:self-certifying-delta}

We prove the certification statement used in Section~\ref{sec:access-model} by adapting the eigenstate-filtering technique of Ref.~\cite{linNearoptimalGroundState2020} to purified access. Here certification means producing, under the bounded-gap promise, a numerical lower bound on \(p_{\min}^+(\boldsymbol{x}_0)\) together with a constant-factor interval containing this eigenvalue. Write \(\Delta_{\mathrm{floor}}>0\) for a known inverse-polynomial search scale guaranteed by the promise. It fixes the maximal search depth, while \(\hat\Delta\) denotes the lower bound determined by the oracle measurements.

\begin{lemma}[Self-certifying the spectral gap]
\label{lemma:self-certifying-delta}
Let \(O_\rho\) be the purification oracle of Eq.~\eqref{eq:purification-oracle-intro}, and suppose the bounded-gap promise of Eq.~\eqref{eq:vartheta-intro} holds with the known search scale \(\Delta_{\mathrm{floor}}\le p_{\min}^+(\boldsymbol{x}_0)\). Then there exists a quantum algorithm that, with success probability at least \(1-\vartheta\), outputs an estimate \(\hat{\Delta}\) satisfying
\begin{equation}
  \hat{\Delta} \;\le\; \min_{n:\, p_n(\boldsymbol{x}_0) > 0} p_n(\boldsymbol{x}_0) \;\le\; 4\,\hat{\Delta},
  \label{eq:self-certified-delta-bound}
\end{equation}
using \(\widetilde{\mathcal{O}}(1/\Delta_{\mathrm{floor}}^2)\) queries to \(O_\rho\), up to polylogarithmic factors in \(\Delta_{\mathrm{floor}}^{-1}\) and \(\vartheta^{-1}\).
\end{lemma}

\begin{proof}
\textit{Part 1: interval filters.}---From the purification oracle \(O_\rho\) we obtain a \((1, a+s, 0)\)-block-encoding \(U_\rho\) of \(\rho\) via the standard construction~\cite[Lemma~45]{gilyenQuantumSingularValue2019}. For each dyadic scale \(\mu_k = 2^{-k}\) (\(k \ge 0\)), standard QSVT constructions~\cite{gilyenQuantumSingularValue2019} provide an even polynomial \(B_k \in \mathbb{R}[x]\), nonnegative on \([0,1]\) and bounded by \(|B_k(x)| \le 1\) on \([-1,1]\), of degree \(d_k = \mathcal{O}\bigl(\frac{1}{\mu_k}\log\frac{1}{\varepsilon'_k}\bigr)\), satisfying
\begin{equation}
\begin{aligned}
  B_k(x) &\ge 1 - \varepsilon'_k && \text{for} \;\; x \in [\mu_k, 2\mu_k],\\
  B_k(x) &\le \varepsilon'_k && \text{for} \;\; x \in [0, \mu_k/2] \cup [4\mu_k, 1],
\end{aligned}
\label{eq:bump-filter-spec}
\end{equation}
with \(\varepsilon'_k = \mu_k/80\). Applying \(B_k\) to \(U_\rho\) via the quantum singular value transformation yields a unitary \(U_{B_k}\) that is a \((1, a+s+1, \varepsilon'')\)-block-encoding of \(B_k(\rho)\), with \(\varepsilon'' = \mu_k/320\). The ideal scalar statistic is
\begin{equation}
  T_k \;=\; \langle \rho|_{SA} \, \bigl( B_k(\rho) \otimes I_A \bigr) \, |\rho\rangle_{SA}
       \;=\; \sum_{n=1}^{N} p_n \, B_k(p_n).
  \label{eq:interval-mass-stat}
\end{equation}
To expose this statistic as an amplitude, let \(R_k\) denote the \((a+s+1)\)-qubit block-encoding ancilla bundle of \(U_{B_k}\), and define
\begin{equation}
  \mathcal V_k
  =\bigl(I_{R_k}\otimes O_\rho^\dagger\bigr)
   \bigl(U_{B_k}\otimes I_A\bigr)
   \bigl(I_{R_k}\otimes O_\rho\bigr).
  \label{eq:self-certification-scalar-amplitude-unitary}
\end{equation}
Writing \(\ket{\boldsymbol{0}}=\ket{0}_{R_k}\ket{0}_S\ket{0}_A\), the block-encoding guarantee gives
\begin{equation}
  \widetilde T_k
  :=\bra{\boldsymbol{0}}\mathcal V_k\ket{\boldsymbol{0}},
  \qquad
  |\widetilde T_k-T_k|\le\varepsilon''.
  \label{eq:self-certification-scalar-amplitude-error}
\end{equation}
Thus, taking the all-zero state as the good subspace, \(\mathcal V_k\) has success amplitude \(\gamma_k=|\widetilde T_k|\). The statistic \(T_k\) detects whether the dyadic interval \([\mu_k, 2\mu_k]\) contains an eigenvalue, through the following self-scaling property: an eigenvalue inside the interval has weight at least the interval's left endpoint. Indeed, if \([\mu_k, 2\mu_k]\) contains an eigenvalue \(p_n\), then, using the nonnegativity of \(B_k\) on \([0,1]\),
\begin{equation}
  T_k \;\ge\; p_n\, B_k(p_n) \;\ge\; \mu_k (1 - \varepsilon'_k) \;\ge\; 0.98\,\mu_k .
  \label{eq:interval-mass-lower}
\end{equation}
Conversely, if no eigenvalue of \(\rho\) lies in the window \([\mu_k/2, 4\mu_k]\), then
\begin{equation}
  T_k \;\le\; \sum_{n=1}^{N} p_n\, \varepsilon'_k \;=\; \varepsilon'_k \;\le\; \frac{\mu_k}{40} .
  \label{eq:interval-mass-upper}
\end{equation}
Together with the block-encoding error \(\varepsilon''\), the two cases are separated by a gap of at least \(\mu_k/4\) around the threshold \(\mu_k/2\): in the first case \(\gamma_k\ge T_k-\varepsilon''>3\mu_k/4\), while in the second \(\gamma_k\le T_k+\varepsilon''<\mu_k/4\). The thresholding primitive of Lemma~\ref{lemma:thresholding}, applied to \(\mathcal V_k\) with \(\mu=\mu_k/2\), \(h=\mu_k/4\), and failure probability \(\eta\), therefore certifies \(b_k = 0\) in the first case and \(b_k = 1\) in the second, each with success probability at least \(1-\eta\); in the intermediate fuzzy region \([\mu_k/2, \mu_k] \cup [2\mu_k, 4\mu_k]\) the output may be arbitrary, and the analysis below is robust to either value. The fractions \(1/80\), \(1/40\), and \(1/320\) are only a convenient fixed allocation of the deterministic approximation errors; their essential feature is their proportionality to \(\mu_k\), which preserves a constant relative separation between the two cases. Majority voting in Lemma~\ref{lemma:thresholding} suppresses only the random failure probability of the threshold measurement and does not reduce these deterministic filter and block-encoding errors.

\textit{Part 2: dyadic search.}---Set \(K_{\max} = \lceil \log_2 (1/\Delta_{\mathrm{floor}}) \rceil + 1\) and run the certified decision of Part 1 at every level \(k = 0, 1, \dots, K_{\max}\), with per-decision failure probability \(\eta = \vartheta/(2(K_{\max}+1))\). Let \(k^\ast\) be the largest level with \(b_k = 0\), and output \(\hat{\Delta} = \mu_{k^\ast}\).

Let \(p^+ := \min_{n:\, p_n > 0} p_n\) be the smallest positive eigenvalue, and let \(j\) be its dyadic index, i.e., \(p^+ \in [\mu_j, 2\mu_j)\). The interval \([\mu_j, 2\mu_j]\) contains \(p^+\), so by Eq.~\eqref{eq:interval-mass-lower} the decision at level \(j\) is certified: \(b_j = 0\), and hence \(k^\ast \ge j\). For every level \(k \ge j+2\), we have \(4\mu_k \le 4\mu_{j+2} = \mu_j \le p^+\). Thus every positive eigenvalue lies in the stopband \([4\mu_k,1]\), while zero eigenvalues lie in \([0,\mu_k/2]\), and Eq.~\eqref{eq:interval-mass-upper} certifies \(b_k = 1\). Hence \(k^\ast \le j+1\). By the union bound over the at most \(K_{\max}+2\) certified decisions, the two conditions hold simultaneously with probability at least \(1-\vartheta\), in which case
\begin{equation}
\begin{aligned}
  \hat{\Delta}=\mu_{k^\ast}
  &\le \mu_j\le p^+<2\mu_j=4\mu_{j+1}\\
  &\le 4\mu_{k^\ast}=4\hat{\Delta},
\end{aligned}
  \label{eq:dyadic-search-conclusion}
\end{equation}
which is Eq.~\eqref{eq:self-certified-delta-bound}. The promised search scale \(\Delta_{\mathrm{floor}} \le p^+\) gives \(j = \lceil \log_2(1/p^+) \rceil \le \lceil \log_2(1/\Delta_{\mathrm{floor}}) \rceil \le K_{\max}\), so the scan always reaches the dyadic interval of \(p^+\).

It remains to account for the query complexity. By Lemma~\ref{lemma:thresholding} with \(h=\mu_k/4\), each decision costs \(\mathcal{O}\!\left(\frac{1}{\mu_k}\log\frac{1}{\eta}\right)\) applications of \(\mathcal V_k\) and \(\mathcal V_k^\dagger\). Each such application uses one call to \(U_{B_k}\) or \(U_{B_k}^\dagger\) and a constant number of calls to \(O_\rho\) and \(O_\rho^\dagger\). Since \(U_{B_k}\) uses \(d_k\) queries to \(U_\rho\), and each query to \(U_\rho\) requires a constant number of purification-oracle queries, the cost per application of \(\mathcal V_k\) remains \(\mathcal{O}(d_k)\) queries to \(O_\rho\). The total is therefore
\begin{align}
  \sum_{k=0}^{K_{\max}} \mathcal{O}\!\left(
  \frac{1}{\mu_k^2}\log\frac{1}{\varepsilon'_k}\log\frac{1}{\eta}\right)
  &= \mathcal{O}\!\left(
  4^{K_{\max}}\,\mathrm{polylog}\!\left(
  \frac{1}{\Delta_{\mathrm{floor}}},\frac{1}{\vartheta}\right)\right)\nonumber\\
  &= \widetilde{\mathcal{O}}\!\left(\frac{1}{\Delta_{\mathrm{floor}}^2}\right),
  \label{eq:self-certification-cost}
\end{align}
as claimed.
\end{proof}

\section{Laguerre Polynomials and Gauss--Laguerre Quadrature}
\label{appendix:laguerre-gl-background}

This appendix collects the elementary facts about Laguerre polynomials and Gauss--Laguerre quadrature used in the main text; see~\cite{szego1939orthogonal} for a comprehensive treatment. We use the standard Laguerre polynomials
\begin{equation}
  L_n(x) = \frac{e^x}{n!}\frac{d^n}{dx^n} \left(e^{-x}x^n\right), \qquad n=0,1,2,\ldots,
  \label{eq:appendix-laguerre-rodrigues}
\end{equation}
so that \(L_n(0)=1\). Equivalently,
\begin{equation}
  L_n(x) = \sum_{r=0}^n (-1)^r {n\choose r}\frac{x^r}{r!}.
  \label{eq:appendix-laguerre-explicit-sum}
\end{equation}
They are orthogonal on \([0,\infty)\) with respect to the weight function \(e^{-x}\):
\begin{equation}
  \int_0^\infty e^{-x}L_m(x)L_n(x)\,dx = \delta_{mn}.
  \label{eq:appendix-laguerre-orthogonality}
\end{equation}
This normalization is convenient because the squared norm of each \(L_n\) is one.

\subsection{Gauss--Laguerre nodes, weights, and exactness}
\label{appendix:gl-nodes-weights-exactness}

Let \(x_{1,K},\ldots,x_{K,K}\) be the \(K\) positive roots of \(L_K(x)\), ordered as
\[
  0<x_{1,K}<x_{2,K}<\cdots<x_{K,K}.
\]
The \(K\)-point Gauss--Laguerre quadrature rule approximates weighted integrals of the form
\begin{equation}
  \int_0^\infty e^{-x}f(x)\,dx
  \approx
  \sum_{k=1}^K w_{k,K}f(x_{k,K}).
  \label{eq:appendix-gl-rule}
\end{equation}
For the normalization in Eq.~\eqref{eq:appendix-laguerre-rodrigues}, the quadrature weights are
\begin{equation}
  w_{k,K}
  =
  \frac{x_{k,K}}{(K+1)^2\bigl[L_{K+1}(x_{k,K})\bigr]^2}.
  \label{eq:appendix-gl-weights}
\end{equation}
They are strictly positive and satisfy
\begin{equation}
  \sum_{k=1}^K w_{k,K}=1,
  \label{eq:appendix-gl-weight-sum-one}
\end{equation}
which follows by applying the quadrature rule to the constant polynomial.

The defining property of Gaussian quadrature is exactness up to degree \(2K-1\): for every polynomial \(p\) with \(\deg p\le 2K-1\),
\begin{equation}
  \int_0^\infty e^{-x}p(x)\,dx
  =
  \sum_{k=1}^K w_{k,K}p(x_{k,K}).
  \label{eq:appendix-gl-exactness}
\end{equation}

For completeness, we recall the standard proof. Divide \(p\) by \(L_K\):
\[
  p(x)=q(x)L_K(x)+r(x),
  \qquad \deg r<K.
\]
Since \(\deg p\le 2K-1\), one has \(\deg q\le K-1\). Orthogonality gives
\[
  \int_0^\infty e^{-x}q(x)L_K(x)\,dx=0,
\]
because \(q\) is a polynomial of degree at most \(K-1\), which is a linear combination of \(L_0,\ldots,L_{K-1}\).
On the quadrature nodes, \(L_K(x_{k,K})=0\), hence
\[
  \sum_{k=1}^K w_{k,K}q(x_{k,K})L_K(x_{k,K})=0.
\]
It remains to integrate the remainder \(r\), which has degree \(<K\). The weights in Eq.~\eqref{eq:appendix-gl-weights} are the Christoffel weights associated with interpolation on the roots of \(L_K\), and therefore the quadrature rule integrates every polynomial of degree \(<K\) exactly. This proves Eq.~\eqref{eq:appendix-gl-exactness}.

For sufficiently smooth functions, the classical remainder formula is
\begin{equation}
\begin{aligned}
  E_K[f]
  &:=\int_0^\infty e^{-x}f(x)\,dx
  -\sum_{k=1}^K w_{k,K}f(x_{k,K})\\
  &=\frac{(K!)^2}{(2K)!}f^{(2K)}(\xi)
\end{aligned}
  \label{eq:appendix-gl-remainder}
\end{equation}
for some \(\xi\in(0,\infty)\). In the main proof we primarily use exactness and Laguerre-series tails rather than this formula, because the Laguerre-tail representation gives a more direct uniform operator bound.

The largest node is needed only when the polynomial degree used to implement an exponential filter depends on the filter time, as it does in the regularized construction.
\begin{lemma}[Asymptotics of the largest root, adapted from~{\cite[Theorem 5]{GATTESCHI20027}}]
  \label{lemma:asymptotic-laguerre-roots-weights}
  Let \(x_{K,K}\) be the largest root of the \(K\)-th Laguerre polynomial \(L_K(x)\). Then
  \begin{equation}
    x_{K,K}=4K+\mathcal{O}(K^{1/3}),\qquad K\to+\infty.
    \label{eq:largest-laguerre-node-theta-K}
  \end{equation}
\end{lemma}

\subsection{Laguerre expansion of decaying exponentials}
\label{appendix:laguerre-expansion-exponential}

The scalar functions appearing after the SLD integral is rescaled are exponentials of the form \(e^{-\mu x}\) with \(\mu\ge0\). Their Laguerre expansion is explicit:
\begin{equation}
  e^{-\mu x}
  =
  \sum_{n=0}^\infty a_n(\mu)L_n(x),
  \qquad
  a_n(\mu)
  =
  \frac{\mu^n}{(1+\mu)^{n+1}}.
  \label{eq:appendix-exp-laguerre-expansion}
\end{equation}
Indeed, by orthogonality,
\begin{equation}
\begin{aligned}
  a_n(\mu)
  &=\int_0^\infty e^{-x}e^{-\mu x}L_n(x)\,dx\\
  &=\int_0^\infty e^{-(1+\mu)x}L_n(x)\,dx.
\end{aligned}
  \label{eq:appendix-laguerre-coefficient-integral}
\end{equation}
Using Eq.~\eqref{eq:appendix-laguerre-explicit-sum}, we obtain
\begin{widetext}
\begin{align}
  \int_0^\infty e^{-(1+\mu)x}L_n(x)\,dx
  &=
  \sum_{r=0}^n
  (-1)^r {n\choose r}\frac{1}{r!}
  \int_0^\infty e^{-(1+\mu)x}x^r\,dx
  \nonumber\\
  &=
  \sum_{r=0}^n
  (-1)^r {n\choose r}\frac{1}{(1+\mu)^{r+1}}
  \nonumber\\
  &=
  \frac{1}{1+\mu}
  \left(1-\frac{1}{1+\mu}\right)^n
  \nonumber\\
  &=
  \frac{\mu^n}{(1+\mu)^{n+1}}.
  \label{eq:appendix-laguerre-coefficient-computation}
\end{align}
\end{widetext}
This proves Eq.~\eqref{eq:appendix-exp-laguerre-expansion}.

The quadrature error for \(e^{-\mu x}\) can then be expressed in terms of the Laguerre tail. Since the rule is exact for all polynomials of degree at most \(2K-1\), only the components with \(n\ge 2K\) contribute. Define
\begin{equation}
  A_n^{(K)}
  :=
  \sum_{k=1}^K w_{k,K}L_n(x_{k,K}).
  \label{eq:appendix-AnK-definition}
\end{equation}
Then
\begin{equation}
  E_K[e^{-\mu x}]
  =
  -\sum_{n=2K}^\infty
  \frac{\mu^n}{(1+\mu)^{n+1}}A_n^{(K)}.
  \label{eq:appendix-exp-quadrature-tail}
\end{equation}
The integral part of each tail term vanishes because
\[
  \int_0^\infty e^{-x}L_n(x)\,dx=0,
  \qquad n\ge1.
\]
This scalar identity is the basis of the operator error analysis in Section~\ref{truncation-error-analysis}.

\subsection{Proof of the coefficient-normalization statement in Theorem~\ref{theorem:quadrature-main}}
\label{appendix:lcu-coefficient-norm}

The LCU assembly uses the positive coefficients
\begin{equation}
  y_k=w_{k,K}e^{x_{k,K}},
  \qquad
  \beta_K=\sum_{k=1}^K y_k.
  \label{eq:appendix-lcu-coefficients}
\end{equation}
The third statement of Theorem~\ref{theorem:quadrature-main} uses the asymptotic estimate of \(\beta_K\) proved here.\footnote{The asymptotic upper-bound argument was developed with assistance from AI tools from OpenAI and DeepSeek under the authors' direction. Its bulk--edge deduction was checked in Lean~4, conditional on the classical zero-distribution and uniform bulk/edge asymptotics cited below. The authors reviewed and verified the argument and take responsibility for its correctness.}

\begin{proof}

    We first prove the lower bound. Using the elementary fact that
    \[e^x \ge \sum_{m=0}^{N-1} \frac{x^m}{m!}\]
    for every \(x > 0\) and \(N \in \mathbb{Z}_+\), we obtain
\begin{align}
    \beta_K
    &= \sum_{k=1}^K w_k \, e^{x_k} \notag \\
    &\ge \sum_{k=1}^K w_k \sum_{m=0}^{2K-1} \frac{x_k^{\,m}}{m!} \\
    &= \sum_{m=0}^{2K-1} \frac{1}{m!} \sum_{k=1}^K w_k \, x_k^{\,m}\\
    &= \sum_{m=0}^{2K-1} \frac{1}{m!} \cdot m!
  \qquad (\text{implied by Eq.~\ref{eq:appendix-gl-exactness}})\\
    &= 2K. \notag
\end{align}

It remains to establish
\begin{equation}
  \lim_{K\to\infty}\frac{\beta_K}{4K}=1,
  \label{eq:appendix-beta-limit-statement}
\end{equation}
which is equivalent to $\beta_K = 4K(1 + o(1))$.
Together with the lower bound $\beta_K\ge 2K$ proved above, Eq.~\eqref{eq:appendix-beta-limit-statement} pins down the leading asymptotic behavior of the LCU coefficient norm.

\paragraph{Zero distribution}

Introduce the scaled nodes
\begin{equation}
  t_{k,K} := \frac{x_{k,K}}{4K}.
  \label{eq:appendix-scaled-variables}
\end{equation}
The distribution of rescaled zeros $\{t_{k,K}\}$ obey the following~\cite{gawronski1987} (Theorem~1, eq.~(0.3))
\begin{equation}
\begin{aligned}
  \lim_{K\to\infty}\frac{1}{K}\,
  \#\{\,\nu : x_{\nu,K}\le 4K\xi\,\}
  &= \frac{2}{\pi}\int_0^\xi
  t^{-1/2}(1-t)^{1/2}\,\dd t,\\
  &\qquad 0<\xi\le1.
\end{aligned}
  \label{eq:appendix-GS87-original}
\end{equation}
Equivalently, the empirical zero measures converge weakly to the probability measure with density
\begin{equation}
  \rho(t) := \frac{2}{\pi}\sqrt{\frac{1-t}{t}},\qquad 0<t<1.
  \label{eq:appendix-MP-density}
\end{equation}
By the Portmanteau theorem, this weak convergence of empirical measures implies convergence of Riemann sums for any continuous test function $g$ on a compact subinterval $[\delta,1-\delta]\subset(0,1)$:
\begin{equation}
\begin{aligned}
  &\lim_{K\to\infty}\frac{1}{K}\!
  \sum_{k\,:\,\delta\le t_{k,K}\le 1-\delta}\!
  g(t_{k,K})\\
  &\qquad=\int_\delta^{1-\delta} g(t)\,\rho(t)\,\dd t.
\end{aligned}
  \label{eq:appendix-GS87-consequence}
\end{equation}

\paragraph{Weight formula via the Christoffel--Darboux kernel}

At a zero $x_{k,K}$ of $L_K$, the Christoffel--Darboux formula and the three-term recurrence combine to give a compact expression for the quadrature weight.  The sign-adjusted Laguerre polynomials $P_n(x):=(-1)^n L_n(x)$ satisfy the symmetric three-term recurrence
\begin{equation}
\begin{aligned}
  x P_n(x)
  &= (n+1)P_{n+1}(x) + (2n+1)P_n(x)\\
  &\quad+ n P_{n-1}(x),
  \qquad n\ge1,
\end{aligned}
  \label{eq:appendix-P-recurrence}
\end{equation}
whose Christoffel--Darboux diagonal identity~\cite[Theorem~3.2.2]{szego1939orthogonal} reads
\begin{equation}
  \sum_{m=0}^{K-1} P_m(x)^2
  = K\bigl[P_K'(x)P_{K-1}(x) - P_K(x)P_{K-1}'(x)\bigr].
  \label{eq:appendix-CD-diag}
\end{equation}
Evaluating~\eqref{eq:appendix-CD-diag} at $x_{k,K}$ and using $P_K(x_{k,K})=0$ (since $P_K$ and $L_K$ share the same zeros) eliminates the second term.  Writing $P_n$ and $P_n'$ in terms of $L_n$ and $L_n'$, the factor $(-1)^{2K-1}=-1$ appears from the product $P_K'(x_{k,K})P_{K-1}(x_{k,K})$.  The recurrence evaluated at the zero~\cite[(5.1.14)]{szego1939orthogonal}, $xL_K'(x) = K L_K(x) - K L_{K-1}(x)$, gives $x_{k,K}L_K'(x_{k,K}) = -K L_{K-1}(x_{k,K})$.  Assembling these algebraic identities yields
\begin{equation}
  \sum_{m=0}^{K-1} L_m(x_{k,K})^2
  = x_{k,K}\,\bigl[L_K'(x_{k,K})\bigr]^2.
  \label{eq:appendix-sum-Lsq}
\end{equation}
The Gauss--Laguerre weights are the reciprocals of the diagonal Christoffel--Darboux kernel~\cite[\S15.3, (15.3.5)]{szego1939orthogonal}: $w_{k,K} = \bigl[\sum_{m=0}^{K-1} L_m(x_{k,K})^2\bigr]^{-1}$, which together with~\eqref{eq:appendix-sum-Lsq} gives the exact identity
\begin{equation}
  w_{k,K}\;x_{k,K}\;\bigl[L_K'(x_{k,K})\bigr]^2 = 1,
  \qquad K\ge1.
  \label{eq:appendix-weight-deriv-identity}
\end{equation}

\paragraph{Weighted-kernel asymptotics in the bulk}

It is convenient to use a kernel theorem directly, avoiding any need to differentiate an asymptotic remainder.  Define the weighted diagonal Christoffel--Darboux kernel
\begin{equation}
  R_K(x):=e^{-x}\sum_{m=0}^{K-1}L_m(x)^2.
  \label{eq:appendix-weighted-CD-kernel}
\end{equation}
By the Christoffel-weight identity above, $y_k=R_K(x_{k,K})^{-1}$.  In the notation of Vanlessen~\cite[Theorem~2.8(a)]{vanlessen2007}, the classical Laguerre case has $Q(x)=x$, $\alpha=0$, Mhaskar--Rakhmanov--Saff number $4K$, and equilibrium density $\widehat\psi(t)=\frac{2}{\pi}\sqrt{(1-t)/t}=\rho(t)$.  Specializing the diagonal bulk-kernel asymptotic gives
\begin{equation}
  R_K(4Kt)
  =\frac{\rho(t)}{4}\bigl(1+\mathcal{O}(K^{-1})\bigr),
  \qquad t\in[\delta,1-\delta],
  \label{eq:appendix-weighted-kernel-bulk}
\end{equation}
uniformly for every fixed $\delta\in(0,1/2)$.  Hence, at the quadrature nodes in this interval,
\begin{equation}
  y_k
  =2\pi\sqrt{\frac{t_{k,K}}{1-t_{k,K}}}\,
   \bigl(1+\mathcal{O}(K^{-1})\bigr).
  \label{eq:appendix-ak-convergence}
\end{equation}

\paragraph{Bulk convergence of the weighted sum}

Equation~\eqref{eq:appendix-ak-convergence} motivates the definition of the weight profile
\begin{equation}
  f(t) := 2\pi\sqrt{\frac{t}{1-t}},\qquad 0<t<1.
  \label{eq:appendix-weight-profile-f}
\end{equation}
With $\rho(t)$ from Eq.~\eqref{eq:appendix-MP-density}, a direct multiplication yields
\begin{equation}
  f(t)\,\rho(t)=4,\qquad 0<t<1.
  \label{eq:appendix-f-rho-product}
\end{equation}
Consequently, for any prescribed $\varepsilon>0$, there exists $K_{\rm bulk}$ such that
\begin{equation}
  |y_k - f(t_{k,K})| < \varepsilon,
  \qquad
  \delta\le t_{k,K}\le 1-\delta,\;\; K\ge K_{\rm bulk}.
  \label{eq:appendix-bulk-convergence-statement}
\end{equation}

\paragraph{Edge control}

The zero-distribution theorem~\eqref{eq:appendix-GS87-original} implies that the fraction of scaled zeros lying outside any compact subinterval $[\delta,1-\delta]$ tends to the $\rho$-measure of the complement:
\begin{equation}
\begin{aligned}
  &\lim_{K\to\infty}\frac{1}{K}\,
  \#\{\,k : t_{k,K}<\delta\;\text{or}\;t_{k,K}>1-\delta\,\}\\
  &\quad= \int_0^\delta\!\rho(t)\,\dd t
    + \int_{1-\delta}^1\!\rho(t)\,\dd t\\
  &\quad= \frac{4}{\pi}\arcsin\sqrt{\delta}
  = \mathcal{O}(\sqrt{\delta}).
\end{aligned}
  \label{eq:appendix-edge-zero-fraction}
\end{equation}
The count alone does not control the weighted edge sum, because the largest soft-edge modified weights grow on the Airy scale.  We now spell out the uniform-integrability consequence of the matched Bessel, bulk, and Airy asymptotics that is needed below.  Vanlessen's polynomial and kernel asymptotics~\cite[Theorem~2.4(c,d) and Theorem~2.8(b,c)]{vanlessen2007}, together with the coordinate expansions in Remarks~3.20 and~3.26 therein, give constants $C,u_0>0$, independent of $K$ and $u$, such that
\begin{subequations}\label{eq:appendix-edge-local-bounds}
\begin{align}
  \#\{k:u\le t_{k,K}<2u\}
  &\le C\bigl(1+K\sqrt{u}\bigr),\nonumber\\
  y_k&\le C\sqrt{u},
  \qquad K^{-2}\le u\le u_0,
  \label{eq:appendix-hard-shell-bounds}\\
  \#\{k:u\le 1-t_{k,K}<2u\}
  &\le C\bigl(1+Ku^{3/2}\bigr),\nonumber\\
  y_k&\le Cu^{-1/2},\nonumber\\
  &\qquad K^{-2/3}\le u\le u_0.
  \label{eq:appendix-soft-shell-bounds}
\end{align}
Moreover, the microscopic windows satisfy
\begin{align}
  \#\{k:t_{k,K}<K^{-2}\}&\le C,\nonumber\\
  y_k&\le CK^{-1},
  \label{eq:appendix-hard-micro-bounds}\\
  \#\{k:1-t_{k,K}<K^{-2/3}\}&\le C,\nonumber\\
  y_k&\le CK^{1/3}.
  \label{eq:appendix-soft-micro-bounds}
\end{align}
\end{subequations}
Each bound on $y_k$ applies to the nodes in the shell on the same line.
Here is how the powers arise and how the quoted results are used.  At the hard edge, Remark~3.26 gives a Bessel argument of order $K\sqrt{t}$.
The bounded spacing of Bessel zeros then gives the count $C(1+K\sqrt{u})$.  Theorem~2.8(c) gives $R_K(4Kt)\ge C^{-1}K$ in the microscopic window; in the matching annuli, substituting Theorem~2.4(d) for the two neighboring polynomial degrees into the Christoffel--Darboux identity and using the large-argument Bessel formulas gives $R_K(4Kt)\ge C^{-1}u^{-1/2}$.  Together these are $R_K(4Kt)\ge C^{-1}\min\{K,u^{-1/2}\}$.  At the soft edge, Remark~3.20 gives an Airy argument of order $K^{2/3}(1-t)$.  Since the number of Airy zeros up to magnitude $a$ is $\mathcal{O}(a^{3/2})$, a shell contains at most $C(1+Ku^{3/2})$ zeros.  Theorem~2.8(b) gives $R_K(4Kt)\ge C^{-1}K^{-1/3}$ in the microscopic window, while the same Christoffel--Darboux substitution using Theorem~2.4(c) and the large-argument Airy formulas gives $R_K(4Kt)\ge C^{-1}\sqrt{u}$ in the matching annuli.  Thus $R_K(4Kt)\ge C^{-1}\max\{K^{-1/3},\sqrt{u}\}$.  Taking reciprocals through $y_k=R_K(x_{k,K})^{-1}$ gives Eqs.~\eqref{eq:appendix-edge-local-bounds}.  The large-argument Airy/Bessel formulas agree with the bulk estimates on the overlaps, so one constant can be chosen throughout the indicated ranges.

We next derive the required edge estimate rather than invoking it directly.
For the hard edge, set $u_j=2^jK^{-2}$ and sum Eq.~\eqref{eq:appendix-hard-shell-bounds} over the shells with $u_j\lesssim\delta$, adding the microscopic window separately.  Then
\begin{align}
 \frac1K\sum_{t_{k,K}<\delta}y_k
 &\le \mathcal{O}(K^{-2})
   +C\sum_{u_j\lesssim\delta}
      \left(u_j+\frac{\sqrt{u_j}}{K}\right)\nonumber\\
 &\le C\delta+\frac{C\sqrt{\delta}}{K}
 +\mathcal{O}(K^{-2}).
 \label{eq:appendix-hard-edge-sum}
\end{align}
Likewise, with $v_j=2^jK^{-2/3}$, the soft-edge bounds give
\begin{align}
 \frac1K\sum_{1-t_{k,K}<\delta}y_k
 &\le \mathcal{O}(K^{-2/3})
   +C\sum_{v_j\lesssim\delta}
      \left(v_j+\frac{1}{K\sqrt{v_j}}\right)\nonumber\\
 &\le C\delta+\mathcal{O}(K^{-2/3}).
 \label{eq:appendix-soft-edge-sum}
\end{align}
Both sums are geometric, and their $o(1)$ terms are uniform for each fixed $\delta\le u_0$.  Therefore
\begin{equation}
\begin{aligned}
  &\lim_{\delta\to0^+}\;\limsup_{K\to\infty}\frac{1}{K}\\
  &\qquad\times
  \sum_{k\,:\,t_{k,K}\notin[\delta,1-\delta]} y_k=0.
\end{aligned}
  \label{eq:appendix-edge-control}
\end{equation}

\paragraph{Completion of the proof}

Fix $\varepsilon>0$ and set $\varepsilon_4:=\varepsilon/4$.
Choose $\delta>0$ sufficiently small that $8\delta<\varepsilon_4$ and, by the edge control~\eqref{eq:appendix-edge-control}, choose $K_{\rm edge}$ so that the normalized edge sum is below $\varepsilon_4$ for every $K\ge K_{\rm edge}$.  Let $K_0$ be the maximum of $K_{\rm edge}$, $K_{\rm bulk}$ (from Eq.~\eqref{eq:appendix-bulk-convergence-statement} with tolerance $\varepsilon_4$), and the threshold from the Riemann sum convergence~\eqref{eq:appendix-GS87-consequence} (also with tolerance $\varepsilon_4$).

For any $K\ge K_0$, partition the index set $\{1,\dots,K\} = B \cup E$ where $B := \{k : \delta\le t_{k,K}\le 1-\delta\}$ and $E := \{k : t_{k,K}<\delta \;\text{or}\; t_{k,K}>1-\delta\}$.
Decompose the average:
\begin{equation}
\begin{aligned}
  \frac{\beta_K}{K}
  &= \underbrace{\frac{1}{K}\sum_{k\in B} f(t_{k,K})}_{\text{(I)}}\\
  &\quad+ \underbrace{\frac{1}{K}\sum_{k\in B}
     \bigl(y_k-f(t_{k,K})\bigr)}_{\text{(II)}}\\
  &\quad+ \underbrace{\frac{1}{K}\sum_{k\in E} y_k}_{\text{(III)}}.
\end{aligned}
  \label{eq:appendix-decomposition}
\end{equation}

The triangle inequality and the four estimates (Eqs.~\eqref{eq:appendix-GS87-consequence}, \eqref{eq:appendix-bulk-convergence-statement}, \eqref{eq:appendix-edge-control}, and $|\!\int_\delta^{1-\delta}\!f\rho - 4| = 8\delta<\varepsilon_4$) give
\begin{equation}
  \Bigl|\frac{\beta_K}{K} - 4\Bigr|
  < \varepsilon_4 + \varepsilon_4 + \varepsilon_4 + \varepsilon_4
  = \varepsilon.
  \label{eq:appendix-eps4-final}
\end{equation}
Since $\varepsilon>0$ was arbitrary, $\lim_{K\to\infty} \beta_K/K = 4$, which is equivalent to $\lim_{K\to\infty} \beta_K/(4K) = 1$.
This completes the proof of the coefficient-normalization statement in Theorem~\ref{theorem:quadrature-main}.
\end{proof}

\section{Analysis of the SLD Quadrature}\label{app:quadrature-analysis}

Section~\ref{sec:gapped-quadrature-outline} derives the canonical SLD quadrature formula and identifies the Laguerre tail that controls its convergence. This appendix supplies the technical remainder estimates under the bounded-gap promise and then analyzes the regularized construction. Both arguments use the access and conditioning assumptions of Section~\ref{sec:access-model}.

\subsection{Truncation Error Analysis}\label{truncation-error-analysis}

We now quantify the error \(\mathcal{R}_j=L_j-\widehat L_j\) for the approximation derived in Eq.~\eqref{eq:approximate-solution-to-sld-in-matrix-form}. The analysis is most transparent in the eigenbasis of \(\mathcal M\), where the quadrature remainder is diagonal term by term. In vectorized form,

\begin{equation}
\begin{aligned}
  \ket{\mathcal{R}_j} &= -\frac{1}{\Delta} \sum_\ell \sum_{n=2K}^{\infty}
    \frac{1}{1+\mu_\ell} \left(\frac{\mu_\ell}{1+\mu_\ell}\right)^n \\
    &\qquad \times \left(\sum_{k=1}^{K} w_{k,K} L_n(x_{k,K})\right)
    \braket{\varphi_\ell | \partial_j \rho} \ket{\varphi_\ell}
\end{aligned}
  \label{eq:vectorized-error-term}
\end{equation}

The Frobenius norm of \(\mathcal{R}_j\) is the Euclidean norm of \(\ket{\mathcal{R}_j}\), so we can write

\begin{equation}
\begin{aligned}
\|\mathcal{R}_j\|_F
&= \|\ket{\mathcal{R}_j}\|_2 = \frac{1}{\Delta} \Bigg[ \sum_\ell \Bigg|
   \sum_{n=2K}^{\infty} \frac{1}{1+\mu_\ell}
   \left(\frac{\mu_\ell}{1+\mu_\ell}\right)^n \\
  &\qquad \times \left(\sum_{k=1}^{K} w_{k,K} L_n(x_{k,K})\right)
   \braket{\varphi_\ell | \partial_j \rho} \Bigg|^2 \Bigg]^{1/2}
\end{aligned}
\label{eq:error-norm-expression}
\end{equation}
and therefore
\begin{equation}
\begin{aligned}
\|\mathcal{R}_j\|_F &\le \frac{1}{\Delta}
  \Bigg(\max_\ell \Bigg|
   \sum_{n=2K}^{\infty} \frac{1}{1+\mu_\ell}
   \left(\frac{\mu_\ell}{1+\mu_\ell}\right)^n \\
  &\qquad \times \left(\sum_{k=1}^{K} w_{k,K} L_n(x_{k,K})\right) \Bigg|\Bigg)
   \cdot \|\partial_j \rho\|_F
\end{aligned}
\label{eq:error-norm-upper-bound}
\end{equation}

where \(\|\partial_j \rho\|_F = \left(\sum_\ell| \braket{\varphi_\ell | \partial_j \rho}|^2\right)^{1/2}\) is the Frobenius norm of \(\partial_j \rho\).
Define \(A_n^{(K)} = \sum_{k=1}^{K} w_{k,K} L_n(x_{k,K})\), \(S_j = \sum_{n=2K}^{\infty} \frac{1}{1+\mu_\ell} \left(\frac{\mu_\ell}{1+\mu_\ell}\right)^n A_n^{(K)}\), and \(q_j = \frac{\mu_\ell}{1+\mu_\ell}\). Since \(\mu_\ell \in [0, \frac{2-\Delta}{\Delta}]\), we have \(q_j \in [0, 1-\frac{\Delta}{2}]\). Then we can rewrite \(S_j\) as
\[ S_j = \sum_{n=2K}^{\infty} (1-q_j) q_j^n A_n^{(K)} = (1-q_j)q_j^{2K} \sum_{m=0}^{\infty} q_j^m A_{2K+m}^{(K)} \]
Thus the problem reduces to a uniform bound on \(A_n^{(K)}\). We first upper-bound the absolute value of \(A_n^{(K)}\) using the triangle inequality:
\[ |A_n^{(K)}| = \left| \sum_{k=1}^{K} w_{k,K} L_n(x_{k,K}) \right| \le \sum_{k=1}^{K} w_{k,K} |L_n(x_{k,K})| \]

Since the weights \(w_{k,K}\) are positive and the Laguerre polynomials satisfy \(|L_n(x)| \le e^{x/2}\) for \(x \ge 0\)~\cite{szego1939orthogonal}, we obtain \( |A_n^{(K)}| \le \sum_{k=1}^{K} w_{k,K} e^{x_{k,K}/2} \).

For \(f(x) = e^{x/2}\), we have \(f^{(n)}(x) = \left(\frac{1}{2}\right)^n e^{x/2} > 0\). Then by the remainder formula Eq.~\eqref{eq:appendix-gl-remainder}, the quadrature error is strictly positive, indicating that the quadrature sum underestimates the integral. Hence

\[ \sum_{k=1}^{K} w_{k,K} e^{x_{k,K}/2} < \int_{0}^{\infty} e^{-x} e^{x/2} dx = 2 \]

So we obtain a crude yet sufficient uniform bound on the quadrature sums:

\begin{equation}
  |A_n^{(K)}| < 2
  \label{eq:bound-on-A_n_K}
\end{equation}

Substituting this into the definition of \(S_j\) gives
\begin{align}
|S_j| &\le (1-q_j) q_j^{2K} \sum_{m=0}^{\infty} q_j^m |A_{2K+m}^{(K)}| \nonumber \\
&< 2 (1-q_j) q_j^{2K} \sum_{m=0}^{\infty} q_j^m \nonumber \\
&= 2 (1-q_j) q_j^{2K} \cdot \frac{1}{1-q_j} \nonumber \\
&= 2 q_j^{2K}
\label{eq:bound-on-S_j}
\end{align}

Combining the previous bounds yields

\begin{align}
\|\mathcal{R}_j\|_F &\le \frac{1}{\Delta} \left(\max_\ell |S_j|\right) \cdot \|\partial_j \rho\|_F \nonumber \\
&\le \frac{1}{\Delta} \cdot 2 (1-\tfrac{\Delta}{2})^{2K} \cdot \|\partial_j \rho\|_F \nonumber \\
&= \frac{(1-\frac{\Delta}{2})^{2K}}{\Delta/2} \|\partial_j \rho\|_F
\label{eq:error-norm-final-bound}
\end{align}

Since \(\|M\| \le \|M\|_F \le \sqrt{\mathrm{rank}(M)} \|M\|\) for any matrix \(M\), we arrive at the operator-norm estimate

\begin{equation}
  \|\mathcal{R}_j\| \le \|\mathcal{R}_j\|_F \le \frac{(1-\frac{\Delta}{2})^{2K}}{\Delta/2} \|\partial_j \rho \|_F
  \label{eq:final-error-bound-in-operator-norm}
\end{equation}

Therefore, to guarantee a desired truncation error \(\|\mathcal{R}_j\|\le \epsilon_{\mathrm{trun}}\), it suffices that

\[ \frac{(1-\frac{\Delta}{2})^{2K}}{\Delta/2} \|\partial_j \rho \|_F \le \epsilon_{\mathrm{trun}} \]

Solving for the minimum required quadrature order \(K\), we obtain
\begin{equation}
  K \ge \frac{\log\left(\frac{2\|\partial_j \rho \|_F}{\epsilon_{\mathrm{trun}} \Delta}\right)}{-2 \log\left(1 - \frac{\Delta}{2}\right)} .
\end{equation}
Using the elementary inequality \(-\log(1-x)\ge x\) for \(x\in(0,1)\), with \(x=\Delta/2\), we have
\begin{equation}
  -2\log\left(1-\frac{\Delta}{2}\right)\ge \Delta .
\end{equation}
Therefore,
\begin{equation}
  \frac{\log\left(\frac{2\|\partial_j \rho \|_F}{\epsilon_{\mathrm{trun}} \Delta}\right)}{-2 \log\left(1 - \frac{\Delta}{2}\right)}
  \le
  \frac{1}{\Delta}\log\left(\frac{2\|\partial_j \rho \|_F}{\epsilon_{\mathrm{trun}} \Delta}\right),
\end{equation}
so it suffices to choose
\begin{equation}
  K = \left\lceil \frac{1}{\Delta} \log\left(\frac{2\|\partial_j \rho \|_F}{\epsilon_{\mathrm{trun}} \Delta}\right) \right\rceil
  = \Theta\left(\frac{1}{\Delta} \log\left(\frac{\|\partial_j \rho \|_F}{\epsilon_{\mathrm{trun}} \Delta}\right)\right) .
  \label{eq:required-number-of-quadrature-nodes}
\end{equation}

For convenience, we assume that the quadrature order \(K\) is a power of two, which can be achieved by padding the smallest sufficient order to the next power of two. This changes \(K\) by at most a factor of two and simplifies the notation in the later block-encoding construction.

\subsection{Quadrature under Tikhonov Regularization}\label{sec:reg:quadrature}

The truncation analysis of Section~\ref{truncation-error-analysis} relies on the bounded-gap promise \(\Delta \ge 1/\operatorname{poly}(s)\).
When \(\Delta\) may be exponentially small, we turn to the regularized SLD \(L_j^{(\theta)}\) defined in Eq.~\eqref{eq:reg:sld_def}, whose regularization bias is controlled by \(|\mathcal{F}_{jj}^{(\theta)}-\mathcal{F}_{jj}| \le 8M\theta\) (Eq.~\eqref{eq:reg:bias_bound}).  Applying \(K\)-point Gauss--Laguerre quadrature (Appendix~\ref{appendix:gl-nodes-weights-exactness}) to the regularized integral representation~\eqref{eq:reg:integral} gives
\begin{equation}
  \tilde{L}_j^{(\theta)} = \frac{1}{\theta} \sum_{k=1}^{K}
    w_{k,K} \; e^{-\frac{x_{k,K}}{\theta}\rho} \,
    (\partial_j \rho) \, e^{-\frac{x_{k,K}}{\theta}\rho}.
  \label{eq:reg:GL_discrete_quad}
\end{equation}

The spectrum of \(\mathcal{M} := \mathcal{L}_\rho/\theta\) is \(\{\mu_\ell = (p_m+p_n)/\theta\}\) with \(\mu_\ell \in [0, 2/\theta]\).  Following the Laguerre-expansion analysis of Section~\ref{truncation-error-analysis}, the quadrature error satisfies \(|A_K(\mu)| \le 2\,q(\mu)^{2K}\) with \(q(\mu) = \mu/(1+\mu) \le (1+\theta/2)^{-1}\).
The Frobenius-norm error is therefore
\begin{equation}
  \|\mathcal{R}_j\|_F \le \frac{1}{\theta} \cdot 2 \,
  \Bigl(1 + \frac{\theta}{2}\Bigr)^{-2K} \cdot \|\partial_j \rho\|_F.
  \label{eq:reg:error_F_quad}
\end{equation}
To achieve \(\|\mathcal{R}_j\|_F \le \epsilon_{\mathrm{trun}}\), we require
\begin{equation}
  K = \Bigl\lceil \frac{1+\theta/2}{\theta} \,
  \log\!\Bigl(\frac{2\|\partial_j\rho\|_F}{\epsilon_{\mathrm{trun}}\,\theta}\Bigr) \Bigr\rceil
  = \Theta\!\Bigl(\frac{1}{\theta}\log\frac{\|\partial_j\rho\|_F}{\epsilon_{\mathrm{trun}}\,\theta}\Bigr).
  \label{eq:reg:K_selection_quad}
\end{equation}

\textit{Remark:} In the gapped construction Eq.~\eqref{eq:expression-for-sld-after-gauss-laguerre-quadrature}, the integral carries a \(1/\Delta\) prefactor and the \(e^{-x}\) weight is compensated by factors \(e^{x_{k,K}}\) in the quadrature sum, requiring the coefficient-normalization statement of Theorem~\ref{theorem:quadrature-main} to bound the LCU coefficient sum.  In the regularized formula~\eqref{eq:reg:GL_discrete_quad}, the weight \(e^{-x}\) is natively provided by the regularized resolvent, and the LCU coefficient sum is simply \(1/\theta\) (since \(\sum_k w_{k,K} = \int_0^\infty e^{-x}\,dx = 1\)).  No compensation factors, no asymptotic lemmas required.

\section{SLD Block-Encoding Implementation and Error Budgets}\label{app:sld-implementation}

Section~\ref{algorithm-design-for-solving-the-lyapunov-equation} gave the construction outline that turns the finite quadrature sum~\eqref{eq:approximate-solution-to-sld-in-matrix-form} into a block-encoded operator. Here we prove the polynomial guarantees and give the complete register, finite-precision, gate-complexity, and error-budget analysis using the primitives collected in Appendix~\ref{app:implementation-tools}.

\subsection{Constructing the Block-Encoding of the SLD}

Assume that we have purified access to \(\rho(\boldsymbol{x}_0)\) as shown in Eq.~\eqref{eq:purification-oracle-intro}, and access to a block-encoding of \(\partial_j \rho\) as in Eq.~\eqref{eq:derivative-block-encoding-intro}. Denote the gate complexity of the purification oracle by \(T_\rho\) and the gate complexity of the derivative block-encoding by \(T_j\). We count arbitrary single-qubit rotations and two-qubit gates as elementary gates.
From the purification oracle we obtain the \((1,a+s,0)\)-block-encoding \(U_\rho\) of \(\rho\) given in Eq.~\eqref{eq:block-encoding-of-rho-using-purification}~\cite[Lemma 45]{gilyenQuantumSingularValue2019}, where \(S'\) is an additional \(s\)-qubit system register initialized in \(\ket{0}_{S'}\).
The next ingredient is a realization of the block encoding of \(e^{- \frac{x_{k,K}}{\Delta}\rho}\). For the bounded-gap case, we expect a polynomial approximation of the exponential filter on the gapped spectral domain, and the following lemma provides the necessary approximation guarantee.
\begin{lemma}[Bounded Even Polynomial Approximation on a Gapped Domain]
\label{lem:gapped_exponential_anchor_tt}
Let $t \ge 0$, $\Delta \in (0, 1]$, and $\epsilon \in (0, 1)$. There exists an efficiently computable, even polynomial $P \in \mathbb{R}[x]$ such that:
\begin{enumerate}
    \item $|P(0) - 1| \le \epsilon$
    \item $\|P(x) - e^{-t|x|}\|_{[-1, 1] \setminus (-\Delta, \Delta)} \le \epsilon$
    \item $\|P(x)\|_{[-1, 1]} \le 1$
\end{enumerate}
The degree of the polynomial is bounded by $d = \mathcal{O}\left(\frac{1}{\Delta} \log\left(\frac{1}{\epsilon}\right)\right)$.
\end{lemma}

\begin{proof}
We construct the desired polynomial by combining one local approximant around the origin with one local approximant for the right branch \(e^{-tx}\) on \([\Delta,1]\), and then symmetrizing. It suffices to consider \(0<\Delta\le1/2\): for \(\Delta>1/2\), apply the construction with effective gap \(1/2\), which gives the required approximation on a larger domain and preserves the degree bound up to a factor of two.

Fix \(\epsilon' := \frac{\epsilon}{6}.\) We first apply \cite[Corollary 66]{gilyenQuantumSingularValue2019} to the constant function
\[
f_0(x)\equiv 1
\]
with center \(x_0=0\), radius \(r_0=\Delta/8\), and buffer width \(\vartheta_0=\Delta/8\).
Since \(f_0\) is entire and its Taylor series at the origin has only one nonzero coefficient, we may take \(B_0=1\).
Hence there exists an efficiently computable polynomial \(Q_0\) of degree \( \mathcal{O}\!\left(\frac{1}{\Delta}\log\frac{1}{\epsilon}\right) \), such that
\begin{align}
|Q_0(x)-1| &\le \epsilon' \qquad \text{for } x\in[-\Delta/8,\Delta/8], \label{eq:Q0_local}\\
|Q_0(x)| &\le 1+\epsilon' \qquad \text{for } x\in[-1,1], \label{eq:Q0_global}\\
|Q_0(x)| &\le \epsilon' \qquad \text{for } x\in[-1,1]\setminus[-3\Delta/16,3\Delta/16]. \label{eq:Q0_outside}
\end{align}
Now define \( S_0(x):=\frac{Q_0(x)+Q_0(-x)}{2}.
\) Then \(S_0\) is an even polynomial of the same degree, and it satisfies the same bounds:
\begin{align}
|S_0(x)-1| &\le \epsilon' \qquad \text{for } x\in[-\Delta/8,\Delta/8], \label{eq:S0_local}\\
|S_0(x)| &\le 1+\epsilon' \qquad \text{for } x\in[-1,1], \label{eq:S0_global}\\
|S_0(x)| &\le \epsilon' \qquad \text{for } x\in[-1,1]\setminus[-3\Delta/16,3\Delta/16]. \label{eq:S0_outside}
\end{align}

Next we apply \cite[Corollary 66]{gilyenQuantumSingularValue2019} to
\[
f_+(x):=e^{-tx}
\]
with center \( x_+ := \frac{1+\Delta}{2}, \) radius \( r_+ := \frac{1-\Delta}{2}, \) and buffer width \( \vartheta_+ := \frac{\Delta}{2}.
\) These choices satisfy \(0<\vartheta_+\le r_+\). Since \(f_+\) is entire and
\[
f_+(x_+ + y)= e^{-t x_+}\sum_{\ell=0}^{\infty}\frac{(-t y)^\ell}{\ell!},
\]
we have
\[
\sum_{\ell=0}^{\infty}(r_+ + \vartheta_+)^\ell |a_\ell|
=
e^{-t x_+} e^{t(r_+ + \vartheta_+)}
=
e^{-t\Delta/2}
\le 1,
\]
because \(r_+ + \vartheta_+ = 1/2\). Thus we may take \(B_+=1\), uniformly for all \(t\ge 0\).
Therefore there exists an efficiently computable polynomial \(P_+\) of degree \( \mathcal{O}\!\left(\frac{1}{\Delta}\log\frac{1}{\epsilon}\right) \) such that
\begin{align}
|P_+(x)-e^{-tx}| &\le \epsilon' \qquad \text{for } x\in[\Delta,1], \label{eq:Pplus_local}\\
|P_+(x)| &\le 1+\epsilon' \qquad \text{for } x\in[-1,1], \label{eq:Pplus_global}\\
|P_+(x)| &\le \epsilon' \qquad \text{for } x\in[-1,3\Delta/4]. \label{eq:Pplus_left}
\end{align}

Define \( \widetilde{P}(x):= S_0(x)+P_+(x)+P_+(-x). \) Then \(\widetilde{P}\) is an even polynomial of degree \( \mathcal{O}\!\left(\frac{1}{\Delta}\log\frac{1}{\epsilon}\right).
\) We now verify its properties.

\paragraph{(i) Value at the origin.}
Since \(0\in[-\Delta/8,\Delta/8]\), by \eqref{eq:S0_local} we have \( |S_0(0)-1|\le \epsilon'.
\) Moreover, \(0\le 3\Delta/4\), so \eqref{eq:Pplus_left} gives \( |P_+(0)|\le \epsilon'.
\) Hence
\[
\begin{aligned}
|\widetilde{P}(0)-1|
&\le |S_0(0)-1| + |P_+(0)| + |P_+(-0)|\\
&\le 3\epsilon'.
\end{aligned}
\]

\paragraph{(ii) Approximation on the right support interval.}
Let \(x\in[\Delta,1]\). Then \(x\notin[-3\Delta/16,3\Delta/16]\), so by \eqref{eq:S0_outside}, \( |S_0(x)|\le \epsilon'.
\) Also \(-x\le -\Delta \le 3\Delta/4\), so by \eqref{eq:Pplus_left}, \( |P_+(-x)|\le \epsilon'.
\) Combining these with \eqref{eq:Pplus_local}, we obtain
\[
\begin{aligned}
|\widetilde{P}(x)-e^{-tx}|
&\le |P_+(x)-e^{-tx}| + |S_0(x)|\\
&\quad+ |P_+(-x)|\\
&\le 3\epsilon'.
\end{aligned}
\]

\paragraph{(iii) Approximation on the left support interval.}
Let \(x\in[-1,-\Delta]\). Since \(-x\in[\Delta,1]\), the same argument gives
\[
|\widetilde{P}(x)-e^{-t|x|}| \le 3\epsilon'.
\]

\paragraph{(iv) Global boundedness.}
We show that \(|\widetilde{P}(x)|\le 1+3\epsilon'\) for all \(x\in[-1,1]\).

If \(|x|\le \Delta/8\), then \eqref{eq:S0_global} and \eqref{eq:Pplus_left} imply
\[
\begin{aligned}
|\widetilde{P}(x)|
&\le |S_0(x)| + |P_+(x)| + |P_+(-x)|\\
&\le (1+\epsilon') + \epsilon' + \epsilon'\\
&= 1+3\epsilon'.
\end{aligned}
\]

If \(x\in[\Delta,1]\), then \eqref{eq:S0_outside}, \eqref{eq:Pplus_global}, and \eqref{eq:Pplus_left} give
\[
\begin{aligned}
|\widetilde{P}(x)|
&\le \epsilon' + (1+\epsilon') + \epsilon'\\
&= 1+3\epsilon'.
\end{aligned}
\]
By symmetry the same holds for \(x\in[-1,-\Delta]\).

It remains to consider the gap region \(x\in[-\Delta,\Delta]\setminus[-\Delta/8,\Delta/8]\).
If \(|x|\le 3\Delta/4\), then both \(x\) and \(-x\) lie in \((-\infty,3\Delta/4]\), so \eqref{eq:Pplus_left} gives
\[
|P_+(x)|\le \epsilon',
\qquad
|P_+(-x)|\le \epsilon',
\]
and thus
\[
|\widetilde{P}(x)| \le |S_0(x)| + 2\epsilon' \le 1+3\epsilon'
\]
by \eqref{eq:S0_global}.
If instead \(3\Delta/4 \le |x| \le \Delta\), then exactly one of \(x\) and \(-x\) can lie in \([3\Delta/4,\Delta]\), while the other lies in \((-\infty,-3\Delta/4]\subseteq(-\infty,3\Delta/4]\). Hence one flank is bounded by \(1+\epsilon'\), the opposite flank is bounded by \(\epsilon'\), and \(S_0(x)\le \epsilon'\) by \eqref{eq:S0_outside}. Therefore again
\[
|\widetilde{P}(x)| \le 1+3\epsilon'.
\]

Thus
\[
\|\widetilde{P}\|_{[-1,1]} \le 1+3\epsilon'.
\]

Define
\[
P(x):=\frac{\widetilde{P}(x)}{1+3\epsilon'}.
\]
Then \(P\) is still even, efficiently computable, and has the same degree as \(\widetilde{P}\). Moreover,
\[
\|P\|_{[-1,1]} \le 1.
\]

At the origin,
\[
\begin{aligned}
|P(0)-1|
&\le \frac{|\widetilde{P}(0)-1|}{1+3\epsilon'}
+ \left|1-\frac{1}{1+3\epsilon'}\right|\\
&\le 3\epsilon' + 3\epsilon'\\
&= 6\epsilon' \le \epsilon.
\end{aligned}
\]

For \(x\in[-1,1]\setminus(-\Delta,\Delta)\), we have \(|e^{-t|x|}|\le 1\), so
\[
\begin{aligned}
|P(x)-e^{-t|x|}|
&\le \frac{|\widetilde{P}(x)-e^{-t|x|}|}{1+3\epsilon'}\\
&\quad+ \left|1-\frac{1}{1+3\epsilon'}\right|\,|e^{-t|x|}|\\
&\le 3\epsilon' + 3\epsilon'\\
&= 6\epsilon' = \epsilon.
\end{aligned}
\]

Therefore \(P\) satisfies
\begin{enumerate}
    \item \( |P(0)-1| \le \epsilon\),
    \item \( \|P(x)-e^{-t|x|}\|_{[-1,1]\setminus(-\Delta,\Delta)} \le \epsilon\),
    \item \( \|P(x)\|_{[-1,1]} \le 1\),
\end{enumerate}
and has degree
\[
\mathcal{O}\!\left(\frac{1}{\Delta}\log\frac{1}{\epsilon}\right).
\]
This proves the lemma.
\end{proof}

Since the approximating polynomial is even, Lemma~\ref{lemma:qsvt-general} applies directly to the block-encoding \(U_\rho\) of the Hermitian operator \(\rho\), and the even singular value transformation reduces to the standard eigenvalue transformation. For the \(k\)-th quadrature node, write \(t_{k,K}=x_{k,K}/\Delta\), and Lemma~\ref{lem:gapped_exponential_anchor_tt} gives a bounded even polynomial approximation on the gapped spectral domain. Applied to the block-encoding \(U_\rho\), this produces a QSVT implementation of \(e^{-t_{k,K}\rho}\) that respects the support gap and leaves the kernel sector controlled at \(x=0\), which is a \((1,a+s+1,\epsilon_\mathrm{p})\)-block-encoding of \(e^{- \frac{x_{k,K}}{\Delta} \rho}\) using a total of \(d=\mathcal O\!\left(\frac{1}{\Delta}\log\frac{1}{\epsilon_\mathrm{p}}\right)\) queries to \(U_\rho\) and \(U_\rho^\dagger\).

To construct the full operator \(e^{- \frac{x_{k,K}}{\Delta} \rho} (\partial_j \rho) e^{- \frac{x_{k,K}}{\Delta} \rho}\), we apply Lemma~\ref{lemma:block-encoding-product} twice.
Let \(a_{\exp}:=a+s+1\). For each quadrature node \(k\), let \(U_k\) be the \((1,a_{\exp},\epsilon_{\mathrm p})\)-block encoding of \(e^{-t_k\rho}\), and let \(U_{\partial_j\rho}\) be an \((\alpha_j,m_j,0)\)-block encoding of \(\partial_j \rho\) as in Eq.~\eqref{eq:derivative-block-encoding-intro}.
To compose these block-encodings, we introduce the registers \(A_1\), \(A_\partial\), and \(A_2\) used in Section~\ref{algorithm-design-for-solving-the-lyapunov-equation}. With \(b=\lceil\log_2K\rceil\) and \(\Pi_\perp=I_{\mathrm{idx}}-\sum_{k=1}^K\ket{k}\!\bra{k}\), define
\[
\begin{aligned}
\operatorname{SELECT}_R
&=\sum_{k=1}^K\ket{k}\!\bra{k}\otimes U_k^{(A_1S)}
+\Pi_\perp\otimes I_{\mathrm{work}},\\
\operatorname{SELECT}_L
&=\sum_{k=1}^K\ket{k}\!\bra{k}\otimes U_k^{(A_2S)}
+\Pi_\perp\otimes I_{\mathrm{work}}.
\end{aligned}
\]
Then
\[
\begin{aligned}
\mathcal{V}
&:=\operatorname{SELECT}_L
\bigl(I_{\mathrm{idx}}\otimes U_{\partial_j\rho}^{(A_\partial S)}\bigr)
\operatorname{SELECT}_R\\
&=\sum_{k=1}^K\ket{k}\!\bra{k}\otimes V_{j,k}
+\Pi_\perp\otimes U_{\partial_j\rho}^{(A_\partial S)},
\end{aligned}
\]
where
\[
V_{j,k}
=U_k^{(A_2S)}U_{\partial_j\rho}^{(A_\partial S)}U_k^{(A_1S)}.
\]
By two applications of Lemma~\ref{lemma:block-encoding-product}, \(V_{j,k}\) is an \((\alpha_j,2a_{\exp}+m_j,2\alpha_j\epsilon_{\mathrm p})\)-block-encoding of \( e^{-t_{k,K}\rho}(\partial_j\rho)e^{-t_{k,K}\rho} \). The gate complexity of this construction is \(\mathcal{O}(Kd+d(T_\rho+a_{\exp})+T_j)\), where the first term comes from the gate complexity of the multiplexed phase rotations in the Select unitaries, the second term comes from the \(d\) queries to \(U_\rho\) and \(U_\rho^\dagger\), respectively, and the last term comes from the block-encoding of \(\partial_j\rho\).

To combine these \(K\) node-dependent blocks, we utilize the LCU construction (Lemma~\ref{lem:lcu-symmetric}) with a selection register of \(b=\lceil \log_2 K\rceil\) qubits. Let \(y \in \mathbb{R}_+^K\) be the coefficient vector with \(y_k = w_{k,K} e^{x_{k,K}}\). We define \(\beta = \|y\|_1\) and assume an \((\beta,b,\epsilon_{\text{prep}})\)-state-preparation unitary such that \footnote{While the coefficients $y_k$ are stored with finite precision $\tilde{y}_k$ such that $\|y - \tilde{y}\|_1 \le \epsilon$, the resulting normalization error remains $\mathcal{O}(\epsilon)$ as long as $\epsilon < \beta$. We directly use the effective state-preparation error $\epsilon_{\text{prep}}$ defined in Definition~\ref{def:state-preparation-unitary} for simplicity.},
\begin{equation}
\mathcal P \ket{0}^{\otimes b} = \sum_{k=1}^{K} c_k \ket{k}, \quad \sum_{k=1}^{K} |\beta |c_k|^2 - y_k| \le \epsilon_{\text{prep}}.
\end{equation}
To remain agnostic of any specific structure in the Gauss--Laguerre weights, we employ a general-purpose amplitude-loading circuit via standard uniformly controlled rotations. This requires \(\mathcal{O}(K)\) single-qubit rotations and two-qubit gates. Specifying each rotation angle to precision \(\delta = \mathcal{O}(\epsilon_{\mathrm{prep}}/(\beta K))\) suffices for the preparation error budget, while the elementary-gate count is \(T_{\mathrm{prep}} = \mathcal{O}(K)\).

By Lemma~\ref{lem:lcu-symmetric}, with the state preparation unitary \(\mathcal P\) (error \(\varepsilon_1 = \epsilon_{\mathrm{prep}}\)) and the Select unitary \(\mathcal{V}\) (each \(V_{j,k}\) is an \((\alpha_j, 2a_{\exp}+m_j, 2\alpha_j\epsilon_{\mathrm p})\)-block encoding, giving node-wise error \(\varepsilon_2 = 2\alpha_j\epsilon_{\mathrm p}\)), we obtain an \((\alpha_j \beta, 2a_{\exp}+m_j+b, \alpha_j \epsilon_{\text{prep}}+2\alpha_j \beta \epsilon_{\mathrm{p}})\)-block encoding of \(\Delta \cdot \hat{L}_j\).

This is precisely the PREPARE--SELECT--UNPREPARE unitary in Eq.~\eqref{eq:block-encoding-unitary-for-L}, shown in Fig.~\ref{fig:sld-block-encoding}.
Its projected block obeys
\[
\begin{aligned}
&\|\Delta \cdot \hat L_j - \alpha_j \beta (\bra{0}^{2a_{\exp}+m_j+b}\otimes I) U_{L_j} \\
&\qquad (\ket{0}^{2a_{\exp}+m_j+b}\otimes I)\| \le \alpha_j \epsilon_{\text{prep}} + 2\alpha_j \beta \epsilon_{\mathrm{p}}.
\end{aligned}
\]

Denote \(\alpha_{L_j} = \alpha_j \beta / \Delta\), \(\epsilon_{\hat{L}_j} = \dfrac{\alpha_j}{\Delta}\bigl(\epsilon_{\mathrm{prep}} + 2\beta\epsilon_{\mathrm{p}}\bigr)\) and \(a_{L_j}=2a+2s+m_j+\lceil \log_2 K \rceil + 2\).
Therefore the total error in the realized block-encoding of \(L_j\) is
\begin{align}
&\bigl\|L_j - \alpha_{L_j} (\langle 0|^{\otimes a_{L_j}} \otimes I) \,
U_{L_j} (|0\rangle^{\otimes a_{L_j}} \otimes I)\bigr\| \nonumber \\
&\le \|L_j - \hat{L}_j\|
   + \bigl\|\hat{L}_j - \alpha_{L_j} (\langle 0|^{\otimes a_{L_j}} \otimes I) \, \nonumber\\
&\qquad U_{L_j} (|0\rangle^{\otimes a_{L_j}} \otimes I)\bigr\| \nonumber \\
&\le \epsilon_{\mathrm{trun}} + \frac{\alpha_j}{\Delta}\bigl(\epsilon_{\mathrm{prep}} + 2\beta\epsilon_{\mathrm{p}}\bigr)
\label{eq:total-error-bound-for-block-encoding-of-L}
\end{align}

The total gate complexity of the block-encoding construction is
\begin{equation}
\begin{aligned}
T_L = \mathcal{O}\bigl(&K + d (T_\rho + a_{\exp}) \\
&+ K d + T_{\partial_j \rho} \bigr),
\end{aligned}
\label{eq:total-gate-complexity-for-block-encoding-of-L}
\end{equation}
where \(K = \Theta\!\left(\frac{1}{\Delta} \log\left(\frac{\|\partial_j \rho \|_F}{\epsilon_{\mathrm{trun}} \Delta}\right)\right)\) is the required number of quadrature nodes to achieve the quadrature error \(\epsilon_{\mathrm{trun}}\), and \(d = \mathcal{O}\!\left(\frac{1}{\Delta} \log\left(\frac{1}{\epsilon_{\mathrm p}}\right)\right)\).

To make the block encoding error upper bounded by a prescribed tolerance \(\epsilon_j\), it suffices to choose

\[
\begin{aligned}
\epsilon_{\mathrm{trun}} &\le \frac{\epsilon_j}{3},\\
\epsilon_{\mathrm{prep}} &\le \frac{\epsilon_j \Delta}{3\alpha_j},\\
\epsilon_{\mathrm{p}} &\le \frac{\epsilon_j\Delta}{6\alpha_j\beta}
\le \frac{\epsilon_j\Delta}{12\alpha_j K}
\end{aligned}
\]
where we invoked the coefficient-normalization statement of Theorem~\ref{theorem:quadrature-main}.
Concretely, we take the number of quadrature nodes and errors as
\begin{equation}
    K = \left\lceil \frac{1}{\Delta} \log\left(\frac{6\|\partial_j \rho \|_F}{\epsilon_j \Delta}\right) \right\rceil
    \label{eq:required-number-of-quadrature-nodes-for-total-error-tolerance}
\end{equation}

\begin{equation}
\begin{aligned}
  \epsilon_{\mathrm{trun}} &= \frac{\epsilon_j}{3},\\
  \epsilon_{\mathrm{prep}}
  &=\frac{\epsilon_j \Delta}{3\alpha_j},\\
  \epsilon_{\mathrm p}
  &=\frac{\epsilon_j\Delta}{6\alpha_j\beta_K}
\end{aligned}
\label{eq:choice-of-internal-errors-for-L-block-encoding}
\end{equation}

With these choices, to construct a block-encoding of \(L_j\) with error \(\epsilon_j\) requires

\begin{equation}
\begin{aligned}
\log\frac{1}{\epsilon_{\mathrm p}}
&=\log\!\left(\frac{6\alpha_j\beta_K}{\epsilon_j\Delta}\right),\\
Q_\rho^{(j)}
&=\mathcal{O}\!\left(\frac{1}{\Delta}\log\frac{1}{\epsilon_{\mathrm p}}\right)\\
&=\mathcal{O}\!\left(
\frac{1}{\Delta}\left[
\log\!\left(\frac{\alpha_j}{\epsilon_j\Delta}\right)
+\log\log\!\left(\frac{\|\partial_j\rho\|_F}{\epsilon_j\Delta}\right)
\right]\right)
\end{aligned}
\label{eq:rho-query-complexity-for-L-block-encoding}
\end{equation}
Here \(\beta_K=\Theta(K)\) by Theorem~\ref{theorem:quadrature-main}, and the last line absorbs fixed multiplicative constants and fixed powers inside the logarithms into the asymptotic notation. The quantity \(Q_\rho^{(j)}\) counts queries to \(O_\rho\) and \(O_\rho^\dagger\); the different quadrature nodes share the same QSVT oracle backbone and differ only in their multiplexed phase rotations.
The total number of queries to the block-encoding of \(\partial_j\rho\) is \(Q_{\partial_j} = \mathcal{O}(1)\), and the total number of ancilla qubits required for the block-encoding of \(L_j\) is
\begin{equation}
\begin{aligned}
a_{L_j} &= 2a + 2s + m_j + 2 + \lceil \log_2 K \rceil \\
    &= 2a + 2s + m_j + 2 + \mathcal{O}\!\left(\log\frac{1}{\Delta} + \log\log\frac{\|\partial_j\rho\|_F}{\epsilon_j\Delta}\right).
\end{aligned}
\label{eq:total-ancilla-qubits-for-block-encoding-of-L}
\end{equation}
\subsection{Block-Encoding of the Regularized SLD}\label{sec:reg:be_construction}

The block-encoding construction of Section~\ref{algorithm-design-for-solving-the-lyapunov-equation} builds the gapped SLD from the quadrature formula, Eq.~\eqref{eq:expression-for-sld-after-gauss-laguerre-quadrature}, using three primitives: QSVT exponential filters, products with \(\partial_j\rho\), and LCU assembly.  We now construct the corresponding block-encoding for the regularized quadrature formula, Eq.~\eqref{eq:reg:GL_discrete_quad}.  The modular structure is identical; the differences arise in the spectral domain over which the QSVT polynomials must approximate the exponential, and in the LCU coefficient normalization.

\smallskip\noindent \textit{QSVT exponential filter.}---%
In the gapped construction, Lemma~\ref{lem:gapped_exponential_anchor_tt} supplies an even polynomial approximating \(e^{-t|x|}\) away from the origin, avoiding the derivative cusp of \(e^{-t|x|}\) at~\(x=0\) and producing \(e^{-t\rho}\) via a single QSVT call with degree \(\mathcal{O}(\Delta^{-1}\log\epsilon_{\mathrm{p}}^{-1})\) independent of~\(t\).

When the spectral gap is absent, the eigenvalues of~\(\rho\) fill~\([0,1]\) and the cusp can no longer be avoided by an even polynomial.  We instead drop the parity requirement and use the following standard lemma.

\begin{lemma}[Bounded approximation of \(e^{-t x}\) on {\([0,1]\)},
  adapted from~{\cite[Corollary~23]{tangCSGuideQuantum2023a}}]\label{lem:reg:exp_approx} Let \(t \ge 0\) and \(\epsilon \in (0,1)\).  There exists an efficiently computable polynomial \(P \in \mathbb{R}[x]\) of degree
\begin{equation}
 d=\begin{cases}
 \mathcal{O}\!\left(\log\dfrac{2}{\epsilon}\right), & 0\le t<1,\\[4pt]
 \mathcal{O}\!\left(t\log\dfrac{t}{\epsilon}\right), & t\ge1,
 \end{cases}
\end{equation}
such that \(|P(x) - e^{-t x}| \le \epsilon\) for \(x \in [0,1]\) and \(\sup_{x\in[-1,1]}|P(x)|=\mathcal{O}(1)\), uniformly in \(t\) and \(\epsilon\). For \(t\ge1\), this follows from the cited corollary after replacing \(x\) by \(-x\); for \(0\le t<1\), a truncated Taylor series gives the stated bound.
  The polynomial \(P\) carries no definite parity.
\end{lemma}

Since \(P\) has no definite parity, choose a common normalization \(\alpha_{\exp}=\mathcal{O}(1)\), with \(\alpha_{\exp}\ge1\) and \(|P|\le\alpha_{\exp}/2\) on \([-1,1]\), and apply Lemma~\ref{lemma:qsvt-arbitrary-parity} to \(Q=P/\alpha_{\exp}\). This implements \(P(\rho)/\alpha_{\exp}\) using \(\mathcal{O}(d)\) queries to~\(U_\rho\), with \(d\) chosen from the two cases of Lemma~\ref{lem:reg:exp_approx} at \(t=t_k\) and \(\epsilon=\epsilon_{\mathrm p}\).
Thus each exponential filter is an \((\alpha_{\exp},a+s+2,\epsilon_{\mathrm p})\)-block encoding of \(e^{-t_k\rho}\).

The key difference from the gapped setting is that the degree now \emph{does} depend on \(t_k\). Since \(x_{K,K}\ge\sum_k w_{k,K}x_{k,K}=1\) and \(0<\theta\le1\), the largest time \(t_{\max}=x_{K,K}/\theta\ge1\) falls in the second case of Lemma~\ref{lem:reg:exp_approx}. By Lemma~\ref{lemma:asymptotic-laguerre-roots-weights}, \(x_{K,K}=4K+\mathcal{O}(K^{1/3})\), so the dominant degree across all \(K\) nodes is
\begin{equation}
  d_{\max}^{(\theta)} = O\!\Bigl(\frac{K}{\theta}
                         \log\frac{K}{\theta\epsilon_{\mathrm{p}}}\Bigr).
  \label{eq:reg:qsvt_degree}
\end{equation}

\smallskip\noindent \textit{LCU assembly.}---%
The quadrature sum~\eqref{eq:reg:GL_discrete_quad} is a linear combination of \(K\) terms, each of the form \(e^{-t_k\rho}(\partial_j\rho)e^{-t_k\rho}\). Two applications of the product lemma show that the node circuit is a \((\mathcal{O}(\alpha_j),2a+2s+m_j+4,2\alpha_{\exp}\alpha_j\epsilon_{\mathrm p})\)-block encoding of this filtered derivative. By the LCU lemma (Lemma~\ref{lem:lcu-symmetric}), the coefficient sum is \(\frac{1}{\theta}\sum_k w_{k,K} = 1/\theta\), native from the quadrature weights summing to unity---no compensation factors appear.

\smallskip\noindent \textit{Normalization and error budget.}---%
Assembling the three primitives (two QSVT filters, product with \(U_{\partial_j}\) via Lemma~\ref{lemma:block-encoding-product}, and LCU over \(K\) nodes), the resulting block-encoding has normalization
\begin{equation}
  \alpha_{L_j^{(\theta)}} = \mathcal{O}\!\left(\frac{\alpha_j}{\theta}\right),
  \label{eq:reg:alpha_L_be}
\end{equation}
where \(\alpha_j\) is the normalization of \(U_{\partial_j}\), and the chosen normalization is \(\alpha_{\exp}^2\alpha_j/\theta\). The total block-encoding error \(\epsilon_{\mathrm{BE}}\)
\begin{equation}
  \epsilon_{\mathrm{BE}} \le \epsilon_{\mathrm{trun}}
              + \alpha_{L_j^{(\theta)}}
                (\epsilon_{\mathrm{prep}} + 2\epsilon_{\mathrm{p}})
  \label{eq:reg:epsilon_j_be}
\end{equation}
receives contributions from quadrature truncation (\(\epsilon_{\mathrm{trun}}\), Section~\ref{sec:reg:quadrature}), polynomial approximation (\(\epsilon_{\mathrm{p}}\)), and LCU state preparation (\(\epsilon_{\mathrm{prep}}\)).

To make the block-encoding error bounded by a prescribed tolerance \(\epsilon_j\), we balance the three contributions equally:
\begin{equation}
  \epsilon_{\mathrm{trun}} = \frac{\epsilon_j}{3},\qquad
  \epsilon_{\mathrm{prep}} = \frac{\epsilon_j}{3\alpha_{L_j^{(\theta)}}},\qquad
  \epsilon_{\mathrm{p}} = \frac{\epsilon_j}{6\alpha_{L_j^{(\theta)}}}.
  \label{eq:reg:error_allocation}
\end{equation}
With these choices, Eq.~\eqref{eq:reg:K_selection_quad} gives \(K = \Theta\bigl(\frac{1}{\theta}\log\frac{\|\partial_j\rho\|_F}{\epsilon_j\theta}\bigr)\).

At the query level, the construction uses \(\widetilde{\mathcal{O}}(d_{\max}^{(\theta)})\) queries to \(O_\rho\) and \(\mathcal{O}(1)\) queries to \(U_{\partial_j}\).
Substituting \(K = \widetilde{\Theta}(1/\theta)\) and \(d_{\max}^{(\theta)} = \mathcal{O}\bigl(\frac{K}{\theta}\log\frac{K}{\theta\epsilon_{\mathrm{p}}}\bigr)\) gives
\begin{equation}
  Q_\rho^{(j)} = \widetilde{\mathcal{O}}\!\Bigl(\frac{1}{\theta^2}\Bigr),\qquad
  Q_{\partial_j} = \mathcal{O}(1),
  \label{eq:reg:query_complexity_be}
\end{equation}
where the tildes suppress polylogarithmic factors in \(1/\theta\) and \(1/\epsilon_{\mathrm{p}}\).
The ancilla count is \(a_{L_j^{(\theta)}} = 2a + 2s + m_j + 4 + \lceil\log_2 K\rceil\).
Relative to the gapped construction, the additional two ancillas arise from the arbitrary-parity transformations used by the two exponential filters.

The elementary-gate cost \(T_L^{(\theta)}\) is given by Eq.~\eqref{eq:total-gate-complexity-for-block-encoding-of-L} with \(d=d_{\max}^{(\theta)}\), \(a_{\exp}=a+s+2\), and the regularized quadrature order above.

\section{QFI and QGT Estimation: Proofs and Resource Analysis}\label{appendix:qfim-proofs}

This appendix follows the three capabilities developed in Section~\ref{sec:accessing-qfim} in the same order. Section~\ref{app:diagonal-analysis} applies the amplitude-estimation interface of Appendix~\ref{app:qpe-free-amplitude-estimation} to the complete bounded-gap diagonal-estimation chain, Section~\ref{app:qgt-block-encoding-details} supplies the projected two-ancilla derivation for the full QGT and QFIM construction, and Section~\ref{app:regularized-estimation} collects the bounded-response error and resource analysis.

\subsection{Bounded-Gap Diagonal QFI Estimation}\label{app:diagonal-analysis}

Since \(\mathcal{F}_{jj}=4\mathcal{N}_j^2\), with \(\mathcal{N}_j=\|(L_j\otimes I_A)\ket{\rho}_{SA}\|\), diagonal QFI estimation reduces to estimating the norm of the SLD-applied purification.  For the gapped construction, each \(U_{L_j}\) is an \((\alpha_{L_j},a_{L_j},\epsilon_j)\)-block encoding of \(L_j\):
\begin{equation}
\begin{aligned}
  &\left\|L_j-\alpha_{L_j}
  (\langle0|^{\otimes a_{L_j}}\otimes I)U_{L_j}\right.\\
  &\left.\qquad\times
  (|0\rangle^{\otimes a_{L_j}}\otimes I)\right\|\le\epsilon_j,
\end{aligned}
  \label{eq:block-encoding-of-Lj}
\end{equation}
where
\begin{equation}
\begin{aligned}
  \alpha_{L_j}&=\frac{\alpha_j\beta_j}{\Delta},
  \\
  \beta_j&=\sum_{k=1}^{K_j}w_{k,K_j}e^{x_{k,K_j}},\\
  a_{L_j}&=2a+2s+2+m_j+\lceil\log_2K_j\rceil,
  \\
  K_j&=\left\lceil\frac{1}{\Delta}
  \log\!\left(\frac{6\|\partial_j\rho\|_F}{\epsilon_j\Delta}\right)\right\rceil.
\end{aligned}
\label{eq:parameters-of-block-encoding-of-Lj}
\end{equation}
The corresponding gate complexity is
\begin{equation}
\begin{aligned}
T_{L_j}=\mathcal O\bigl(&K_j\\
 &+d_j(T_\rho+a_{\exp})+K_jd_j+T_j\bigr),
\end{aligned}
\label{eq:gate-complexity-for-block-encoding-of-Lj}
\end{equation}
where \(d_j=\mathcal O(\Delta^{-1}\log(1/\epsilon_{\mathrm p}))\), with the internal tolerances in Eq.~\eqref{eq:choice-of-internal-errors-for-L-block-encoding}.

Applying this block encoding to the purification gives
\begin{equation}
\begin{split}
  (U_{L_j}\otimes I_A)|0\rangle^{\otimes a_{L_j}}|\rho\rangle_{SA}
  &=\frac{1}{\alpha_{L_j}}|0\rangle^{\otimes a_{L_j}}
    (\widetilde L_j\otimes I_A)|\rho\rangle_{SA}\\
  &\quad+|\perp'\rangle,
\end{split}
\label{eq:preparation-of-psi-j-using-block-encoding}
\end{equation}
where \(|\perp'\rangle\) is orthogonal to the zero-ancilla sector and \(\|\widetilde L_j-L_j\|\le\epsilon_j\).  Writing \(\widetilde{\mathcal N}_j=\|(\widetilde L_j\otimes I_A)|\rho\rangle\|\), the success probability is
\begin{equation}
  p_j=\left(\frac{\widetilde{\mathcal N}_j}{\alpha_{L_j}}\right)^2,
\end{equation}
and
\begin{equation}
  |\widetilde{\mathcal N}_j-\mathcal N_j|
  \le\|\widetilde L_j-L_j\|\le\epsilon_j.
\label{eq:error-bound-for-norm-of-approximated-state}
\end{equation}
Thus, with \(\mathcal W_j=(U_{L_j}\otimes I_A)(I_{a_{L_j}}\otimes O_\rho)\),
\begin{equation}
\begin{aligned}
  \mathcal W_j|0\rangle^{\otimes a_{L_j}}|0\rangle_{SA}
  &=\sqrt{p_j}|0\rangle^{\otimes a_{L_j}}|\widehat\psi_j\rangle\\
  &\quad+\sqrt{1-p_j}|\perp\rangle,
\end{aligned}
\label{eq:normalized-state-after-block-encoding}
\end{equation}
and the approximate norm is encoded in \(\gamma_j=\sqrt{p_j}=\widetilde{\mathcal N}_j/\alpha_{L_j}\).

\subsubsection{Norm bounds}

\begin{lemma}[Bounds on the block-encoded state norm]
\label{lemma:norm-bounds}
Let \(\rho\) satisfy the bounded-gap promise, and let \(L_j\) be the SLD of Eq.~\eqref{eq:definition-of-sld}. Then
\begin{equation}
  \frac{1}{\sqrt2}\|\partial_j\rho\|_F
  \le\mathcal N_j\le
  \frac{1}{\sqrt{2\Delta}}\|\partial_j\rho\|_F.
\label{eq:norm-frobenius-bounds}
\end{equation}
\end{lemma}

\begin{proof}
We evaluate both norms in the eigenbasis of the density matrix, \(\rho = \sum_n p_n |\phi_n\rangle\langle \phi_n|\). The squared Frobenius norm of the derivative is the sum of its squared matrix elements, \(||\partial_j \rho||_F^2 = \sum_{n,m} |\bra{\phi_n}\partial_j \rho\ket{\phi_m}|^2\). The squared state norm expands as:
\begin{equation}
\begin{aligned}
    \mathcal{N}_j^2
    &= \frac{\mathcal{F}_{jj}}{4}\\
    &= \frac{1}{2} \sum_{\substack{m,n=1 \\ p_m + p_n > 0}}^{N}
    \frac{|\bra{\phi_n}\partial_j \rho\ket{\phi_m}|^2}{p_m + p_n}
\end{aligned}
    \label{eq:app-squared-state-norm-expansion}
\end{equation}
We first establish the lower bound by partitioning this sum into off-diagonal (\(n \neq m\)) and diagonal (\(n = m\)) contributions, and then establish the upper bound by minimizing the denominator.

\paragraph{Lower bound on off-diagonal elements (\(n \neq m\)).}
Since \(\rho\) is a normalized quantum state, its eigenvalues satisfy \(p_k \ge 0\) and \(\sum_k p_k = 1\). For any two distinct indices \(n \neq m\), it strictly holds that \(p_n + p_m \le 1\). Therefore, the weighting coefficient satisfies \(\frac{1}{2(p_n + p_m)} \ge \frac{1}{2}\). Consequently:
\begin{widetext}
\begin{equation}
    \sum_{\substack{m,n=1 \\ m \neq n, p_m + p_n > 0}}^{N} \frac{1}{2(p_n + p_m)} |\bra{\phi_n}\partial_j \rho\ket{\phi_m}|^2 \ge \frac{1}{2} \sum_{\substack{m,n=1 \\ m \neq n, p_m + p_n > 0}}^{N} |\bra{\phi_n}\partial_j \rho\ket{\phi_m}|^2
    \label{eq:app-off-diagonal-bound}
\end{equation}
\end{widetext}

\paragraph{Lower bound on diagonal elements (\(n = m\)).}
Let \(\Delta_0 = \min_{p_n > 0} p_n\) be the minimal non-zero eigenvalue of \(\rho\).
For the diagonal terms, the sum simplifies to \(\frac{1}{4} \sum_n \frac{x_n^2}{p_n}\), where \(x_n = \bra{\phi_n}\partial_j \rho\ket{\phi_n}\). By the Cauchy-Schwarz inequality, \(\left(\sum_n \frac{x_n^2}{p_n}\right) \left(\sum_n p_n\right) \ge \left(\sum_n |x_n|\right)^2\). Since \(\text{Tr}(\rho) = 1\), we have \(\sum_n p_n = 1\).
Furthermore, trace-preservation requires \(\text{Tr}(\partial_j \rho) = \sum_n x_n = 0\). This implies that the sum of the positive elements, \(S\), equals the absolute sum of the negative elements. Thus, \(\sum_n |x_n| = 2S\), and \(\left(\sum_n |x_n|\right)^2 = 4S^2\).
Because \(a^2 + b^2 \le (a+b)^2\) for non-negative numbers, the sum of squares of the positive elements is bounded by \(S^2\), and similarly for the negative elements. Hence, \(\sum_n x_n^2 \le 2S^2\).
Chaining these inequalities together yields:
\begin{equation}
    \sum_{\substack{n=1 \\ p_n > 0}}^{N} \frac{x_n^2}{p_n} \ge 4S^2 \ge 2 \sum_{\substack{n=1 \\ p_n > 0}}^{N} x_n^2
\end{equation}
Dividing by 4, we obtain the bound for the diagonal contribution:
\begin{widetext}
\begin{equation}
    \sum_{\substack{n=1 \\ p_n > 0}}^{N} \frac{1}{4p_n} |\bra{\phi_n}\partial_j \rho\ket{\phi_n}|^2 \ge \frac{1}{2} \sum_{\substack{n=1 \\ p_n > 0}}^{N} |\bra{\phi_n}\partial_j \rho\ket{\phi_n}|^2
    \label{eq:app-diagonal-bound}
\end{equation}
\end{widetext}
Combining Eq.~\eqref{eq:app-off-diagonal-bound} and Eq.~\eqref{eq:app-diagonal-bound}, every term in the spectral expansion of \(\mathcal{N}_j^2\) is bounded by at least half of the corresponding term in the Frobenius norm:
\begin{equation}
\begin{aligned}
    \mathcal{N}_j^2
    &\ge \frac{1}{2} \sum_{\substack{m,n=1 \\ p_m + p_n > 0}}^{N}
    |\bra{\phi_n}\partial_j \rho\ket{\phi_m}|^2\\
    &= \frac{1}{2} ||\partial_j \rho||_F^2
\end{aligned}
\end{equation}
The final equality holds as guaranteed by Eq.~\eqref{eq:condition-for-sld-to-be-well-defined}, which ensures that \(\bra{\phi_n}\partial_j \rho\ket{\phi_m} = 0\) whenever \(p_n + p_m = 0\).

\paragraph{Upper bound.}
To establish the upper bound, we minimize the denominator \(p_m + p_n\) in Eq.~\eqref{eq:app-squared-state-norm-expansion}. The kernel condition in Eq.~\eqref{eq:condition-for-sld-to-be-well-defined} guarantees that \(|\bra{\phi_n}\partial_j \rho\ket{\phi_m}|^2 = 0\) whenever both \(p_m = 0\) and \(p_n = 0\). Thus, for every non-vanishing term in the sum, at least one index must belong to the support of \(\rho\), implying \(\max(p_m, p_n) \ge \Delta_0\). Consequently, the denominator is bounded from below by \(p_m + p_n \ge \Delta_0 + 0 = \Delta_0\).
Applying this inequality to every term in the sum gives:
\begin{equation}
\begin{aligned}
    \mathcal{N}_j^2
    &\le \frac{1}{2\Delta_0}
    \sum_{\substack{m,n=1 \\ p_m + p_n > 0}}^{N}
    |\bra{\phi_n}\partial_j \rho\ket{\phi_m}|^2\\
    &= \frac{1}{2\Delta_0} ||\partial_j \rho||_F^2\\
    &\le \frac{1}{2\Delta} ||\partial_j \rho||_F^2
\end{aligned}
\end{equation}
Taking the square root of the final bounds yields the strict inequality in Eq.~\eqref{eq:norm-frobenius-bounds}.

\end{proof}

Set \(\vartheta_{\mathrm{amp}}=\vartheta\).  Lemma~\ref{lem:qpe-free-additive-amplitude-estimation} then gives \(\widehat{\mathcal N}_j=\alpha_{L_j}\widehat\gamma_j\) such that
\[
  |\widehat{\mathcal N}_j-\widetilde{\mathcal N}_j|
  \le \alpha_{L_j}\epsilon_{\mathrm{amp}}
\]
with probability at least \(1-\vartheta\).  Combining this estimate with Eq.~\eqref{eq:error-bound-for-norm-of-approximated-state}, define
\[
  E_j:=\epsilon_j+\alpha_{L_j}\epsilon_{\mathrm{amp}},
  \qquad
  |\widehat{\mathcal N}_j-\mathcal N_j|\le E_j.
\]
The difference of squares therefore obeys the unconditional bound
\begin{equation}
\begin{aligned}
  |\widehat{\mathcal F}_{jj}-\mathcal F_{jj}|
  &=4|\widehat{\mathcal N}_j^2-\mathcal N_j^2|\\
  &\le 4E_j(2\mathcal N_j+E_j).
\end{aligned}
\label{eq:diagonal-square-propagation}
\end{equation}
Write \(B_j=\|\partial_j\rho\|_F/\sqrt{2\Delta}\), so that \(\mathcal N_j\le B_j\) and \(\mathcal F_{jj}\le4B_j^2\). If \(4B_j^2\le\epsilon\), output zero. Otherwise \(B_j>0\), and any allocation satisfying \(E_j\le\epsilon/(12B_j)\) gives \(E_j<B_j\). Equation~\eqref{eq:diagonal-square-propagation} therefore gives \(|\widehat{\mathcal F}_{jj}-\mathcal F_{jj}|\le12B_jE_j\le\epsilon\), without a lower bound on the signal. Substituting \(B_j\) and \(\alpha_{L_j}=\alpha_j\beta_{K_j}/\Delta\) from Proposition~\ref{prop:canonical-sld-block-encoding} yields
\begin{equation}
\begin{aligned}
|\widehat{\mathcal F}_{jj}-\mathcal F_{jj}|
&\le \frac{12}{\sqrt{2\Delta}}\,\|\partial_j\rho\|_F\\
&\quad\times\left(
\epsilon_j+\frac{\alpha_j\beta_j}{\Delta}\epsilon_{\mathrm{amp}}
\right),
\end{aligned}
\label{eq:diagonal-error-final}
\end{equation}
where \(\beta_j:=\beta_{K_j}=\sum_{k=1}^{K_j}w_{k,K_j}e^{x_{k,K_j}}\).  To expose its dependence on the internal accuracy, write
\begin{equation}
  \Lambda_j:=\log\!\left(
  \frac{6\|\partial_j\rho\|_F}{\epsilon_j\Delta}
  \right).
  \label{eq:diag-budget-definitions}
\end{equation}
The quadrature order in Proposition~\ref{prop:canonical-sld-block-encoding} and Theorem~\ref{theorem:quadrature-main} imply \(K_j=\Theta(\Delta^{-1}\Lambda_j)\) and \(\beta_j=\Theta(\Delta^{-1}\Lambda_j)\).  Thus, up to constants independent of the internal accuracies, Eq.~\eqref{eq:diagonal-error-final} imposes
\begin{equation}
  \epsilon_j
  +\Theta\!\left(
    \frac{\alpha_j}{\Delta^2}\Lambda_j
    \epsilon_{\mathrm{amp}}
  \right)
  \le \frac{\epsilon\sqrt{2\Delta}}{12\|\partial_j\rho\|_F}.
  \label{eq:diag-asymptotic-constraint}
\end{equation}

For the high-precision asymptotics, the amplitude-estimation cost is polynomial in \(1/\epsilon_{\mathrm{amp}}\), whereas the SLD block-encoding cost is only polylogarithmic in \(1/\epsilon_j\).  Consequently, the two contributions in Eq.~\eqref{eq:diag-asymptotic-constraint} should not simply be set equal.  Write the second coefficient there as \(c_j\Lambda_j\), with \(c_j=\Theta(\alpha_j/\Delta^2)\).  Saturating the constraint gives \(\epsilon_{\mathrm{amp}}=(\frac{\epsilon\sqrt{2\Delta}}{12\|\partial_j\rho\|_F}-\epsilon_j)/(c_j\Lambda_j)\), so the dominant amplitude-estimation factor is proportional to \(c_j\Lambda_j/(\frac{\epsilon\sqrt{2\Delta}}{12\|\partial_j\rho\|_F}-\epsilon_j)\).  Since \(d\Lambda_j/d\epsilon_j=-1/\epsilon_j\), its stationary point satisfies \(\frac{\epsilon\sqrt{2\Delta}}{12\|\partial_j\rho\|_F}=\epsilon_j(1+\Lambda_j)\).  The additional polylogarithmic factors in the per-call SLD cost do not change the resulting \(\Theta\)-scalings.  Equivalently,
\begin{equation}
\begin{aligned}
  \epsilon_j
  &=\Theta\!\left(
    \frac{\epsilon\sqrt{\Delta}}
    {\|\partial_j\rho\|_F\log(1/\epsilon)}
  \right),\\
  \epsilon_{\mathrm{amp}}
  &=\Theta\!\left(
    \frac{\epsilon\Delta^{3/2}}
    {\alpha_j\beta_j\|\partial_j\rho\|_F}
  \right).
  \label{eq:diag-explicit-optimal-params}
\end{aligned}
\end{equation}
Suppressing these logarithmic factors, the allocation takes the compact form
\begin{equation}
\begin{aligned}
\epsilon_j
&=\tilde\Theta\!\left(
\frac{\epsilon\sqrt{\Delta}}{\|\partial_j\rho\|_F}
\right),\\
\epsilon_{\mathrm{amp}}
&=\tilde\Theta\!\left(
\frac{\epsilon\Delta^{5/2}}
{\alpha_j\|\partial_j\rho\|_F}
\right),
\end{aligned}
\label{eq:diag-optimal-params}
\end{equation}
where the second relation uses \(\beta_j=\Theta(\Delta^{-1}\Lambda_j)\).  These choices give
\[
  E_j=\mathcal O\!\left(
  \frac{\epsilon\sqrt{\Delta}}{\|\partial_j\rho\|_F}
  \right).
\]
For completeness, a uniform allocation on the nontrivial branch is obtained by setting
\[
 \delta_j=\frac{\epsilon}{24B_j},\qquad
 \ell_j=1+\log\!\left(e+\frac{\alpha_j}{\delta_j\Delta}\right),
\]
and choosing \(\epsilon_j=\min\{\delta_j/\ell_j,\alpha_j\Delta/12\}\) and \(\epsilon_{\mathrm{amp}}=\min\{1/4,\delta_j/\alpha_{L_j}\}\). Then \(E_j\le2\delta_j=\epsilon/(12B_j)\), as required. These choices have the same high-precision scalings as above, up to logarithmic factors. An available upper bound on \(\|\partial_j\rho\|_F\) may replace the norm in both the branch test and the allocation.

It remains to substitute these precisions into the resource bounds.  Lemma~\ref{lem:qpe-free-additive-amplitude-estimation} gives
\begin{equation}
  Q_{\mathrm{amp}}
  =\widetilde{\mathcal O}(\epsilon_{\mathrm{amp}}^{-1})
  =\widetilde{\mathcal O}\!\left(
  \frac{\alpha_j\|\partial_j\rho\|_F}
  {\epsilon\Delta^{5/2}}
  \right).
  \label{eq:diag-amplitude-query-substitution}
\end{equation}
Each query to \(\mathcal W_j\) invokes one SLD block encoding and a constant number of additional purification operations.  Hence
\begin{equation}
\begin{aligned}
  Q_\rho^{jj}
  &=\mathcal O\!\left(
    Q_{\mathrm{amp}}(Q_\rho^{(j)}+1)
  \right),\\
  Q_{\partial_j}^{jj}
  &=\mathcal O\!\left(
    Q_{\mathrm{amp}}Q_{\partial_j}^{(j)}
  \right).
\end{aligned}
\label{eq:diag-query-composition}
\end{equation}
Using Proposition~\ref{prop:canonical-sld-block-encoding}, Eq.~\eqref{eq:diag-query-composition} becomes
\begin{equation}
\begin{aligned}
  Q_\rho^{jj}
  &=\widetilde{\mathcal O}\!\left(
  \frac{\alpha_j\|\partial_j\rho\|_F}
  {\epsilon\Delta^{7/2}}
  \right),\\
  Q_{\partial_j}^{jj}
  &=\widetilde{\mathcal O}\!\left(
  \frac{\alpha_j\|\partial_j\rho\|_F}
  {\epsilon\Delta^{5/2}}
  \right).
\end{aligned}
  \label{eq:app-diag-final-query-complexities}
\end{equation}
Each threshold-circuit query embeds \(\mathcal W_j\), so its total gate complexity is \(T_{\mathrm{amp}}=\mathcal O\!\left(Q_{\mathrm{amp}}(T_{L_j}+T_\rho+a_{L_j})\right)\).  Finally, the SLD block-encoding ancillas, the purification reference and system registers, and the additional QSVT ancilla give
\begin{equation}
\begin{aligned}
  a_{\mathrm{tot}}
  &=a_{L_j}+a+s+1\\
  &=3a+3s+m_j+3+\lceil\log_2K_j\rceil,
\end{aligned}
\label{eq:app-diag-total-qubits}
\end{equation}
which completes the proof of Theorem~\ref{thm:diagonal-qfi-estimation}.

The following elementary rank bound will be used below to pass between Frobenius-norm and operator-norm estimates without introducing unnecessary support factors.

\begin{lemma}[Rank bound of the state derivative]
\label{lemma:rank-bound-derivative}
Let \(\rho(\boldsymbol{x})\) be a density matrix of rank \(r\) on an \(N\)-dimensional Hilbert space. If the kernel condition \(P_{\mathcal{K}}\, \partial_j \rho(\boldsymbol{x})\, P_{\mathcal{K}} = 0\) of Eq.~\eqref{eq:condition-for-sld-to-be-well-defined} holds---so that the SLD is well defined---then the rank of the derivative operator is bounded by
\begin{equation}
    \operatorname{rank}(\partial_j \rho(\boldsymbol{x})) \le \min\{2r, N\} .
\end{equation}
\end{lemma}

\begin{proof}
We analyze the rank of \(\partial_j \rho(\boldsymbol{x})\) by expanding it in terms of the orthogonal projectors onto the support (\(P_{\mathcal{S}}\)) and the kernel (\(P_{\mathcal{K}}\)). Since \(P_{\mathcal{S}} + P_{\mathcal{K}} = \mathbb{I}\), we can expand the derivative as:
\begin{equation}
\begin{aligned}
    \partial_j \rho &= (P_{\mathcal{S}} + P_{\mathcal{K}}) \partial_j \rho (P_{\mathcal{S}} + P_{\mathcal{K}}) \\
    &= P_{\mathcal{S}} \partial_j \rho P_{\mathcal{S}} + P_{\mathcal{S}} \partial_j \rho P_{\mathcal{K}}
       + P_{\mathcal{K}} \partial_j \rho P_{\mathcal{S}} + P_{\mathcal{K}} \partial_j \rho P_{\mathcal{K}} .
\end{aligned}
\end{equation}
By the kernel condition required for a well-defined SLD (Eq.~\eqref{eq:condition-for-sld-to-be-well-defined}), the term entirely within the kernel vanishes (\(P_{\mathcal{K}} \partial_j \rho P_{\mathcal{K}} = 0\)). This allows us to factor the remaining terms as the sum of two operators:
\begin{equation}
    \partial_j \rho = P_{\mathcal{S}} \partial_j \rho + P_{\mathcal{K}} \partial_j \rho P_{\mathcal{S}} .
\end{equation}
By the subadditivity of the matrix rank (\(\text{rank}(A + B) \le \text{rank}(A) + \text{rank}(B)\)), we have:
\begin{equation}
    \text{rank}(\partial_j \rho) \le \text{rank}(P_{\mathcal{S}} \partial_j \rho) + \text{rank}(P_{\mathcal{K}} \partial_j \rho P_{\mathcal{S}}) .
\end{equation}
The rank of any matrix product is bounded by the minimum rank of its factors. Since \(P_{\mathcal{S}}\) projects onto the \(r\)-dimensional support \(\mathcal{S}\), \(\text{rank}(P_{\mathcal{S}}) = r\). Consequently:
\begin{equation}
    \text{rank}(P_{\mathcal{S}} \partial_j \rho) \le \text{rank}(P_{\mathcal{S}}) = r,
\end{equation}
and similarly,
\begin{equation}
    \text{rank}(P_{\mathcal{K}} \partial_j \rho P_{\mathcal{S}}) \le \text{rank}(P_{\mathcal{S}}) = r .
\end{equation}
Summing these yields \(\text{rank}(\partial_j \rho) \le 2r\). Because \(\partial_j \rho\) is an operator acting on the full \(N\)-dimensional Hilbert space, its rank trivially cannot exceed \(N\). Intersecting these two constraints yields the global bound \(\text{rank}(\partial_j \rho) \le \min\{2r, N\}\).
\end{proof}

\subsection{Full QGT and QFIM Block Encoding}\label{app:qgt-block-encoding-details}

We give the projected state evolution underlying Theorem~\ref{thm:qgt-block-encoding}. Write \(B=A_{L,R}\) and \(C=A_{L,L}\) for the two independent \(a_L\)-qubit SLD ancilla registers. For a matrix-input basis state \(\ket{j}_R\), the initial state is
\begin{equation}
  \ket{\Psi_0}
  =\ket{0}_X\ket{j}_R\ket{0}_B\ket{0}_C
   \ket{0}_S\ket{0}_A .
  \label{eq:app:qgt-initial-state}
\end{equation}
After applying \(O_\rho\) and the uniform superposition gate \(\mathcal D\),
\begin{equation}
  \ket{\Psi_1}
  =\frac{1}{\sqrt m}\sum_{i=1}^m
    \ket{i}_X\ket{j}_R\ket{0}_B\ket{0}_C\ket{\rho}_{SA}.
  \label{eq:app:qgt-after-preparation}
\end{equation}
The \(R\)-controlled multiplexor \(\mathcal U_R\) acts on \(BS\). Using the block-encoding relation for \(\widetilde L_j\), its component in the zero subspace of \(B\) is
\begin{equation}
\begin{aligned}
  &(\bra{0}_B\otimes I)\mathcal U_R\ket{\Psi_1}\\
  &\quad=\frac{1}{\alpha\sqrt m}\sum_{i=1}^m
    \ket{i}_X\ket{j}_R\ket{0}_C
    (\widetilde L_j\otimes I_A)\ket{\rho}_{SA}.
\end{aligned}
  \label{eq:app:qgt-after-right-sld}
\end{equation}
The second multiplexor \(\mathcal U_L\) is controlled by \(X\) and acts on the disjoint ancilla \(C\) together with \(S\). Projecting both SLD ancillas onto zero therefore gives
\begin{equation}
\begin{aligned}
  &(\bra{0}_B\bra{0}_C\otimes I)\mathcal U_L\mathcal U_R\ket{\Psi_1}\\
  &\quad=\frac{1}{\alpha^2\sqrt m}\sum_{i=1}^m
    \ket{i}_X\ket{j}_R
    (\widetilde L_i\widetilde L_j\otimes I_A)\ket{\rho}_{SA}.
\end{aligned}
  \label{eq:app:qgt-after-two-slds}
\end{equation}
The independence of \(B\) and \(C\) is essential here: garbage produced by the first block encoding remains orthogonal to \(\ket{0}_B\) and cannot be mapped back into the selected subspace by the second block encoding.

The register swap maps \(\ket{i}_X\ket{j}_R\) to \(\ket{j}_X\ket{i}_R\).
Applying \(\mathcal D^\dagger\) and selecting \(\bra{0}_X\) contributes the second factor \(1/\sqrt m\). Finally, \(\bra{0}_S\bra{0}_A O_\rho^\dagger=\bra{\rho}_{SA}\), and hence
\begin{equation}
\begin{aligned}
 &(\bra{0}_X\bra{0}_B\bra{0}_C\bra{0}_S\bra{0}_A\otimes I_R)
 U_{\mathcal T}\\
 &\quad\times
 (\ket{0}_X\ket{0}_B\ket{0}_C\ket{0}_S\ket{0}_A\otimes\ket{j}_R)\\
 &\qquad=\frac{1}{m\alpha^2}\sum_{i=1}^m
   \Tr(\rho\,\widetilde L_i\widetilde L_j)\ket{i}_R .
\end{aligned}
\label{eq:app:qgt-projected-block}
\end{equation}
For an unused input index \(j>m\), the swap places \(\ket{j}\) in register \(X\), and \(\bra{0}\mathcal D^\dagger\ket{j}=0\). Thus the selected block is \(\widetilde{\mathcal T}/(m\alpha^2)\), with zeros on the unused matrix-index subspace. The register accounting is the one stated in Theorem~\ref{thm:qgt-block-encoding},
\begin{equation}
  a_{\mathrm{tot}}=w+2a_L+a+s,
  \label{eq:app:qgt-total-ancillas}
\end{equation}
with separate \(a_L\)-qubit registers for the two SLD block encodings.
For each matrix element,
\begin{equation}
  |\widetilde{\mathcal T}_{ij}-\mathcal T_{ij}|
  \le\alpha(\epsilon_i+\epsilon_j)+\epsilon_i\epsilon_j,
  \label{eq:app:qgt-elementwise-error}
\end{equation}
and the Frobenius-to-spectral conversion gives \(\|\widetilde{\mathcal T}-\mathcal T\|\le\epsilon_{\mathrm{tot}}\).
Combining the encodings of \(\mathcal T\) and \(\mathcal T^\top\) as in Corollary~\ref{thm:qfim-block-encoding} preserves the established QFIM parameters \((4m\alpha^2,a_{\mathrm{tot}}+1,4\epsilon_{\mathrm{tot}})\).

\subsection{Bounded-Response Diagonal QFI Estimation}\label{app:regularized-estimation}

The bounded-gap construction of Section~\ref{estimation-of-the-norms} estimated QFI by combining the SLD block-encoding with the QPE-free amplitude estimation routine, under the bounded-gap promise.  We now obtain the corresponding result under the bounded-response promise, using the regularized SLD block-encoding of Section~\ref{sec:reg:be_construction}.  Since \(\mathcal{F}_{jj} = 4(\mathcal{N}_j^{(\theta)})^2\) up to a regularization bias of size \(8M\theta\), estimating QFI reduces to estimating the norm of the regularized SLD-applied state and correcting the bias.  The following parameters, carried over from the block-encoding construction, enter the estimation pipeline.

\bigskip\noindent \textit{Regularized SLD block-encoding parameters (Section~\ref{sec:reg:be_construction}).} Each \(U_{L_j^{(\theta)}}\) is an \((\alpha_{L_j^{(\theta)}}, a_{L_j^{(\theta)}}, \epsilon_j)\)-block-encoding of the regularized SLD \(L_j^{(\theta)}\):
\begin{equation}
  \| L_j^{(\theta)} - \alpha_{L_j^{(\theta)}} (\langle 0|^{\otimes a_{L_j^{(\theta)}}} \otimes I) U_{L_j^{(\theta)}} (|0\rangle^{\otimes a_{L_j^{(\theta)}}} \otimes I) \| \le \epsilon_j,
  \label{eq:reg:block-encoding-of-Lj}
\end{equation}
with
\begin{equation}
\begin{aligned}
  \alpha_{L_j^{(\theta)}} &= \mathcal{O}\!\left(\frac{\alpha_j}{\theta}\right), \\
  a_{L_j^{(\theta)}} &= 2a + 2s + 4 + m_j + \lceil \log_2 K \rceil, \\
  K &= \left\lceil \frac{1+\theta/2}{\theta} \log\!\Bigl(\frac{6\|\partial_j \rho\|_F}{\epsilon_j\,\theta}\Bigr) \right\rceil
      = \Theta\!\Bigl(\frac{1}{\theta}\log\frac{\|\partial_j \rho\|_F}{\epsilon_j\,\theta}\Bigr),
\end{aligned}
  \label{eq:reg:parameters-of-block-encoding-of-Lj}
\end{equation}
where \(\alpha_j\) is the subnormalization of the derivative block-encoding \(U_{\partial_j\rho}\).  Each application of the block-encoding requires
\begin{equation}
  Q_\rho^{(j)} = \widetilde{\mathcal{O}}\!\Bigl(\frac{1}{\theta^2}\Bigr),\qquad
  Q_{\partial_j} = \mathcal{O}(1),
  \label{eq:reg:query-complexity-per-call}
\end{equation}
queries to \(O_\rho\) and to \(U_{\partial_j\rho}\), respectively.

The following lemma bounds the norm of the regularized SLD-applied state in terms of the derivative Frobenius norm and the regularization parameter, which controls the success probability of all subsequent state-preparation and amplitude-estimation routines.

\begin{lemma}[Norm bounds for the regularized state]
  \label{lemma:reg:norm_bounds}
  Under the bounded-response promise as in Eq.~\eqref{eq:derivative-bound-intro}, the norm of the regularized SLD-applied state \(\ket{\psi_j^{(\theta)}} = (L_j^{(\theta)}\otimes I_A)\ket{\rho}_{SA}\) satisfies
  \begin{equation}
    \sqrt{\max\{0,\tfrac{1}{2}\|\partial_j \rho\|_F^2 - 2M\theta\}}
    \;\le\; \mathcal{N}_j^{(\theta)} \;\le\; \sqrt{M}.
    \label{eq:reg:N_bounds_qfi}
  \end{equation}
  The lower bound approaches the unregularized bound \(\mathcal{N}_j \ge \|\partial_j\rho\|_F/\sqrt{2}\) of Lemma~\ref{lemma:norm-bounds} as \(\theta \to 0\); the upper bound is independent of \(\theta\).
\end{lemma}

\begin{proof}
  Using the spectral expression~\eqref{eq:reg:L_theta}, \((\mathcal{N}_j^{(\theta)})^2 = \frac{1}{2}\sum_{p_m+p_n>0} |\langle\phi_m|\partial_j\rho|\phi_n\rangle|^2 (p_m+p_n)/(p_m+p_n+\theta)^2\).
  For the upper bound, expand
  \[
  \begin{aligned}
  \frac{p_m+p_n}{(p_m+p_n+\theta)^2} &= \frac{1}{p_m+p_n} - \frac{2\theta}{(p_m+p_n)^2} \\
  &\qquad \left(1+\frac{\xi}{p_m+p_n}\right)^{-3}
  \end{aligned}
  \]
  (\(0 < \xi < \theta\)).  Dropping the non-negative correction and using the symmetric inequality \(1/(p_m+p_n) \le \mathbf{1}_{p_m>0}/p_m + \mathbf{1}_{p_n>0}/p_n\) together with the Hermiticity of \(\partial_j\rho\), \((\mathcal{N}_j^{(\theta)})^2 \le \frac{1}{2}\sum_{p_m+p_n>0} \frac{|\langle\phi_m|\partial_j\rho|\phi_n\rangle|^2}{p_m+p_n} \le \|\rho^+\partial_j\rho\|_F^2 \le M\).
  For the lower bound, retaining the negative correction, \((\mathcal{N}_j^{(\theta)})^2 \ge \frac{1}{2}\sum |\langle\phi_m|\partial_j\rho|\phi_n\rangle|^2/(p_m+p_n) - \theta\sum |\langle\phi_m|\partial_j\rho|\phi_n\rangle|^2/(p_m+p_n)^2\).
  The first term equals \(\mathcal{N}_j^2 \ge \frac{1}{2}\|\partial_j\rho\|_F^2\) by Lemma~\ref{lemma:norm-bounds}; the second term is bounded by \(2M\). Combining these gives the lower bound in Eq.~\eqref{eq:reg:N_bounds_qfi}.
\end{proof}

The block-encoding of \(L_j^{(\theta)}\) gives direct access to an approximation of \(\ket{\psi_j^{(\theta)}}\).  Applying \(U_{L_j^{(\theta)}}\) to the purification of \(\rho\), the postselected state takes the form
\begin{equation}
\begin{split}
  \mathcal{W}_j^{(\theta)} |0\rangle^{\otimes a_{L_j^{(\theta)}}} |0\rangle_{SA}
  &= \gamma_j^{(\theta)} |0\rangle^{\otimes a_{L_j^{(\theta)}}} |\hat{\psi}_j^{(\theta)}\rangle \\
  &\quad + \sqrt{1 - (\gamma_j^{(\theta)})^2} |\perp\rangle,
\end{split}
  \label{eq:reg:normalized-state-after-block-encoding}
\end{equation}
where \(\gamma_j^{(\theta)} = \tilde{\mathcal{N}}_j^{(\theta)}/\alpha_{L_j^{(\theta)}}\) is the success amplitude, \(|\tilde{\mathcal{N}}_j^{(\theta)} - \mathcal{N}_j^{(\theta)}| \le \epsilon_j\) by the block-encoding error, and \(|\hat{\psi}_j^{(\theta)}\rangle\) is the normalized approximation to the target state. Lemma~\ref{lem:qpe-free-additive-amplitude-estimation} applies unchanged to this block-encoding-like relation: it returns an estimate \(\hat{\gamma}_j^{(\theta)}\) with \(|\hat{\gamma}_j^{(\theta)} - \gamma_j^{(\theta)}| \le \epsilon_{\mathrm{amp}}\) using \(\widetilde{\mathcal{O}}(1/\epsilon_{\mathrm{amp}})\) queries to \(\mathcal{W}_j^{(\theta)}\) and one additional ancilla qubit.

Propagating the two norm-estimation errors through \(\hat{\mathcal{F}}_{jj}^{(\theta)} = 4(\hat{\mathcal{N}}_j^{(\theta)})^2\) and using \(\mathcal{N}_j^{(\theta)} \le \sqrt{M}\) from Lemma~\ref{lemma:reg:norm_bounds} gives the implementation error, with \(E_j=\epsilon_j+\alpha_{L_j^{(\theta)}}\epsilon_{\mathrm{amp}}\),
\begin{equation}
  |\hat{\mathcal{F}}_{jj}^{(\theta)} - \mathcal{F}_{jj}^{(\theta)}|
  \le 8\sqrt{M}E_j+4E_j^2.
  \label{eq:reg:impl_error_qfi}
\end{equation}
Combining this with the regularization bias \(|\mathcal{F}_{jj}^{(\theta)} - \mathcal{F}_{jj}| \le 8M\theta\) (Eq.~\eqref{eq:reg:bias_bound}) yields the total error constraint
\begin{equation}
  8\sqrt{M}E_j+4E_j^2+8M\theta \;\le\; \epsilon.
  \label{eq:reg:total_error_qfi}
\end{equation}

The amplitude-estimation subroutine contributes a factor \(1/\epsilon_{\mathrm{amp}}\) to the total query count (Lemma~\ref{lem:qpe-free-additive-amplitude-estimation}), while the SLD block-encoding requires \(Q_\rho^{(j)} = \widetilde{\mathcal{O}}(1/\theta^2)\) queries to \(O_\rho\) per call (Eq.~\eqref{eq:reg:query-complexity-per-call}) and depends only logarithmically on \(\epsilon_j\).  The total query counts to the two oracles are therefore
\begin{equation}
  Q_\rho^{\mathrm{total}} = \frac{Q_\rho^{(j)}}{\epsilon_{\mathrm{amp}}}
    = \widetilde{\mathcal{O}}\!\Bigl(\frac{1}{\theta^2\,\epsilon_{\mathrm{amp}}}\Bigr),\qquad
  Q_{\partial_j}^{\mathrm{total}} = \widetilde{\mathcal{O}}\!\Bigl(\frac{1}{\epsilon_{\mathrm{amp}}}\Bigr),
  \label{eq:reg:total-query-counts}
\end{equation}
and the total number of qubits is \(3a + 3s + 5 + m_j + \lceil \log_2 K \rceil\), comprising the block-encoding ancilla register of Eq.~\eqref{eq:reg:parameters-of-block-encoding-of-Lj} together with the thresholding ancilla and the purification registers, as in Theorem~\ref{thm:diagonal-qfi-estimation}. Since \(\mathcal F_{jj}\le4M\), outputting zero suffices when \(\epsilon\ge4M\). Otherwise, choosing \(\theta=\epsilon/(16M)\), \(\epsilon_j=\epsilon/(64\sqrt M)\), and \(\epsilon_{\mathrm{amp}}=\epsilon/(64\alpha_{L_j^{(\theta)}}\sqrt M)\) satisfies Eq.~\eqref{eq:reg:total_error_qfi} and gives the allocation
\begin{equation}
  \theta = \Theta\!\Bigl(\frac{\epsilon}{M}\Bigr),\qquad
  \epsilon_j = \Theta\!\Bigl(\frac{\epsilon}{\sqrt{M}}\Bigr),\quad
  \epsilon_{\mathrm{amp}} = \Theta\!\Bigl(\frac{\epsilon^2}{\alpha_j M^{3/2}}\Bigr).
  \label{eq:reg:optimal_params_qfi}
\end{equation}
Substituting the allocation into Eq.~\eqref{eq:reg:total-query-counts} and into the qubit count yields the query complexities and circuit resources stated in Theorem~\ref{thm:reg:diagonal_qfi} in the main text. Since \(\mathcal{N}_j^{(\theta)} \to \mathcal{N}_j \ge \|\partial_j\rho\|_F/\sqrt{2}\) as \(\theta \to 0\) (Lemma~\ref{lemma:reg:norm_bounds}), the total estimation error remains small relative to the signal, \(E_j/\mathcal{N}_j^{(\theta)} = \Theta(\epsilon/\|\partial_j\rho\|_F^2) \to 0\) as \(\epsilon \to 0\).
\section{Physical Examples Supporting the Bounded-Response Promise}
\label{app:reg:physical}

We give structured examples satisfying the bounded-response promise under the conditions stated below, allowing exponentially small gaps, and a full commuting Gibbs example in which the promise fails.

\smallskip\noindent\textit{Example 1: Low-rank unitary encoding.}~%
Let \(\rho(x) = e^{-ixG} \rho_0\, e^{ixG}\) with a Hermitian generator \(G\) satisfying \(\|G\|=\mathcal{O}(\operatorname{poly}(s))\) and \(\rho_0 = \sum_{i=1}^{r} p_i \ket{\phi_i}\bra{\phi_i}\) of rank \(r=\mathcal{O}(\operatorname{poly}(s))\).
Since \(\partial_x \rho = -i[G,\rho]\) and the eigenbasis of \(\rho\) is the rotated basis \(\{e^{-ixG}\ket{\phi_i}\}\), the spectral form of \(M\) reads
\begin{equation}
  \|\rho^+ \partial_x \rho\|_F^2
  = \sum_{i=1}^{r} \sum_{n=1}^{N}
      |\langle \phi_i | G | \phi_n \rangle|^2
      \Bigl(\frac{p_n}{p_i} - 1\Bigr)^2,
  \label{eq:unitary-encoding-M}
\end{equation}
where \(p_n = 0\) for \(n > r\). For a pure state (\(r=1\)) this is the variance of \(G\) in \(\ket{\psi}\): \(\|\rho\,\partial_x\rho\|_F^2 = \langle\psi|G^2|\psi\rangle - \langle\psi|G|\psi\rangle^2 \le \|G\|^2\), which is \(\mathcal{O}(\operatorname{poly}(s))\) under the assumed generator-norm bound. For \(r>1\), suppose \(G\) approximately preserves the spectral shells of \(\rho_0\): there is a constant \(C\ge1\) such that \((p_n/p_i - 1)^2 \le C\) for every pair \((i,n)\) with \(\langle\phi_i|G|\phi_n\rangle \neq 0\) (matrix elements into the kernel of \(\rho_0\) contribute the factor \(1\)). Then
\begin{equation}
\begin{aligned}
  \|\rho^+ \partial_x \rho\|_F^2
  &\le C \sum_{i=1}^{r} \|G\ket{\phi_i}\|^2\\
  &\le C\, r\, \|G\|^2\\
  &= \mathcal{O}\bigl(r \cdot \operatorname{poly}(s)\bigr).
\end{aligned}
  \label{eq:unitary-encoding-M-rank-r}
\end{equation}
The gap, however, can be exponentially small: take \(r = \operatorname{poly}(s)\) eigenvalues \(p_i = e^{-\beta E_i}/Z\) spanning an energy range of width \(\Theta(s)/\beta\)---a thermal mixture on a polynomially large subspace, with \(\Delta = p_r = e^{-\Theta(s)}\)---and let \(G\) obey the stated spectral-shell condition.

\smallskip\noindent\textit{Example 2: Truncated commuting Gibbs ensemble.}~%
Let \(H(\mathbf{x}) = \sum_{j=1}^{m} x_j h_j\) with \([h_j, h_k] = 0\) and \(\|h_j\| \le 1\), written as \(H(\mathbf{x}) = \sum_z E_z(\mathbf{x}) \ket{\phi_z}\bra{\phi_z}\) in the common eigenbasis. Take \(\beta=\mathcal{O}(\operatorname{poly}(s))\) and truncate the ensemble to a fixed, parameter-independent window \(W\) of polynomially many configurations, \(|W| = r = \operatorname{poly}(s)\), spanning an energy range of width \(\Theta(s)/\beta\), and set \(\rho(\mathbf{x}) = P_W e^{-\beta H(\mathbf{x})} P_W / Z_W(\mathbf{x})\) with \(Z_W = \sum_{z \in W} e^{-\beta E_z(\mathbf{x})}\). Since the eigenbasis is parameter-independent and the state is supported on \(W\), \(\partial_j \rho\) is diagonal with \((\partial_j \rho)_{zz} = -\beta\, (E_{z,j} - \langle \partial_j H\rangle_\rho)\, p_z\), where \(E_{z,j} = \partial_j E_z = \langle \phi_z | h_j | \phi_z \rangle\), so that
\begin{equation}
\begin{aligned}
  \|\rho^+ \partial_j \rho\|_F^2
  &= \beta^2 \sum_{z \in W}
      \bigl(E_{z,j} - \langle \partial_j H\rangle_\rho\bigr)^2
  \\
  &\le 4\beta^2 r \|h_j\|^2\\
  &= \mathcal{O}(\operatorname{poly}(s)).
\end{aligned}
  \label{eq:truncated-commuting-M}
\end{equation}
At the same time the smallest positive eigenvalue is the smallest Boltzmann weight in the window, \(\Delta = e^{-\beta \max_{z\in W} E_z}/Z_W = e^{-\Theta(s)}\) by the width of \(W\). The corresponding QFI is \(\mathcal{F}_{jj} = \beta^2 \operatorname{Var}_\rho(E_{\cdot,j})\le\beta^2\).
For comparison, consider the full untruncated ensemble (\(W = \{0,1\}^s\)) with \(h_j = Z_1\) and fixed \(\beta>0\). The same computation gives \(\|\rho^+\partial_j\rho\|_F^2 = \beta^2 2^s (1 + \langle Z_1\rangle_\rho^2) = e^{\Theta(s)}\), which violates the bounded-response promise even though \(\mathcal{F}_{jj}=\beta^2\operatorname{Var}_\rho(Z_1)\le\beta^2\).

\smallskip\noindent\textit{Example 3: Product encoding with a structured probe.}~%
Let the parameter act only on a register \(A\), \(\rho(\mathbf{x}) = \rho_A(\mathbf{x}) \otimes \ket{\psi}\bra{\psi}_B\), with the environment register \(B\) frozen in a pure state. Then \(\rho^+ = \rho_A^+ \otimes \ket{\psi}\bra{\psi}\) and \(\partial_j \rho = (\partial_j \rho_A) \otimes \ket{\psi}\bra{\psi}\), so that
\begin{equation}
  \|\rho^+ \partial_j \rho\|_F^2
  = \|\rho_A^+ \partial_j \rho_A\|_F^2
  = \mathcal{O}(\operatorname{poly}(s)),
  \label{eq:product-encoding-M}
\end{equation}
whenever \(\rho_A(\mathbf{x})\) satisfies the promise, for instance under the conditions of Examples 1 and 2. The global gap is the gap of the probe, \(\Delta = \Delta_A\), which may be exponentially small.

\smallskip\noindent\textit{Compatibility with purified access.}~%
The examples are used under the access assumptions of Section~\ref{sec:access-model}. For a truncated commuting ensemble, a purification can be prepared by coherently loading the square roots of the Boltzmann weights on \(W\), provided those weights and the common eigenbasis states are efficiently accessible. Unitary encodings inherit a purification circuit when the encoding unitary and a purification circuit for \(\rho_0\) are available; product encodings append the fixed pure environment to a probe purification.

\section{Robustness of the Quadrature Construction}
\label{appendix:robustness}

This appendix records the stability of the construction under perturbations of the input state, complementing the robustness discussion in Section~\ref{sec:discussion}.  All notation follows the main text.

\begin{lemma}[Purification gauge invariance]
\label{lemma:purification-gauge-invariance}
Let \(O_\rho\) be any purification oracle of \(\rho\) as in Eq.~\eqref{eq:purification-oracle-intro}, and let \(U_\rho\) be the block encoding constructed from \(O_\rho\) by the construction of Eq.~\eqref{eq:block-encoding-of-rho-using-purification}.  Then the block \(\langle 0|_A U_\rho |0\rangle_A = \rho\) depends only on \(\rho\), and not on the choice of purification: for any two purifications \(\ket{\psi_1}, \ket{\psi_2} \in \mathcal{H}_S \otimes \mathcal{H}_A\) of the same \(\rho\), related by a unitary \(I_S \otimes U_A\) on the ancilla register, the constructed block encodings coincide.
Likewise, every measurement statistic consumed by the estimation layers is of the form \(\mathrm{Tr}[\rho M]\) for a system-register observable \(M\).
\end{lemma}

\begin{proof}
The SWAP construction evaluates the block through the partial trace over the ancilla register, and \(\mathrm{Tr}_A\) is invariant under \(I_S \otimes U_A\); hence the \(|0\rangle\)-block of \(U_\rho\) is unchanged by the ancilla gauge.
For the estimation layers, the purified-state contractions take the form \(\langle \rho|_{SA} (M \otimes I_A) |\rho\rangle_{SA} = \mathrm{Tr}[\rho M]\), which is again gauge-invariant.
\end{proof}

\begin{lemma}[Exponential operator Lipschitz]
\label{lemma:exponential-operator-lipschitz}
For positive semidefinite Hermitian \(\rho_1, \rho_2\) and any \(t \ge 0\),
\begin{equation}
  \bigl\| e^{-t\rho_1} - e^{-t\rho_2} \bigr\|
  \;\le\; t\, \|\rho_1 - \rho_2\| .
  \label{eq:exponential-lipschitz}
\end{equation}
\end{lemma}

\begin{proof}
Integrating the derivative of \(e^{-(t-u)\rho_1}e^{-u\rho_2}\) over \(u\in[0,t]\) gives
\[
 e^{-t\rho_1}-e^{-t\rho_2}
 =-\int_0^t e^{-(t-u)\rho_1}(\rho_1-\rho_2)e^{-u\rho_2}\,\mathrm du.
\]
Since \(\rho_1,\rho_2\succeq0\), both exponential factors have operator norm at most one. Taking norms therefore gives
\[
 \|e^{-t\rho_1}-e^{-t\rho_2}\|
 \le\int_0^t\|\rho_1-\rho_2\|\,\mathrm du
 =t\|\rho_1-\rho_2\|.
\]
\end{proof}

\begin{lemma}[Lipschitz stability of the quadrature approximation]
\label{lemma:quadrature-lipschitz}
Fix the parameter derivative \(\partial_j \rho\), and let \(\rho_1, \rho_2\) be positive semidefinite Hermitian operators.  For the gapped approximation of Eq.~\eqref{eq:approximate-solution-to-sld-in-matrix-form},
\begin{equation}
  \bigl\| \tilde{L}_j(\rho_1) - \tilde{L}_j(\rho_2) \bigr\|_F
  \;\le\; \frac{2\,\|\partial_j\rho\|_F\, S_1(K)}{\Delta^{2}} \cdot \|\rho_1 - \rho_2\|,
  \label{eq:lipschitz-gapped}
\end{equation}
with \(S_1(K) := \sum_{k=1}^{K} w_{k,K}\, e^{x_{k,K}}\, x_{k,K}\).  For the regularized approximation of Eq.~\eqref{eq:reg:GL_discrete_quad},
\begin{equation}
  \bigl\| \tilde{L}_j^{(\theta)}(\rho_1) - \tilde{L}_j^{(\theta)}(\rho_2) \bigr\|_F
  \;\le\; \frac{2\,\|\partial_j\rho\|_F}{\theta^{2}} \cdot \|\rho_1 - \rho_2\|.
  \label{eq:lipschitz-regularized}
\end{equation}
\end{lemma}

\begin{proof}
Write \(E_i^{(k)} = e^{-a_k \rho_i}\) with \(a_k = x_{k,K}/\lambda\), where \(\lambda = \Delta\) or \(\lambda = \theta\).
Telescoping each quadrature node,
\begin{equation}
\begin{aligned}
  E_1 \partial_j \rho E_1 - E_2 \partial_j \rho E_2
  &= (E_1 - E_2)\, \partial_j \rho E_1\\
  &\quad+ E_2\, \partial_j \rho\, (E_1 - E_2),
\end{aligned}
\end{equation}
and, using Lemma~\ref{lemma:exponential-operator-lipschitz} together with the contraction property \(\|E_i\| \le 1\) of positive semidefinite operators,
\begin{equation}
\begin{aligned}
  \|(E_1 - E_2)\, \partial_j \rho E_1\|_F
  &\le a_k\, \|\rho_1 - \rho_2\|\, \|\partial_j \rho\|_F,\\
  \|E_2\, \partial_j \rho\, (E_1 - E_2)\|_F
  &\le \|\partial_j \rho\|_F\, a_k\, \|\rho_1 - \rho_2\| .
\end{aligned}
\end{equation}
Each node therefore contributes at most \(2 a_k \|\partial_j \rho\|_F \|\rho_1 - \rho_2\|\).  In the gapped case the node coefficients are \(w_{k,K} e^{x_{k,K}}/\Delta\), giving Eq.~\eqref{eq:lipschitz-gapped} with the sum \(S_1(K)\).
In the regularized case the node coefficients are \(w_{k,K}/\theta\), and the resulting sum \(\sum_k w_{k,K} x_{k,K}\) equals \(\int_0^\infty x\, e^{-x}\, dx = 1\) exactly, by the exactness of Gauss--Laguerre quadrature on the linear moment, giving Eq.~\eqref{eq:lipschitz-regularized}.
\end{proof}


\bibliography{ref.bib}
\end{document}